\documentclass[preprint,preprintnumbers,amsmath,amssymb,nofootinbib]{revtex4-2}  % Typos 1 to 6 corrected Sept 2026
\usepackage{amsfonts}    % list packages between braces
\usepackage{amssymb}
\usepackage{amsmath}
\usepackage{latexsym}
\usepackage{eepic}
\usepackage{graphicx}
\usepackage[usenames,dvipsnames]{color}
\usepackage{color}

\allowdisplaybreaks

\usepackage[active]{srcltx}
\usepackage{epstopdf}

\newcommand{\al}{\alpha}
\newcommand{\pa}{\partial}

\newcommand{\veps}{\varepsilon}

\newcommand{\om}{\omega}

\newcommand{\rar}{\rightarrow}

\newcommand{\non}{\nonumber}

\usepackage{etoolbox}
\newcommand{\zerodisplayskips}{%
  \setlength{\abovedisplayskip}{2pt}%
  \setlength{\belowdisplayskip}{2pt}%
  \setlength{\abovedisplayshortskip}{2pt}%
  \setlength{\belowdisplayshortskip}{2pt}}
\appto{\normalsize}{\zerodisplayskips}
\appto{\small}{\zerodisplayskips}
\appto{\footnotesize}{\zerodisplayskips}

\begin{document}

\title{Tremblay-Turbiner-Winternitz (TTW) system at integer index $k$: polynomial algebras of integrals}
\author{Juan Carlos L\'opez Vieyra}
\email{vieyra@nucleares.unam.mx}
\author{Alexander~V.~Turbiner}
\email{turbiner@nucleares.unam.mx}
\affiliation{Instituto de Ciencias Nucleares, Universidad Nacional
Aut\'onoma de M\'exico, Apartado Postal 70-543, 04510 M\'exico,
D.F., Mexico{}}

%\date{January 12, 2025}
\date{\today}

\begin{abstract}
An infinite 3-parametric family of superintegrable and exactly-solvable quantum models 
on a plane, admitting separation of variables in polar coordinates,   
marked by integer index $k$ was introduced in \cite{TTW:2009} and was called in literature 
the TTW system. Its eigenenergies are linear in quantum numbers and correspond 
to 2D harmonic oscillator with frequency ratio $1:k$.
In this paper it is conjectured that the Hamiltonian and both integrals 
of the TTW system have a hidden algebra $g^{(k)}$ - this was checked for $k=1,2,3,4$ -
having a finite-dimensional representation spaces as the invariant subspaces.
It is checked that for $k=1,2,3,4$ the Hamiltonian $H$, its two integrals 
${\cal I}_{1}, {\cal I}_{2}$ and their commutator ${\cal I}_{12} = [{\cal I}_1,{\cal I}_2]$ 
are four generating elements of the polynomial algebra of integrals of the order $(k+1)$: 
$[{\cal I}_1,{\cal I}_{12}] = P_{k+1}(H, {\cal I}_{1}, {\cal I}_{2}, {\cal I}_{12})$, 
$[{\cal I}_2,{\cal I}_{12}] = Q_{k+1}(H, {\cal I}_{1}, {\cal I}_{2}, {\cal I}_{12})$, 
where $P_{k+1},Q_{k+1}$ are polynomials of degree $(k+1)$ written in terms of the 
ordered monomials of $H, {\cal I}_{1}, {\cal I}_{2}, {\cal I}_{12}$. This implies 
that the polynomial algebra of integrals is a subalgebra of $g^{(k)}$. 
It is conjectured that this is true for any integer $k$.
\end{abstract}

\maketitle

\clearpage

\section{Introduction}

The Hamiltonian for the two-dimensional TTW system \cite{TTW:2009} in polar coordinates 
 is given by
\begin{equation}
\label{HTTW}
 H_k (r,\varphi;\om, \al, \beta)\ =\ -\pa_r^2 -
 \frac{1}{r}\pa_r - \frac{1}{r^2}\pa_{\varphi}^2  + \om^2 r^2 +
 \frac{\al k^2}{r^2 \cos^2 {k \varphi}} + \frac{\beta k^2}{r^2 \sin^2 {k \varphi}}\ ,
\end{equation}
where $k$ and $\om > 0$, and $\al, \beta > - \frac{1}{4 k^2}$ are parameters, 
with domain $r \in [0, \infty)$ and $\varphi \in [0, \pi/k]$. 
It is evident that if the index $k$ in (\ref{HTTW}) is integer the Hamiltonian $H_k$ is invariant with respect to dihedral $I_{2k}$ group rotations. At $\om=\al=\beta=0$ the Hamiltonian 
$H_k$ degenerates to two-dimensional flat Laplacian.

The Hamiltonian (\ref{HTTW}) to the best of our knowledge includes {\it all} published superintegrable and exactly-solvable systems in a Euclidean plane $E_2$ that allow
a separation of variables in polar coordinates:
for $k=1$ this system coincides with the Smorodinsky-Winternitz system, for $k=2$ it corresponds to the so-called $BC_2$ rational model, while for $k=3$ it describes the Wolfes model \cite{Wolfes:1974}, or, equivalently, the so-called $G_2/I_6$ rational model in the Hamiltonian reduction method nomenclature, in particular, if $\al=0$ it reduces to the $A_2$/3-body rational/Calogero model \cite{Calogero:1969}. For general integer $k$ the Hamiltonian (\ref{HTTW}) corresponds to the $I_{2k}$ model in the Hamiltonian reduction method \cite{Ol-Per:1977,OP:1983}. Furthermore, for integer $k$, the algebraic and (Lie-)algebraic representation 
of the Hamiltonian (\ref{HTTW}) can be obtained by making a gauge rotation with the ground state function $\Psi_0$ as the gauge factor and taking into account the the ground state energy $E_0$,
\[h_k\ =\ \Psi_0^{-1} (H_k - E_0) \Psi_0\ ,\]  
\[\Psi_0 \ =\ r^{(a+b)k}\ \cos^a {k \varphi}\ \sin^b {k \varphi}\ e^{-\frac{\om r^2}{2}}\ ,\]
\[ E_0\ =\ 2\om [(a + b)k +1]\ ,\]
and the change of variables,
\begin{equation}
\label{coord}
t\ =\ r^2\ ,\ u\ =\ r^{2k} \sin^2 {k\varphi}\ ,
\end{equation}
where the coordinates (\ref{coord}) have a meaning of the invariants of the dihedral 
$I_{2k}$ group. The resulting gauge-rotated Hamiltonian $h_k$ in the coordinates (\ref{coord}) 
takes the algebraic form of the differential operator (without constant term) with polynomial coefficients:
\begin{align}
  h_k\ &=   -4 t \pa^2_t - 8k u \pa^2_{tu} - 4k^2 t^{k-1} u \pa^2_u
     \non \\ & \label{Lie}
    + 4[\om t - (a+b)k -1]\pa_t + [4\om k u - 2 k^2(2 b + 1)
    t^{k-1}]\pa_u \, ,
\end{align}
with eigenvalues,
\begin{equation}
\label{eps-n}
  \veps_{p,q}\ =\ 4\om (p + k q)\ ,
\end{equation}
see \cite{TTW:2009}, 
where $a,b$ are parameters related to $\al,\beta$ as $\al = a(a -1)$ and $\beta = b(b - 1)$ and $p,q=0,1,2,\ldots$ are quantum numbers. Hence, for integer $k$ the formal spectral problem 
for the TTW system is the exactly solvable with energies/eigenvalues linear in the quantum numbers (with vanishing lowest eigenvalue) and with polynomial in $(u,t)$ eigenfunctions. 
Formally, (\ref{eps-n}) corresponds to the spectrum of a two-dimensional anisotropic harmonic oscillator with frequency ratio $1:k$.

The resulting algebraic Hamiltonian $h_k$ preserves the space of polynomials,
\begin{equation}
\label{space_r}
 {\cal P}_{{\cal N}}^{(s)}\ =\ (t^{p} u^{q} | 0\leq (p + s q) \leq {\cal N})\ , \quad {\cal N} = 0,1,2,\ldots  \ ,
\end{equation}
for integer $s \geq (k-1)$ and any integer ${\cal N}$. Hence, for fixed $s \geq (k-1)$ 
the Hamiltonian $h_k$ has infinitely-many finite-dimensional invariant subspaces 
${\cal P}_{{\cal N}}^{(s)}$, which can be ordered by forming an {\it infinite flag},
\begin{equation}
\label{flag}
  {\cal P}_0^{(s)} \subset {\cal P}_1^{(s)} \subset {\cal P}_2^{(s)} \ldots {\cal P}_{\cal N}^{(s)} 
  \ldots\ \equiv\ {\cal P}^{(s)} .
\end{equation}
This implies the very high degeneracy of the operator ${h}_k$. 
In turn, each space ${\cal P}_{\cal N}^{(s)}$ is the representation space of 
the finite-dimensional  irreducible representation of the algebra $g^{(s)}$ (see \cite{TTW:2009,LT:2024})
\footnote{For $s=2$ this algebra was introduced in \cite{RTC:1998} and studied in \cite{Turbiner:2005,SIGMA:2013} in relation to the $G_2$ integrable system.}: 
the algebraic Hamiltonian can be written as a non-linear combination of the $g^{(s)}$ 
generators. It is worth noting that the eigenvalues (\ref{eps-n}) hint the hidden algebra 
of the TTW Hamiltonian can correspond to $s=k$. However, in reality, 
the hidden algebra of the TTW Hamiltonian is defined ambiguously: 
the algebraic Hamiltonian can be rewritten 
in terms of the generators of the algebra $g^{(s)}$ with $s \geq k-1$ \cite{TTW:2009}.

Since the TTW system (1) is superintegrable for integer $k$, see Kalnins-Kress-Miller  
\cite{Miller:2011} and references therein, there must exist two algebraically independent 
integrals, we denote them ${\cal I}_{1}, {\cal I}_{2}$. They have the form of differential 
operators of the 2nd degree for ${\cal I}_{1}$ and of the $(2k)$-th degree for ${\cal I}_{2}$ 
\footnote{It is important to emphasize that the proof of superintegrability presented 
in \cite{Miller:2011}, based on recurrence relations, did not imply 
that the degree of the 2nd integral ${\cal I}_{2}$ should be $2k$, even for $k=1$, 
which is minimal degree as conjectured in \cite{TTW:2009}. Note that the degrees 
of $({\cal I}_{1}, {\cal I}_{2})$ coincide with degrees/dimensions of the $I_{2k}$ invariants 
$(r,u)$, see (\ref{coord}). }. 
For $k=1,2,3,4$ these integrals were written explicitly in \cite{TTW:2009}: 
while the integral ${\cal I}_{1}$ appears to be sufficiently simple for any $k$, the complexity 
of the 2nd integral ${\cal I}_{2}$ grows dramatically with increasing $k$, 
see e.g. \cite{LT:2024}. A natural question arises about the 
 
\pagebreak 

\noindent
spectra of these integrals, 
their hidden algebras: what is their {\it common} hidden algebra. Another question is about
the existence of the algebra of integrals. Both questions will be addressed in this paper. 

We call {\it the polynomial algebra of integrals} the infinite-dimensional, 4-generated, associative 
algebra of ordered monomials
\begin{equation}
\label{PAI}
  H^n\,{\cal I}_1^m\,{\cal I}_2^p\, {\cal I}_{12}^q\ ,
\end{equation}
where $n, m, p, q$ are non-negative integers, with the generating elements 
$(H, {\cal I}_1, {\cal I}_2, {\cal I}_{12})$ obeying the following commutation relations:  
\[
 [H,{\cal I}_1] = [H,{\cal I}_2] = 0\ ,
\]
with
\[
  {\cal I}_2,{\cal I}_1 = -{\cal I}_{12} + {\cal I}_1,{\cal I}_2\ ,
\]
and
\[
  [H,{\cal I}_{12}]\ =\ 0\ .
\]
The ``double" commutators $[{\cal I}_1,{\cal I}_{12}]$ and  $[{\cal I}_2, {\cal I}_{12}]$ 
\begin{equation}
\label{I1-I12abst}
  [{\cal I}_1,{\cal I}_{12}]\ =\ P (H, {\cal I}_1,{\cal I}_2,{\cal I}_{12}) \ ,
\end{equation}
\begin{equation}
\label{I2-I12abst}
  [{\cal I}_2,{\cal I}_{12}]\ =\ Q (H, {\cal I}_1,{\cal I}_2,{\cal I}_{12}) \ ,
\end{equation}
are given by finite-degree polynomials made out of ordered monomials in 
$H, {\cal I}_1,{\cal I}_2,{\cal I}_{12}$. 
This algebra was defined in \cite{Turbiner:2024}.

By taking a planar superintegrable system (in the flat space) one can construct 
the representation theory of (\ref{PAI}) by realizing the generating elements 
$(H, {\cal I}_1, {\cal I}_2, {\cal I}_{12})$ as differential operators and by
identifying them as the Hamiltonian, two integrals and the commutator of integrals.
Such a realization of the polynomial algebra of integrals was proposed in \cite{Miller:2010}.
By calculating the double commutators $[{\cal I}_1,{\cal I}_{12}]$ (\ref{I1-I12abst}) and  
$[{\cal I}_2, {\cal I}_{12}]$ (\ref{I2-I12abst}) one
can find the finite-degree polynomials in $H, {\cal I}_1,{\cal I}_2,{\cal I}_{12}$,
\begin{equation}
\label{I1-I12}
  [{\cal I}_1,{\cal I}_{12}]\ =\ P_n (H, {\cal I}_1,{\cal I}_2,{\cal I}_{12}) \ ,
\end{equation}
\begin{equation}
\label{I2-I12}
  [{\cal I}_2,{\cal I}_{12}]\ =\ Q_m (H, {\cal I}_1,{\cal I}_2,{\cal I}_{12}) \ ,
\end{equation}
which specify a concrete polynomial algebra of integrals. It turns out that for many 
studied planar superintegrable systems the polynomials $P,Q$ are of the same degree, 
$m=n$. It has to be mentioned that in the case of physics systems the Hamiltonian $H$ 
and the integral ${\cal I}_1$, responsible for separation of variable, are defined uniquely.
The second integral ${\cal I}_2$ admits a certain ambiguity: 
(i) in its basic form it should be of the minimal degree differential operator and 
(ii) a combination of $H, {\cal I}_1$, which does not exceed the minimal degree, can be added. 
If ${\cal I}_2$ is of non-minimal degree, resulting polynomial algebra of integrals (\ref{PAI}) 
realized by differential operators appears incomplete. We assume that in this case 
the $P,Q$ polynomials are not of the minimal degree, at least, see \cite{Miller:2011} 
and a discussion in Conclusions.
It must be also emphasized that in four-dimensional phase space there exist at most 
three algebraically independent operators (or integrals of motion), 
where the Hamiltonian is included. 
It implies the existence of the syzygy: differential operators 
$(H, {\cal I}_1, {\cal I}_2, {\cal I}_{12})$ are algebraically/polynomially related! 

The final goal of this paper is to find the algebra of integrals for the TTW system 
for $k=1,2,3,4$ and to conjecture about it for higher $k$. Except for $k=1$\,, 
we anticipate that all calculations are cumbersome and very difficult computationally, 
they will be carried by using a specially designed MAPLE-18 code to tackle differential 
operators with polynomial coefficients. 
The obtained results will be crosschecked with MAPLE-2024 and Mathematica-13 
to ensure the absence of errors and bugs in the symbolic programs and to guarantee 
the coincidence of the final results. 
Note that this code was successfully applied in \cite{LT:2024} to the study 
of the integrability of the Wolfes model, equivalently, $G_2$ rational/$I_6$ model.  

\section{$g^{(s)}$ algebra}
\label{gk}

Around 1880 Sophus Lie studied the (Lie) algebras acting on the Euclidean plane $E_2$
realized as first order differential operators in two variables. He discovered that the non-semi-simple Lie algebra $g\ell(2,{\bf R}) \ltimes {\cal R}^{(s)}$ 
with integer index $s=1,2,3,\ldots$ acts on $E_2$ as the algebra of vector fields, 
see Case 24 in \cite{Lie:1880}.
In \cite{gko:1992,gko:1994} the Lie results were extended to the $g\ell(2,{\bf R})$ algebra of  first order differential operators,
\[
J^1\  =\  \pa_t \ ,
\]
\begin{equation}
\label{gl2r}
  J^2_N\  =\ t \pa_t\ -\ \frac{N}{3} \ ,
 \ J^3_N\  =\ s u\pa_u\ -\ \frac{N}{3}\ ,
\end{equation}
\[
    J^4_N\  =\ t^2 \pa_t \  +\ s t u \pa_u \ - \ N t\ ,
\]
where $N$ is a real number (the case of vector fields corresponds to $N=0$), and
\begin{equation}
\label{R}
   R_{i}\  = \ t^{i}\pa_u\ ,\ i=0,1,\dots, s\ ,\quad {\cal R}^{(s)}\equiv (R_{0},\ldots, R_{s})\ ,
\end{equation}
which spans the commutative algebra ${\cal R}^{(s)}$. For $s=1$ this algebra becomes 
the subalgebra of the algebra $s\ell(3)$. If $N$ is a non-negative integer, there exists a finite-dimensional representation of (\ref{gl2r}), (\ref{R}) i.e. ${\cal P}_{{\cal N}}^{(s)}$ (\ref{space_r}), where the algebra acts reducibly. 

In \cite{Turbiner:1998,TTW:2009} it was shown that adding to (\ref{gl2r}), (\ref{R}) a single higher order differential operator,
\begin{equation}
\label{grT}
 T^{(s)}_0\ =\ u\pa_{t}^s\ ,
\end{equation}
makes the action on ${\cal P}_{{\cal N}}^{(s)}$ irreducible. Furthermore, 
by taking multiple commutators 
\begin{equation}
\label{grT-subalgebra}
    T_i^{(s)}\ =\ \Bigr[ \underbrace{J^4,\bigr[J^4,[ \ldots , [J^4,T_0^{(s)}]]\bigr]}_i \Bigr] =
    u\pa_{t}^{s-i} J_0 (J_0+1)\ldots (J_0+i-1) \ ,
\end{equation}
with $i=1,\ldots s$,\ we generate the differential operators of the degree $s$ acting 
on the space ${\cal P}_{{\cal N}}^{(s)}$ (\ref{space_r}), see \cite{Turbiner:2010} and 
also \cite{ST:2013}. 
Here 
\begin{equation}
\label{j0}
 J_0=t\pa_t \  +\ s u\pa_u \ - \ N\ ,
\end{equation} 
is the Euler-Cartan generator (or number operator), which maps a monomial in $(t,u)$ to itself. Interestingly, there is a property of nilpotency,
\[
  T_i^{(s)} = 0\ ,\ i > s \ .
\]
The non-trivial $T$-operators span the commutative algebra,
\[
  [T_i^{(s)},T_j^{(s)}] = 0\ ,\quad i,j=0,\ldots s\ ,\quad {\cal T}^{(s)}\equiv (T_{0}^{(s)},\ldots, T_{s}^{(s)})\ ,
\]
of dimension $(s+1)$. Let us generate its structure in the form of the Gauss decomposition:
\[ 
{\cal T}^{(s)} \rtimes gl_2 \ltimes {\cal R}^{(s)}
\]
which corresponds to the following diagram
\begin{center}
%\fbox{
\begin{picture}(100,50)
%%%%%%%%%%%%%%%%%%%%%%%%%%%%%%%%
\put(50,0){\vector(1,0) {18}}
\put(50,0){\vector(-1,0){18}}
\put(50,0){\vector(0,1) {15}}
%%%%%%%%%%%%%%%%%%%%%%%%%%%%%%%%
\put(48,32){$g\ell_2$}
\put(30,28){\rotatebox{225}{\Large $\ltimes$}}
\put(68,23){\rotatebox{-45}{\Large $\ltimes$}}
\put(7,0){$\scriptstyle {\cal R}^{(s)}$}
\put(87,0){$\scriptstyle {\cal T}^{(s)}$}
\put(35,-17){${ P}_s{ (g\ell_2)}$}
%%%%%%%%%%%%%%%%%%%%%%%%%%%%%%%%%
\end{picture}
%}
\end{center}
where ${P}_s{ (g\ell_2)}$ is a polynomial of degree $s$ in the $g\ell_2$ generators.
The dimension of the structure is $(2s+6)$. 
For $s=1$ it is the true Gauss decomposition of the $s\ell(3)$ algebra. By definition the $g^{(s)}$ algebra is the infinite-dimensional,\ $(2s+6)$-generated associative algebra of differential operators with ${\cal P}_n^{(s)}$ as its finite-dimensional irreducible representation space. In particular, $g^{(1)} = U_{gl(3)}$.

\section{\bf Complete Integrability.}

The operator
\begin{equation}
\label{X_k}
   {\cal X}_k (\al, \beta)\ =\ -L_3^2 + \frac{\al k^2}{\cos^2 {k \varphi}} + \frac{\beta k^2}{\sin^2 {k \varphi}}\ ,
\end{equation}
where $L_3 = \pa_{\varphi}$ is the 2D angular momentum, is an integral of motion for the TTW \hbox{system~\cite{TTW:2009}}: $[H_k,{\cal X}_k]=0$ for any real $k$. This can be easily 
checked by direct calculation.
Its existence is directly related to the separation of variables of the Schr\"odinger 
equation in polar coordinates. 
Hence, the Hamiltonian (\ref{HTTW}) defines a completely-integrable system
for any real $k \neq 0$. The gauge rotated integral $x_k\ =\ \Psi_0^{-1} ({\cal X}_k - c_k) \Psi_0$ in variables $(t,u)$ takes 
the algebraic form
\begin{equation}
\label{X_k-alg}
    {x}_k = -4k^2 u (t^k - u) \pa_u^2 - 4k^2 [(b+\tfrac{1}{2})t^k - (a+b+1)u]\pa_u \ ,
\end{equation}
(without constant term), for integer $k$, and $\al = a(a -1)$ and $\beta = b(b - 1)$, 
here $c_k=k^2 (a+b)^2$ is the lowest eigenvalue of ${\cal X}_k$. The operator ${x}_k$ can be rewritten in terms of the $g^{(k)}$ generators (\ref{gl2r})-(\ref{j0}). Hence, its hidden 
algebra is  $g^{(k)}$. Note that at $\al=\beta=0$ the Integral ${\cal X}_k$  
degenerates to the angular momentum squared, $L_3^2$.

\section{TTW at $k=1$, or Smorodinsky-Winternitz system}
The TTW system at $k=1$ was found long ago in \cite{Fris:1965, Wint:1966}. It was named the Smorodinsky-Winternitz (SW) system \cite{evans:1990}. The Hamiltonian and the two second order integrals (meaning: the second order differential operators) for the SW system in 
$(t,u)$-variables are given by
\begingroup
\small
\begin{align}
\label{H-k=1}
  H \equiv h_1 &= -4\,t\, \pa^2_{t}  -8\,u \, \pa_t\pa_u
  - 4\,u \pa^{2}_{u}
  + 4 \left( \om\,t -  \,a -  \,b - 1 \right) \pa_{t}
  + 4\left( \om\,u - b - \tfrac{1}{2} \right) \pa_{u}  \ ,
\end{align}
\normalsize
\endgroup
\begin{equation}
  {\cal I}_1 = -4 u (t-u)
  \pa^2_{u}
  - 4 \left( \left(b+\tfrac{1}{2}\right) t-(a+b+1) u\right)
  \partial_{u}\ ,
\end{equation}
\begin{equation}
  {\cal  I}_2 = 4\, \left( t-u \right) \partial^2_{t}
  + 4\, \left( \omega\, \left( u-t \right) + a + \tfrac{1}{2}  \right) \partial_{t}\ .
\end{equation}
It is easy to check that the hidden algebra is $g^{(1)} = gl(3)$, see Section II, which implies that 
$H, {\cal  I}_{1}, {\cal  I}_{2}$ can be rewritten in the $gl(3)$-generators.
The commutator of the integrals
\begin{equation*}
\label{I12-k=1}
 {\cal I}_{12} \equiv [{\cal I}_1,{\cal I}_2]\ ,
\end{equation*}
is also an integral, which is a 3rd order differential operator with polynomial coefficients 
in $(t,u)$ coordinates, see \cite{TTW:2001}. Explicitly, it reads
\begingroup
\small
\begin{align}
\label{i12-k=1}
 {\cal I}_{12} &=
 32 u\,\left( t - \,u \right) \pa^{2}_{t}\pa_{u}
 + 32 u \left( t - \,u \right) \pa_t\pa^{2}_{u}
 %%\nonumber \\ &    %% Typo 1 (In maple the expression coincides with the correction) corrected here:
 - 8 \left( 2a\,u  - 2\,b  (t - u) - \,t + 2\,u \right)
 \partial^{2}_{t}
\nonumber  \\ &
 + 16 (t-u)\left( 1 + 2\,b -2\,\om\,u \right) \pa_t\pa_u
%% \non \\ &
 +8\, \left(
 1 + 2 a -2 \om ( t - u)
 \right) u
\pa^{2}_{u}
 \non \\ &
 %% -t was missing in the line below. Maple file O.K.
 + 8\, \left( 2\,u (a+1) - 2\,b (t-\,u) - t \right) \om\,
 \pa_{t}
 +4\, \left( 1+ 2\,a-2\,\om\,(t-u)  \right) \left( 2\,b+1 \right)
 \pa_{u}\ ,
\end{align}
\normalsize
\endgroup
it can also be rewritten in terms of the $gl(3)$-generators. 

\subsection{Polynomial algebra of integrals for TTW at $k=1$}

By construction
\[
 [H,{\cal I}_1] = [H,{\cal I}_2] = 0\ .
\]
The commutator of the integrals
\[
 [{\cal I}_1,{\cal I}_2] \equiv {\cal I}_{12}\ ,
\]
does not vanish, see (\ref{i12-k=1}), and it is an integral
\begin{equation*}
%\label{I12}
  [H,{\cal I}_{12}]\ =\ 0\ .
\end{equation*}
The double commutators are found to be of the form
\begin{align}
\label{poly112-k=1}
[{\cal I}_1, {\cal I}_{12}] &=
8\,  H  {\cal I}_{{1}}
+ 16\,{\cal I}_{{1}} {\cal I}_{{2}}
%%\nonumber \\ &
+ 8\,\left( 2\,a+1 \right) \left(a+b-1\right) H
+ 16\, \left( (a+b)^2 - 1\right)  {\cal I}_{{2}} - 8\,{\cal I}_{{12}} 
 \nonumber \\ &
 - 16\,\om\, \left( a-b \right) {\cal I}_{{1}}\ ,
\end{align}
where ${\cal I}_{12}$ enters to the rhs in the first degree, and
\begin{align}
\label{poly212-k=1}
 [{\cal I}_2,{\cal I}_{12}] &=
 - 8\,  H  {\cal I}_{{2}}
 - 8\,{{\cal I}_{{2}}}^{2}
 %% \nonumber \\ &
 + 8\,\omega\, \left[
 \left( 2\,a +1 \right)  H
 +2\, \left( a-b \right) {\cal I}_{{2}}
 \right]
 %% \nonumber \\ &
  -16\,{\omega}^{2}{\cal I}_{{1}} \ ,
\end{align}
cf. e.g. \cite{Post:2011}, Section 2.
Thus, the operators $H,{\cal I}_1,{\cal I}_2,{\cal I}_{12}$ are four generating elements 
of the polynomial algebra of degree 2 or quadratic polynomial algebra \cite{Miller:2010}. 
Syzygy is realized via a polynomial relation,
\begin{equation}
\label{syzygy}
  {\cal I}_{12}^2\ =\ {\cal R}_1(H, {\cal I}_{1}, {\cal I}_{2})  \ ,
\end{equation}
where ${\cal R}_1$ can be easily found, cf. e.g. \cite{Post:2011}, Section 2. If the integrals 
(\ref{H-k=1})-(\ref{i12-k=1}) are taken, the syzygy is given by the cubic polynomial
%%%%%%%%%%%%%%%
%\begingroup
%\footnotesize
\[
{\cal I}_{12}^2 \ =\ 4\,(2\,a+1)\,(2\,a-3)\,H^2 + 16\,{\cal I}_1 {\cal I}_2^2 + 16\,H {\cal I}_1 {\cal I}_2
   + 16\,(2 a^2 + 2 a b - a + b - 9)\,H {\cal I}_2
\]  % TTW k=1 Syzygy OK Checked
\[
 - 8\,H {\cal I}_{12} + 16\,((a+b)^2-9)\,{\cal I}_2^2 - 16\,{\cal I}_2 {\cal I}_{12} 
 - 16\,\om\,(2\,a+1)\,H {\cal I}_1 
\]
\[
 - 32\,\om\,(a-b)\,{\cal I}_1 {\cal I}_2 - 32\,\om\,(2 a+1)(2 b-3)\,H
 + 256\,\om\,(a-b)\,{\cal I}_2 + 16\,\om\,(a-b)\,{\cal I}_{12}
\]
\begin{equation}
\label{syzygy_k=1}  
  + 16\,\om^2\,{\cal I}_1^2
- 128\,\om^2\,{\cal I}_1 \ .
\end{equation}
%\endgroup
%%%%%%%%%%%%%%

It must be emphasized that the quadratic polynomial algebra remains for any values 
of the parameters $\om, a, b$ of the TTW model even though it loses its Hermiticity, see (1). 
For vanishing parameters $\om=a=b=0$, when TTW model corresponds to free motion - $H$ degenerates 
to 2D flat Laplacian - the polynomial algebra of integrals simplifies, 
\[
[{\cal I}_1, {\cal I}_{12}] = 8\,  H  {\cal I}_{1} + 16\,{\cal I}_{{1}} {\cal I}_{2}
- 8 H - 16 {\cal I}_{2} - 8\,{\cal I}_{{12}}\ ,
\]
\[
[{\cal I}_2,{\cal I}_{12}] = - 8\,  H  {\cal I}_{{2}} - 8\,{{\cal I}_{{2}}}^{2}\ ,
\]
but remain quadratic, the syzygy is given by a cubic polynomial,
\[
{\cal I}_{12}^2 \ =\ -12 \,H^2 +
  16\,H {\cal I}_1 {\cal I}_2 + 16\,{\cal I}_1 {\cal I}_2^2
   -144\,H {\cal I}_2 - 8\,H {\cal I}_{12} 
  - 144\,{\cal I}_2^2 - 16\,{\cal I}_2 {\cal I}_{12} 
\ .
\]

In general, the polynomial algebra of integrals is embedded into the universal enveloping algebra 
$U_{gl(3)} \equiv g^{(1)}$ being its subalgebra.

\section{TTW at $k=2$, or $BC_2$ rational model}

The TTW system at $k=2$ is equivalent to the $BC_2$ rational model of the Hamiltonian 
reduction \cite{OP:1983}, see \cite{TTW:2009}. The algebraic form of the Hamiltonian for this system in 
$(t,u)$ coordinates is
\begingroup
\small
\begin{align}
\label{H-k=2} 
  H \equiv h_2 &= -4\,t
  \partial^{2}_{t}
  -16\,u
  \pa_t\pa_u
  - 16\,tu
  \pa^{2}_{u}
%%\nonumber \\ & \hspace{20pt}
+ 4\left( \,\om\,t - 2\,a - 2\,b - 1 \right)
  \pa_{ t}
  \mbox{}+  8 \left( \,\om\,u-\, \left( 2\,b+1 \right) t \right)
  \pa_{ u} \,.
\end{align}
\endgroup
This system is characterized by two integrals: one of them is the second order 
differential operator (\ref{X_k}) at $k=2$, while the second integral is a 
4th order differential operator. The second order integral is given by
\begin{equation}
\label{i1-k2}
  {\cal I}_1\ =\ -16\,u \left( {t}^{2}-u \right)
  \pa^{2}_{u}
  -16\, \left(  \left( b+\tfrac{1}{2} \right) {t}^{2}- \left( a+b+1 \right) u \right)
  \pa_{ u} \ ,
\end{equation}
cf.(\ref{X_k-alg}). The 4th order integral  ${\cal I}_2$ (denoted $y_4$ in \cite{TTW:2009}), 
is given by 
\footnote{In the formula for $y_4$, see eq. (31) in \cite{TTW:2009}  there are two misprints in the coefficients: 
the overall factor in front of the second derivative with respect to variable $t$ should be +1 instead of +16, and 
the overall factor in front of the second derivative with respect to variables $t$ and $u$ should be  +4 instead of -4}:

\bigskip

\begingroup
\footnotesize
\begin{align}
\label{i2-k2}
   {\cal I}_2 &= 16 \left[ (t^2-u)\,\pa^4_{t}
     - 8 (t^2-u) u\,\pa^2_{t}\pa^2_{u} + 16 (t^2-u) u^2 \,\pa^4_{u}
     - 2 ( \om (t^2-u)-(2 a+1) t)\,\pa^3_{t}
\right.  
\non \\ &
-4 \left( (2 b+1) t^2- 2 (a+b+1) u\right)\,\pa^2_{t}\pa_u
     +8 u ( \om (t^2 - u) -(2 a+1) t)\,\pa_{t}\,\pa^2_{u}
\non \\ &
  +16 u ( (2 b+3) t^2- 2 (a+b+2) u)\, \pa^3_{u}
%%\non \\ &
 + ( \om^2 (t^2 - u) - 3 (2 a+1)  \omega  t+(2 a+1) (2 a+2 b+1))\,\pa^2_{t}
%%%% \nonumber
\\ &
 + 4  \left( \om\,(2 b+1)\,t^2 - (2 a+1) (2 b+1) t - 2 \om\,(a+b+1)\,u \right)\,\pa_{t} \pa_{u}
 \nonumber \\ &
 +4 \left((2 b+1) (2 b+3) t^2 + \om\,(2 a+1)\,t u - 2 (2 a^2+6 a b+2 b^2+8 a+7 b+5) u \right)\,\pa^2_{u}
 \non\\ & \left.
 +\om\,(2 a+1)\,(\om  t-2 a-2 b-1)\, \pa_{t}\ 
 +\ 2\,(2a+1)\,(2 b+1)\, (\om t-2 a-2 b-1)\,\pa_{u} \right]  \,. \non
\end{align}
\endgroup
%Hidden algebra of (23)-(25) is $g^{(2)}$.
%}
The commutator of integrals
\[
 {\cal I}_{12} \equiv [{\cal I}_1,{\cal I}_2]\ ,
\]
is also integral, which is a 5th order differential operator 
of the form
\begin{equation}
\label{i12-k=2}
  {\cal I}_{12} = \sum_{\substack{i,j=0 \\ i+j\leq 5}}^5 {C}^{(12)}_{ij}(t,u)
  {\frac {\pa^{i+j}}{\pa {t}^{i} {u}^{j}}} \,,
\end{equation}
where the coefficients $ {C}^{(12)}_{ij}(t,u;a,b,\om)$ are polynomials in $(t,u)$ variables,  
${C}^{(12)}_{00}=0$ and $\om,a,b$ are parameters.~In explicit form it is given by
the expression 
\begingroup
\footnotesize
\begin{align}
{\cal I}_{12} &=
512\,u \left( {t}^{2}-u \right)
\pa^4_{t}\pa_u  
%%%%
\, + \,  2048\,tu \left( {t}^{2}-u \right)
\pa^3_{t}\pa^2_{u}
\, - \,8192\,t{u}^{2} \left( {t}^{2}-u \right)
\pa_t\pa^4_{u}
%%%%
\nonumber \\ &
- 8192\,{u}^{3} \left( {t}^{2}-u \right)
\pa^5_{u}
 \,+\,128 \left((2 b+1) t^2-2 u (a+b+1)\right)
\pa^4_{t}
%%%%
%%%\nonumber \\ &
+1024\, \left( {t}^{2}-u \right)  \left( (2\,b+1)t -\omega\,u \right)
\pa^3_{t}\pa_{u}
\non \\ &
\,+\,3072\,u \left( -\omega\,{t}^{3}+ 2\left( a+1 \right) {t}^{2}+\omega\,tu-u \right)
\pa^2_{t}\pa^2_{u}
%%%%
%%\nonumber \\ &
+8192\,tu \Big( - \left( 2\,b+3 \right) {t}^{2}+ 2\left( a+b+2 \right) u \Big)
\pa_{t}\pa^3_{u}
%%%%
\nonumber \\ &
-2048\,{u}^{2} \Big( -2\,\om\,{t}^{3}+ \left( 4\,a+10\,b+29 \right) {t}^{2}+2\,\om\,tu 
- \left( 10\,(a+b) +32 \right) u \Big) \pa^4_{u}
%%%%
\non \\ &
 -256\,\om\, \Big(  \left( 2\,b+1 \right) {t}^{2} - 2 \left(  a+b+1 \right) u \Big)
\pa^3_{t}
%%%%
%%\nonumber  \\ &
%
+\Big( -1536\,\om\, \left( 2\,b+1 \right) {t}^{3}
+ 512 \left( {\om}^{2}u   + 6\, \left( 2\,b+1 \right)  \left( a+1 \right)  \right) {t}^{2}
\non  \\ &
 +1536\,\om\,\left( 2\,b+1 \right) ut -512\,{\om}^{2}{u}^{2}
 -1536 \left( 2\,b+1 \right) u \Big)\pa^2_{t}\pa_{u}
%%%%
%%%\nonumber \\ &  Typo T3 corrected here. - 6\,u\om\,\left( a+1 \right) {t}^{2}
 + 1024 \Big(\left({\om}^2 u - 2\, \left( 2\,b+3 \right)  \left(2\,b+1\right)\right) t^3
   \non \\& 
 - 6\,u\om\,\left( a+1 \right) {t}^{2}
 - \left( {\om}^{2}{u}^{2} - \left( 8\,{a}^{2}+24\,ab
 + 8\,{b}^{2}+40\,a+28\,b+24 \right) u \right) t + 3\,\om \,u^2 \Big) \pa_t \pa^2_{u}
\label{i12-k2}  \\ &
  + 2048\,u \Big(2\om\left(2\,b+3 \right) t^3 - 2 (3 + 2 b) (7 + 2 a + 2 b) t^2
%% \nonumber  \\ &
%% \mbox{}
  -4\om\left(a + b + 2\right) ut
\non \\ &
  + \left(57 + 8 a^2 + 20 a b + 8 b^2 + 46 (a + b) \right) u\Big)\pa^3_{u}
%% \nonumber \\ &
 +128\,{\om}^{2} \Big(  \left( 2\,b+1 \right) {t}^{2}- 2\left( a+b+1 \right) u \Big) \pa^2_{t}
  \non \\ &
 +512\,\left( 2\,b+1 \right)  \Big( {\om}^{2}{t}^{3} -6 \om \left(  a +1 \right) {t}^{2} +
  \left(8 a^2+8 a (b+2)+4 b-u \om^2+6 \right) t +3\,\om\,u \Big) \pa_t\pa_u
 \non \\ &
 + \Big(
    1024\,\omega\, \left( 2\,b+3 \right)  \left( 2\,b+1 \right) {t}^{3}
    + \left( 1024\,{\omega}^{2} \left( a+1 \right) u-512\, \left( 2\,b+3
 \right)  \left( 2\,b+1 \right)  \left( 4\,a+2\,b+9 \right)  \right) {
t}^{2}
 \nonumber  \\ &  
-2048\,\omega\, \left( 2\,{a}^{2}+6\,ab+2\,{b}^{2}+10\,a+7\,b+6
 \right) ut-512\,{\omega}^{2}{u}^{2}
  \nonumber \\&  %% Type T4 fixed. In agreement with Maple file
 +
 512 \left(a^2 (48 b+68)+8 a^3+4 a \left(12 b^2+42 b+35\right)+8 b^3+68 b^2+134 b+75\right)\,u 
 \Big)
       \pa^2_{u}
\nonumber  \\ &
+256\, \left( 2\,b+1 \right)  \Big(
 2 {\omega}^{2} \left( a+1  \right) {t}^{2} -{\omega}^{2}u
 - 2\omega \left( 4\,{a}^{2} + 4\,ab +8\,a + 2\,b + 3 \right) t   \nonumber \\& 
\mbox{}
+(2 a+1) \left(4 a^2+8 a (b+1)+4 b^2+8 b+3\right)
\Big)
\pa_{u}  \ , \non
\end{align}
\endgroup
which is the fifth order differential operator in $(t,u)$ variables. 
Straightforward analysis shows that eqs. (\ref{H-k=2})-(\ref{i12-k2}) can be rewritten in terms of $g^{(2)}$ algebra
generators, see Section II. This is the hidden algebra of the TTW at $k=2$ system.
%%%%

With explicitly known the Hamiltonian $H$ (\ref{H-k=2}) and 
the integrals ${\cal I}_{1},{\cal I}_{2},{\cal I}_{12}$ (\ref{i1-k2})-(\ref{i12-k2}) by making direct calculations
with specially designed MAPLE-18 code we obtain the following expressions for the double commutators
%%%
\begingroup
\footnotesize
\begin{align}
 [{\cal I}_1,{\cal I}_{12}]  &=
- 32\,{H}^{2}{\cal I}_{{1}}
%\nonumber  \\ &
-  128 \left(  2\,{a}^{2} + 2\,ab - \,a + \,b - 1\right) {H}^{2}
+ 64\,{\cal I}_{{1}}{\cal I}_{{2}}
%%  \\ &
+  256\left( \,(a+b)^2  - 1 \right) {\cal I}_{{2}}
- 32\,{\cal I}_{{12}}
\nonumber \\ &
- 128 \om \left[
 \left( 2\,(a + b) + \,1 \right)  H {\cal I}_{{1}}
+ 4 \left( 4a^2 (\,{a} + 2\,b) + \left( 4\,{b}^{2}+2\,b-3 \right) \,a+ \left( 2\,{b}^{2}-b-1 \right)\right)
H\right]
\nonumber \\ &
+  128\,{\om}^{2}\,\left[
{{\cal I}_{{1}}}^{2} + 4 \left( 2\,{a}^{2} + 2\,ab -\,1 \right) {\cal I}_{{1}}\right] \ ,
\label{poly112-k=2}
\end{align}
\endgroup
and
\begin{align}
\label{poly212-k=2}
 [{\cal I}_2,{\cal I}_{12}] &=
  32\,{H}^{2}{\cal I}_{{2}}
  -32\,{{\cal I}_{{2}}}^{2}
  %% \nonumber \\ &
  +128\,\omega\, \left( 2\,a+2\,b+1\right)   H  {\cal I}_{{2}}
    \\ &
  +128\,{\omega}^{2} \left[
    4 \left( 2\,a+1 \right)\left( a-1 \right) {H}^{2}
  - 2\,{\cal I}_{{1}}{\cal I}_{{2}}
  - 4\,  \left( 2\,{a}^{2}+2\,ab-1 \right) {\cal I}_{{2}}
 + {\cal I}_{{12}}
 \right]
\nonumber \\ &
+ 2048\,{\om}^{3} \left( 2\,a+1 \right)  \left( a-1 \right)  \left( 2\,a+2\,b+1 \right)  H
%%\\ &
- 4096\,{\om}^{4} \left( {a}^{2}-1 \right) {\cal I}_{{1}} \ ,
 \non
\end{align}
%%%
where ${\cal I}_{12}$ enters to the rhs of (31)-(32) linearly. 
There is a syzygy relation between the integrals (\ref{H-k=2})-(\ref{i12-k2}) 
given by
\[
   {\cal I}_{12}^2\ =
\]
\begingroup
\footnotesize
\begin{align*}  %%%%% OK Checked w maple file
% {\cal I}_{12}^2  & =
 &
 %%%%%%% 4th degree
   64\,\left( 2\,a+1 \right)\left( 2\,a-3 \right)\,H^4 - 64\,H^2\,{\cal I}_1\,{\cal I}_{2}
  %%%%%%% 3rd degree
  -256 \left(2 a^2 + 2 a b - a + b -9 \right) H^2 {\cal I}_{2} + 32\,H^2\,{\cal I}_{12} 
  \\ &
  + 64\,{\cal I}_1\,{\cal I}_{2}^2
  + 256\, \left( (a+b)^2 -9  \right)   {\cal I}_{2}^2 - 64\,{\cal I}_{2}  {\cal I}_{12}
  \\ &
  + 512\,\om\,\left( 2\,a+1\right)\left(2\,a-3 \right)\left(2\,a+2\,b+1\right)\,H^3 
  - 256\,\om\, \left( 2\,a+2\,b+1 \right) H {\cal I}_{1}  {\cal I}_{2}
  %\\ &
  \\ & 
- 1024\,{\om}^{2} \left( 2\,a+1 \right)  \left( a-1 \right) H^2 {\cal I}_{1} 
+ 256\,{\om}^{2}{\cal I}_{1}^2  {\cal I}_{2}
\\ &
  %%%%%%% 2nd degree
 +1024\,{\om}^{2} \left(2\,a+1\right) \left(8\,{a}^{3}+4\left( 4\,b - 1 \right)a^{2}+2\,\left( 2\,b-1 \right)
 \left(2\,b-3 \right) a-12\,b^2-20\,b-15 \right) H^2
\\ &
 -4096\,{\om}^{3} \left( 2\,a+1 \right) \left( a-1 \right)  \left( 2\,a+2\,b+1 \right) H {\cal I}_{1}
%%\\ &
 - 1024\,\om\, \left( 2\,a+2\,b+1 \right)  \left( 2\,{a}^{2}+2\,ab-a+b-9 \right) H {\cal I}_{2}
 \\ &
+ 128\,\om\, \left( 2\,a+2\,b+1 \right) H  {\cal I}_{12}
+ 1024\,{\om}^{2} \left( 2\,{a}^{2}+2\,a b-9 \right) {\cal I}_{1}  {\cal I}_{2}
- 256\,{\om}^{2}{\cal I}_{1}  {\cal I}_{12}
\\ &
+ 4096\,{\om}^{4} \left( a^2-1 \right)   {\cal I}_{1}^2
% \\ &
 %%%%%%% 1st degree
  +16384\,{\om}^{3} \left( 2\,a+1 \right)\left( 2\,a+2\,b+1 \right)\left( 4\,a-2\,b-3 \right) H
  \\&
   - 131072\,{\om}^{4} \left( a^2 - 1 \right)   {\cal I}_{1}
   - 32768\,{\om}^{2} \left( a^2-b^2 \right)   {\cal I}_{2}
   - 512\,{\om}^{2} \left( 2\,{a}^{2}+2\,ab-9 \right) {\cal I}_{12}\ ,
\end{align*}
\endgroup
which is the tenth order differential operator in $(t,u)$ variables, 
while it is the fourth degree polynomial in $(H,{\cal I}_1,{\cal I}_2,{\cal I}_{12})$.

Hence, we arrive at the cubic algebra of integrals generated by the operators 
$H,{\cal I}_1,{\cal I}_2,{\cal I}_{12}$~ 
%\footnote{Syzygy has the form (\ref{syzygy}), it is not shown}. 
This algebra is embedded into $g^{(2)}$ being its subalgebra. It must be emphasized that 
the cubic polynomial algebra of integrals remains for any values of the parameters 
$\om, a, b$ of the TTW model even if it loses its Hermiticity, see (1). 
%\flushbottom

\section{TTW $k=3$, or $G_2$ rational model}

The TTW system for $k=3$ is equivalent to the $G_2$ rational integrable model, see e.g. \cite{LT:2024}. The algebraic Hamiltonian of the TTW system for $k=3$ in $(u,t)$ 
coordinates (\ref{coord}) is given by
\begin{align}
\label{H-k3}
  {\cal H}\ \equiv \ h_3\ =\ &
-4\,t {\pa_t^{2}}
-24\,u {\frac {\pa ^{2}}{\pa u\pa t}}
-36\,{t}^{2}u{\frac {\pa ^{2}}{\pa {u}^{2}}} \\[3pt] &
+ 4 \left( \,\omega\,t - 3\,(a+b) - 1 \right) {\frac {\pa }{\pa t}}
 + 6 ( 2\om\,u - 3 (2 b+1) t^2)  {\frac {\pa }{\pa u}} \non \ ,
\end{align}
%%%%%%%%%%%%%%%  H formatted OK
%
which can be written in terms of the $g^{(2)}$-algebra generators,
and the first integral ${\cal I}_1$ (\ref{X_k}), (\ref{X_k-alg}),  which commutes with the Hamiltonian, is
\begin{equation}
{\cal I}_1\ =\  - 36\,u \left( {t}^{3}-u \right)
{\frac {\pa ^{2}}{\pa {u}^{2}}}
- 36\, \left(  \left( b+\frac{1}{2} \right) {t}^{3}
- \left( a+b+1 \right) u \right) \pa_u \ ,
\end{equation}
%%%%%%%%%%%%%%%  I1 formatted OK
which can be written in terms of the $g^{(3)}$-algebra generators, see Section II.

The second integral ${\cal I}_2$, commuting with the algebraic Hamiltonian $h_3$, should be 
the 6th order algebraic 
differential operator, presented in \cite{TTW:2009}, see also \cite{LT:2024} for the particular 
case $\om=0$ 
\footnote{The version of ${\cal I}_2$, which was found in \cite{TTW:2009} (therein called $y_6$), 
was printed with some typos; the  corrected form of $y_6$  was used in the present article to 
find ${\cal I}_2$ in the form (\ref{I2-k=3}).}.

\noindent
{\it Remark.} The integral ${\cal I}_2$ is defined ambiguously, up to the addition of terms 
${\cal H}^n {\cal I}_1^m$, $n+m\leq  3$, which do not change the order of the integral. 
By adding some of these terms we have found a form for ${\cal I}_2$ for which the double 
commutators $[{\cal I}_1,{\cal I}_{12}]$,  $[{\cal I}_2,{\cal I}_{12}]$ are represented 
as 4th degree polynomials in the ordered monomials of
${\cal H},{\cal I}_{1},{\cal I}_{2},{\cal I}_{12}$
(see below). 

We can write the integral ${\cal I}_2$ in the form
\begin{equation}
\label{I2-k=3}
 {\cal I}_{2} = \sum_{\substack{0\leq i,j \leq 6\\[3pt] i+j\leq 6 }}  C^{(2)}_{{i,j}}(t,u; a,b,\omega)
 {\frac {\pa^{i+j}}{\pa {t}^{i} {u}^{j}}}\ ,
\end{equation}
where the coefficient functions $C^{(2)}_{{i,j}}(t,u; a,b,\om)$, are polynomials in $(t,u)$ variables, $C^{(2)}_{0,0}=0$. The explicit form of these coefficients is given in Appendix \ref{I2k3coeffs}.

In turn, the integral
\[
{\cal I}_{12} \equiv [{\cal I}_1,{\cal I}_2]\,,
\]
is a 7th-order algebraic differential operator.
Unlike ${\cal I}_2$, the integral ${\cal I}_{12}$ is unchanged 
if extra terms in ${\cal H}, {\cal I}_1$ are added to ${\cal I}_2$.
This can be written as
\begin{equation}
\label{I12-k=3}
 {\cal I}_{12} = \sum_{\substack{0\leq i,j \leq 7\\[3pt] i+j\leq 7 }}  C^{(12)}_{{i,j}}(t, u; a,b,\omega)
 {\frac {\pa^{i+j}}{\pa {t}^{i} {u}^{j}}}\,,
\end{equation}
where the  coefficient functions $C^{(12)}_{{i,j}}(t,u; a,b,\om)$, are polynomials in $(t,u)$ variables with $C^{(12)}_{0,0}=0$, their explicit form is given in Appendix \ref{I12k3coeffs}.

The double commutator $[{\cal I}_1,{\cal I}_{12}]$ is quite complicated being a 8th order differential operator with polynomial coefficients. However, by taking ${\cal I}_2$ in the form (\ref{I2-k=3}) and ${\cal I}_{12}$ in the form (\ref{I12-k=3}), the double commutator is also represented (unexpectedly) as a 4th degree polynomial in the ordered monomials of ${\cal H},{\cal I}_1,{\cal I}_2,{\cal I}_{12}$:
\begingroup
\footnotesize
\begin{align*}
[{\cal I}_1,{\cal I}_{12}] &=
72\,{\cal H}^{3}   {\cal I}_1
\,+\, 648\,  \left( 2\,a+1 \right) \left(a+b-1 \right) {\cal H}^{3} + 144\,  {\cal I}_1   \,{\cal I}_2
+ 1296\, \left( (a+b)^2 - 1 \right)   {\cal I}_2
\,-\, 72\,{\cal I}_{12}
\\ &
%%% omega
\,+\, 432\,\om\, \left( 3\,(a+b) + 1 \right) \, \Big[ {\cal H}^{2} {\cal I}_1
%%\\ &
+  9\, \left( 2\,a+1 \right)  \left( a+b-1 \right) {\cal H}^{2}
\Big]
\\ &
%%% omega^2
\,-\, 144\,{\om}^{2} \Big[  3\, H {{\cal I}_1}^{2}
%%\\ &
+  \left(19\,{a}^{2}+54\,a b - 81 b^{2}-28\,a + 72\,b-23 \right)  {\cal H}   {\cal I}_1
\\ &
\hspace{20pt} - 18\, \left( a+b-1 \right)
\Big(31\,{a}^{3}+31\,{a}^{2}b+27\,a{b}^{2}+27\,{b}^{3}
+ 36\,{a}^{2}-22\,a b +11\,a-37\,b-2 \Big)
{\cal H} \Big]
%%%\\ &
\\ &
\,-\, 1728\,{\om}^{3 } \left( 3\,(a+b) +1 \right)
\Big[
{{\cal I}_1}^{2}
 %%\\ &
+ \Big( 12\,{a}^{2}+18\,a b + 6\,{ b}^{2}-a+b-9 \Big) {\cal I}_1
\Big ] \ .
\end{align*}
\endgroup
%%%
In a similar way the second double commutator $[{\cal I}_2,{\cal I}_{12}]$, which is 
quite complicated, 
being now a 12th order differential operator, can be written as a 4th degree polynomial 
of the ordered monomials in ${\cal H},{\cal I}_1,{\cal I}_2,{\cal I}_{12}$ (also quite unexpectedly): 
%%%
\begingroup
 \footnotesize
 \begin{align*}
[{\cal I}_2,{\cal I}_{12}] &=
- 72\,{\cal H}^{3}  {\cal I}_2 - 72 \,{{\cal I}_2 }^{2}
\\[3pt] &
%%%% omega
- 432\,\om\, \Big( 3\,(a+b)+1 \Big) {\cal H}^{2} {\cal I}_2
\\[3pt] &
%%%% omega^2
+ 72\,{\om}^{2} \, \Big[
\Big( 100\,{a}^{2}+54\,a b -54\,{b}^{2}-73\,a+81\,b-52 \Big) {\cal H}^{4} + 12 {\cal H}  {\cal I}_1 \,  {\cal I}_2
\\ &
\hspace{40pt} + 2 \left( 19\,{a}^{2}+54\,a b - 81\,{b}^{2}-28\,a+72\,b-23 \right) {\cal H} \, {\cal I}_2
- 6   {\cal H}  {\cal I}_{12}
\Big]
\\[3pt] &
%%%% omega^3
+ 432\,{\om}^{3}\, \Big( 3\,(a+b)+1 \Big) \Big[ 2 {\cal H}^{3}\, {\cal I}_1
%%%\\ &
+\Big( 148\,{a}^{2}+66\,a b -54\,{b}^{2}-107\,a+87\,b-81 \Big) {\cal H}^{3}
\\ &
+ 8  {\cal I}_1  \,  {\cal I}_2
+ 4 \Big( 12\,{a }^{2}+18\,a b +6\,{b}^{2}-a+b-9 \Big)  {\cal I}_2 - 4\,   {\cal I}_{12}
\Big]
\\[3pt] &
%%% omega^4
+ 72\,{ \om}^{4}   \Big[  9 \,{\cal H}^{2}{{\cal I}_1}^{2}
+ 6 \, \Big( 25\,{a}^{2}+162\,a b +225\,{b} ^{2}+128\,a-72\,b+163 \Big) {\cal H}^{2} \, {\cal I}_1
   \\ &
+ 4 \, \Big( 5438\,{a}^{4}+11394\,{a}^{3}b+7452\,{a}^{2}{b}^{2}+3402\,a{b}^ {3}
- 1458\,{b}^{4}-139\,{a}^{3}-1953\,{a}^{2}b
 \\ &
\hspace{20pt} -3969\,a{b}^{2}+3645\,{b} ^{3}-4824\,{a}^{2}
-4986\,a b  - 4320\,{b}^{2}-1879\,a-423\,b-48 \Big)
{\cal H}^{2}
\Big]
   \\[3pt] &
%%% omega^5
- 1728 \,{\om}^{5} \,\Big( 3\,(a+b)+1 \Big) \Big[3\, {\cal H} {{\cal I}_1}^{2}
 + 2\,  \Big( 46\,{a}^{2 }+18\,a b -72\,{b}^{2}-43\,a+63\,b-91 \Big) {\cal H} {\cal I}_1
\\ &
- 2\, \Big( 372\,{a}^{4}+558\,{a}^{3}b+510\,{a }^{2}{b}^{2}+486\,a{b}^{3}+162\,{b}^{4}
 +29\,{a}^{3} -203\,{a}^{2}b-483 \,a{b}^{2}
   \\ &
 \hspace{20pt}
 - 135\,{b}^{3}-212\,{a}^{2}+332\,a b
 - 330\,{b}^{2}+45\,a+473\,b +82 \Big) {\cal H}
 \Big]
\\[3pt] &
%%% omega^6
- 1152\,{\om}^{6} \, \Big[ 4\, {{\cal I}_1}^{3}
+  3 \Big( 72\,{a}^{2}+180\, a b +72\,{b}^{2}+54\,a+54\,b+73 \Big) {{\cal I}_1}^{2}
\\ &
+ 2\, \Big( 1296\,{a}^{4}+3888\,{a}^{3}b+4212\, {a}^{2}{b}^{2}+1944\,a{b}^{3}+324\,{b}^{4} + 648\,{a}^{3}+1620\,{a}^{2}b
\\ &
\hspace{20pt} +1296\,a{b}^{2}+324\,{b}^{3}-639\,{a}^{2}-270\,a b
- 531\,{b}^{2}-504\,a- 468\,b+271 \Big) \,  {\cal I}_1 \Big] \ .
\end{align*}
\endgroup
%%%
Both double commutators depend on ${\cal I}_{12}$ linearly. 

Hence, we arrive at the quartic algebra of integrals generated by the operators 
$H,{\cal I}_1,{\cal I}_2,{\cal I}_{12}$. Syzygy has the form (\ref{syzygy}), 
being 14th order differential operator, 
\begingroup
\footnotesize
\begin{align}
{\cal I}_{12}^2 & = 324 (2 a+1) (2 a-3) H^6 +144 H^3 {\cal I}_1 {\cal I}_2 -72 H^3\,{\cal I}_{12} +144 {\cal I}_1 {\cal I}_2^2
 +3888 \om\, (2 a+1) (2 a-3) (3 (a+b)+1) H^5  
\non \\ &
 -144 \om^2 \Big(2 a^2\big(-27 b^2 +27 b +50\big)
 + a\big(54 b -73\big) + \big(81 b -52\big)\Big)\,H^4 {\cal I}_1 - 864 \om^3 (3 (a+b)+1) H^3 {\cal I}_1^2
\non \\ &
 - 432 \om^4 H^2 {\cal I}_1^3
%%%%%%%%%%%%%%%%%%%%%%%%%%%
 + 1296 \om^2 \Big(448\, a^4 + a^3\big(386 b -75\big) + a^2\big(432 b^2 -674 b -211\big)   
\non \\
&\qquad
 + a\big(54 b^3 -351 b^2 -267 b -247\big)+ \big(54 b^3 -891 b^2 +430 b -165\big)\Big)\,H^4 
\non\\&
 -864 \om^3 (3 (a+b)+1) \Big(a^2\big( -54 b^2 +66 b +148\big)
 - a\big(54 b^2 - 66 b +107\big) - \big(54 b^2 -87 b +153\big)\Big)\,H^3 {\cal I}_1 
\non \\ &
 + 1296 \Big(2a^2+2ab -a + b -9\Big)\,H^3 {\cal I}_2
 - 432 \om^4 \Big(25 a^2 + a\big(162 b +128\big)
 + \big(225 b^2 -72 b +91\big) \Big)H^2 {\cal I}_1^2 
\non \\ &
 +864 \om\, (3 (a+b)+1) H^2 {\cal I}_1 {\cal I}_2
 +3456 \om^5 (3 (a+b)+1) H {\cal I}_1^3 -864 \om^2 H  {\cal I}_1^2 {\cal I}_2
 +2304 \om^6 {\cal I}_1^4 \non \\ &
%%%%%%%%%%%%%%%%%%%%%%%%%%%
 +15552 \om^3 (3(a + b)+1) \Big(62 a^4 + a^3\big(62b -21\big) + a^2\big(54 b^2 -109 b +160\big)  
 + a\big(54 b^3 -177 b^2 +44 b -187\big)
\non \\ &\quad
 + \big(54 b^3 -106 b^2 +91 b -102\big)\Big)\,H^3 -576 \om^4 \Big(5438a^4
 + a^3\big(11394 b -139\big) + 9 a^2\big(828 b^2 -217\big) 
\non \\ &\qquad
+ a\big(9 b\big(63 b(-7+6 b)-1418\big)-8791\big)
 -9 b(-385 + 3 b(592 + 27 b(-5 + 2 b))) -9336\Big)\,H^2 {\cal I}_1
\non \\ &
 + 7776 \om\, (3 (a+b)+1) \Big(2 a^2 + a\big(2 b -1\big) + \big(b -9\big)
 \Big) H^2 {\cal I}_2 -432 \omega\, (3 (a+b)+1) H^2 {\cal I}_{12} 
\non \\ &
 +3456 \omega^5 (3 (a+b)+1) \Big(46 a^2 + a\big(18 b -43\big)
 + \big(-72 b^2 +63 b -163\big)\Big) H {\cal I}_1^2
\non \\ &
-288 \om^2 \Big(19 a^2 + a\big(54 b -28\big) + \big(-81 b^2 +72 b -239\big)
\Big) H {\cal I}_1 {\cal I}_2
+864 \om^2 H {\cal I}_1 {\cal I}_{12}
\non \\ &
 + 2304 \om^6 \Big(72 a^2 + a\big(18 +180 b\big)
 + \big(72 b^2 +54 b +25\big)\Big) {\cal I}_1^3
 - 3456 \om^3 (3 (a+b)+1) {\cal I}_1^2 {\cal I}_2
%%%%%%%%%%%%%%%%%%%%%%%%%%%
\non \\ &
 +5184 \om^4 \Big(961 a^6 + a^5\cdot 62(5 + 31 b) + a^4\big(12916 + 527 b(-2 + 5 b)\big)
 + a^3\big(-1642 + 2 b(4419)    
\non \\ & 
 + 2 b^2(-692 + 837 b)\big) + a^2\big(-11969 + b(-10578 + b(-3916 + 27 b(-96 + 89 b)))\big) 
\non \\ &\qquad
 + a\big( -2364 + 2 b(-4493) + 2 b^2(-14981) + 54 b^3(66 + b(-49 + 27 b))\big)
\non \\ &\qquad
 + \big(b(-3868 + b(-19313 + 27 b(1146 + b(-668 + 27(-2 + b) b)))) + 348\big)\Big) H^2
\non \\ &
 -6912 \omega^5 (3 (a+ b)+1) \Big(372 a^4 + a^3\big(29 + 558 b\big)
 + a^2\big(510 b^2 -203 b +1120\big)  
\non \\&\qquad
 + a\big(486 b^3 -483 b^2 +332 b -1503\big)
 + \big(162 b^4 -135 b^3 -3246 b^2 +2741 b -2870\big)
\Big) H {\cal I}_1 
\non \\ &
 +5184 \omega^2 \Big(31 a^4 + a^3\big(62 b +5\big)
 + a^2\big(58 b^2 -17 b -57\big)  + a\big(54 b^3 -49 b^2 -4 b -125\big) 
\non \\ &
+ \big(27 b^4 -27 b^3 -469 b^2 +323 b +18\big)
\Big) H {\cal I}_2 +144 \om^2 \Big(19 a^2 + a\big(54 b -28\big) - \big(81 b^2 -72 b +239\big)
\Big) H {\cal I}_{12} 
\non \\ &
+2304 \om^6 \Big(1296 a^4 + a^3\big(3888 b +648\big)
+ a^2\big(4212 b^2 +1620 b -4527\big)  + a\big(1944 b^3 +1296 b^2 -10638 b -4392\big)
\non \\ &
 + \big(324 b^4 +324 b^3 -4419 b^2 -4356 b -8009\big)\Big) {\cal I}_1^2 
 - 3456 \om^3 (3 (a+b)+1) \Big(12 a^2 + a\big(18 b -1\big)
\non \\ &
+ \big(6 b^2 + b -81\big)
\Big) {\cal I}_1 {\cal I}_2
+3456 \omega^3 (3 (a+b)+1) {\cal I}_1 {\cal I}_{12}
+1296 (a+b+3) (a+b-3) {\cal I}_2^2
-144 {\cal I}_2 {\cal I}_{12}
%%%%%%%%%%%%%%%%%%%%%%%%%%%
\non\\ &
+497664 \om^5 (3 (a+b)+1) \Big(93 a^4 + a^3\big(-16 b^2 -146 a^2 b -81 b^4\text{)}\big)
\non \\ &\qquad
- a^2\big(12 b^2 - 208 b + 35 \big) + a\big(106 + 208 b -138 b^2\big)
+ \big(-81 b^4 +108 b^3 +102 b +48\big)
\Big) H 
\non \\ &
-331776 \om^6 \Big(324 a^4 + a^3\big(486 b +81\big)
- a^2\big(162 b^2 -81 b +108\big)
+ a\big(-486 b^3 -81 b^2 +450 b -9\big)
\non \\ &
- \big(162 b^4 +81 b^3 +54 b^2 -9 b -248\big)
\Big) {\cal I}_1
+248832 \om^3 (3 (a+b) -1 ) (3 (a+ b)+1) (a-b) {\cal I}_2
\non \\ &
+1728 \omega^3 (3 (a+b)+1) \Big(12 a^2 + a\big(18 b -1\big)
+ \big(6 b^2 + b -81\big)\Big) {\cal I}_{12}\ ,
\end{align}
\endgroup
it is the sixth degree polynomial in $(H,{\cal I}_1,{\cal I}_2,{\cal I}_{12})$. 
At $\om=0$ the syzygy dramatically simplifies, its explicit expression reads
\begin{align}
{\cal I}_{12}^2 & =  324 (2a+1)(2a-3) H^6  + 144  H^3 {\cal I}_1 {\cal I}_2 
   - 72  H^3 {\cal I}_{12} + 1296 (2a^2 + 2 a b -a + b -9)H^3 {\cal I}_2     
\non \\ &
 + 144\,{\cal I}_1\,{\cal I}_2^2 - 144\,{\cal I}_2\,{\cal I}_{12}   
 + 1296\,((a+b)^2 - 9)\,{\cal I}_2^2\ ,
\end{align}
but it remains the sixth degree polynomial.

 This algebra of integrals is embedded into $g^{(3)}$ being its subalgebra. 
It must be emphasized that the quartic polynomial algebra of integrals 
remains for any values of the parameters $\om, a, b$ of the TTW model at $k=3$ even if it loses 
its Hermiticity, see (1). 

%%%%%%%%%%%%%%%%%  

\section{TTW at $k=4$, or $I_8$ rational model}

The algebraic Hamiltonian of the TTW system (3) for $k=4$ in $(u,t)$ coordinates is given by
\begin{align}
{\cal H}\ \equiv\ h_4\ =\ &
-4\,t{\frac {\pa ^{2}}{\pa {t}^{2}}}
-32\,u{\frac {\pa ^{2}}{\pa u\pa t}}
-64\,{t}^{3}u{\frac {\pa ^{2}}{\pa {u}^{2}}} \\ &
+ 4\left( \,\om\,t - 4\,(a+b)-1 \right) {
\frac {\pa }{\pa t}}  + 16 \left( \,\om\,u - 2\, \left( 2\,b+1 \right) {t}^{3} \right) {\frac {\pa
}{\pa u}} \nonumber \,,
\end{align}
which can be written in terms of the $g^{(3)}$-algebra generators.
The first integral ${\cal I}_1$, which commutes with the Hamiltonian, is
\begin{equation}
\label{I1-k=4}
{\cal I}_1 =  -64\,u \left( {t}^{4}-u \right)
{\frac {\pa^{2}}{\pa {u}^{2}}}
- 64\, \left(  \left( b+\frac{1}{2} \right) {t}^{4}
- \left( a+b+1 \right) u \right)\,{\frac {\pa}{\pa u}}
 \ ,
\end{equation}
which can be written in terms of the $g^{(4)}$-algebra generators.
This integral appears as the result of the separation of variables of the original 
TTW Hamiltonian (1) in polar coordinates. Both, the Hamiltonian $h_4$ and 
the first integral ${\cal I}_1$ are second order algebraic differential operators.

The second integral ${\cal I}_2$ commuting with the Hamiltonian is an 8th order algebraic differential operator. 
It can be written in terms of the $g^{(4)}$-algebra generators. 
A certain version of ${\cal I}_2$ was printed in the TTW paper \cite{TTW:2009}
(thereby it was called $y_8$, it contained a number of misprints). 

\noindent
{\it Remark.}
The integral ${\cal I}_2$ is defined ambiguously, up to the addition of a combinations 
of monomials ${\cal H}^n {\cal I}_1^m$, $n+m\leq  4$.  By adding some of these terms 
we have found a form for ${\cal I}_2$ for which the double commutators
$[{\cal I}_1,{\cal I}_{12}]$,  $[{\cal I}_2,{\cal I}_{12}]$ are represented as 5th degree polynomials in ordered monomials ${\cal H},{\cal I}_{1},{\cal I}_{2},{\cal I}_{12}$
(see below). 

We present the integral ${\cal I}_2$ in the form
\begin{equation}
\label{I2-k=4}
 {\cal I}_{2} = \sum_{\substack{0\leq i,j \leq 8\\[3pt] i+j\leq 8 }}  C^{(2)}_{{i,j}}(t,u,a, b,\om)
 {\frac {\pa^{i+j}}{\pa {t}^{i} {u}^{j}}}\ ,
\end{equation}
where the coefficient functions $C^{(2)}_{{i,j}}(t,u,a,b,\om)$ are polynomials in 
$(t,u)$ variables with $C^{(2)}_{{0,0}}=0$, they are given explicitly in Appendix \ref{I2k4coeffs}.

In turn, the commutator of the integrals
\[
{\cal I}_{12} \equiv [{\cal I}_1,{\cal I}_2]\ ,
\]
is a 9th order algebraic differential operator. Unlike ${\cal I}_2$, this integral remains 
unchanged if extra terms in ${\cal H}, {\cal I}_1$ are added to ${\cal I}_2$. It can be written as
\begin{equation}
\label{I12-k=4}
 {\cal I}_{12} = \sum_{\substack{0\leq i,j \leq 9\\[3pt] i+j\leq 9 }}  C^{(12)}_{{i,j}}(t,u,a,b,\om)
 {\frac {\pa^{i+j}}{\pa {t}^{i} {u}^{j}}}\,,
\end{equation}
where the coefficient functions $C^{(12)}_{{i,j}}(t,u;a,b,\om)$ are polynomials in $(t,u)$ variables
with $C^{(12)}_{{0,0}}=0$ and their explicit form is given in Appendix \ref{I12k4coeffs}.

Now we can proceed to calculating the first (I) double commutator
$[{\cal I}_1,{\cal I}_{12}]$ which is a 10th order differential operator. 
With our choice of ${\cal I}_2$, see (\ref{I2-k=4}), 
the double commutator is represented unexpectedly(!) as a 5th degree polynomial in the ordered 
monomials in ${\cal H},{\cal I}_1,{\cal I}_2,{\cal I}_{12}$:
\begingroup
\footnotesize
\begin{align}
  [{\cal I}_1,{\cal I}_{12}] &\ =\
-128\,{\cal H}^4{\cal I}_1
-2048\,(2 a +1) (a+b-1)\,{\cal H}^4 + 256\,{\cal I}_1{\cal I}_2 +4096\,((a+b)^2-1)\,{\cal I}_2
-128 \,{\cal I}_{12} 
%  [{\cal I}_1,{\cal I}_{12}] &\ =\ 
% -128\,{\cal H}^4{\cal I}_1 +
%{\scriptstyle C^{\rm (I)}_{4,0,0,0}}\,{\cal H}^4 +
%{\scriptstyle C^{\rm (I)}_{3,1,0,0}}\,{\cal H}^3{\cal I}_1 +
%{\scriptstyle C^{\rm (I)}_{2,2,0,0}}\,{\cal H}^2{\cal I}_1^2 
\non \\ & + 
{\scriptstyle C^{\rm (I)}_{3,0,0,0}}\,{\cal H}^3 +
{\scriptstyle C^{\rm (I)}_{0,3,0,0}}\,{\cal I}_1^3 +
{\scriptstyle C^{\rm (I)}_{2,1,0,0}}\,{\cal H}^2{\cal I}_1 +
{\scriptstyle C^{\rm (I)}_{1,2,0,0}}\,{\cal H}{\cal I}_1^2 
\label{doublecI1I12} \\ & + 
{\scriptstyle C^{\rm (I)}_{2,0,0,0}}\,{\cal H}^2 +
{\scriptstyle C^{\rm (I)}_{0,2,0,0}}\,{\cal I}_1^2 +
 256\,{\cal I}_1{\cal I}_2 +
{\scriptstyle C^{\rm (I)}_{1,1,0,0}}\,{\cal H}{\cal I}_1  \non 
\\ & +
{\scriptstyle C^{\rm (I)}_{1,0,0,0}}\,{\cal H} +
{\scriptstyle C^{\rm (I)}_{0,1,0,0}}\,{\cal I}_1 +
{\scriptstyle C^{\rm (I)}_{0,0,1,0}}\,{\cal I}_2 -
 128\,{\cal I}_{12} \ , 
\non
\end{align}
\endgroup
in total 16 terms, denoted as $C^{\rm(I)}_{p, q, r, s} {\cal H}^p \,{\cal I}_1^q \,{\cal I}_2^r \,{\cal I}_{12}^s$, 
where the coefficients are:
\begingroup
\small
\begin{align*}
C^{\rm(I)}_{3, 1, 0, 0}  &=  -1024\,\om\,  (4 (a+b)+1)\ ,                  \\
C^{\rm(I)}_{2, 2, 0, 0}  &=   1024\,\om^2\ ,                               \\
C^{\rm(I)}_{3, 0, 0, 0}  &=  -16384\,\om\,  (2 a+1) (4 (a+b)+1) (a+b-1)\ , \\
C^{\rm(I)}_{0, 3, 0, 0}  &=  -2048\,\om^4\ ,                               \\
C^{\rm(I)}_{2, 1, 0, 0}  &=  -2048\,\om^2\, ( 32 a^2 + 4 b + 16 b^2 - 4 a (3 - 4 b) -9)\ , \\
C^{\rm(I)}_{1, 2, 0, 0}  &=   4096 \,\om^3\,(1 + 4 (a + b))\ ,    \\
C^{\rm(I)}_{2, 0, 0, 0}  &=  -32768 \,\om^2\,(1 + 2 a) (-1 + a + b) 
               (-17 - 12 b + 16 b^2 + 32 a^2 + 4a (1 + 12 b))\ ,   \\
C^{\rm(I)}_{0, 2, 0, 0}  &=   8192 \,\om^4\,(178 + 161 b + 31 b^2 + 80 a^2 + 8 a (14 + 3 b))\ , \\
% C^{\rm(I)}_{0, 1, 1, 0}  &=   256\ ,   \\
C^{\rm(I)}_{1, 1, 0, 0}  &=  8192\,\om^3\, \Big(37 + 57 b + 19 b^2 - 8 b^3 - 928 a^5 + 4 a^3 (215 + 117 b + 35 b^2) +
                \\ 
     & \hspace{70pt}  2 a^2 (79 - 13 b + 29 b^2 + 6 b^3) - a (52 + 104 b - 56 b^2) \Big)\ ,      \\
C^{\rm(I)}_{1, 0, 0, 0}  &=  131072\,\om^3\, (a + b -1)  \Big(45 + 132 b + 96 b^2 - 5 b^3 - 8 b^4 -
                928 a^6 - 928 a^5 (1 + b)
               \\ & \hspace{50pt}
               + 4 a^4 (199 + 117 b + 35 b^2) + 2 a^3 (477 + 555 b + 333 b^2 + 76 b^3) 
               \\ & \hspace{30pt}
         + 2 a^2 (67 - 58 b - 52 b^2 + 35 b^3 + 6 b^4) + a (27 - 51 b - 125 b^2 - 16 b^3) \Big)\ ,    
                           \\
C^{\rm(I)}_{0, 1, 0, 0}  & = 131072\,\om^4\, \Big(-174 - 161 b + 140 b^2 + 161 b^3 + 31 b^4 + 88 a^4 
                           +  16 a^3 (7 + 13 b)
                            \\ & \hspace{30pt}
              + a^2 (89 + 385 b + 183 b^2) +  2 a (-56 + 158 b + 217 b^2 + 47 b^3)\Big)\ ,   
\end{align*}
\endgroup
In a similar way, the second (II) double commutator $[{\cal I}_2,{\cal I}_{12}]$ is 
a 16th order differential operator, and it is also written as a 5th degree polynomial 
in the ordered monomials in ${\cal H},{\cal I}_1,{\cal I}_2,{\cal I}_{12}$ totally unexpectedly(!):
\begin{align}
  [{\cal I}_2,{\cal I}_{12}] &=
{\scriptstyle C^{\rm (II)}_{5, 0, 0, 0}}\,  {\cal H}^5  +
{\scriptstyle C^{\rm (II)}_{4, 1, 0, 0}}\,  {\cal H}^4 {\cal I}_1 +
 128\,  {\cal H}^4  {\cal I}_2  
\non \\ & +
{\scriptstyle C^{\rm (II)}_{4, 0, 0, 0}}\,  {\cal H}^4  +
{\scriptstyle C^{\rm (II)}_{3, 1, 0, 0}}\,  {\cal H}^3 {\cal I}_1 +
{\scriptstyle C^{\rm (II)}_{3, 0, 1, 0}}\,  {\cal H}^3  {\cal I}_2 +
{\scriptstyle C^{\rm (II)}_{2, 2, 0, 0}}\,  {\cal H}^2 {\cal I}_1^2  +
{\scriptstyle C^{\rm (II)}_{2, 1, 1, 0}}\,  {\cal H}^2 {\cal I}_1 {\cal I}_2   
\label{doublecI2I12} \\ & \hspace{20pt} +
{\scriptstyle C^{\rm (II)}_{3, 0, 0, 0}}\,  {\cal H}^3  +
{\scriptstyle C^{\rm (II)}_{2, 1, 0, 0}}\,  {\cal H}^2 {\cal I}_1 +
{\scriptstyle C^{\rm (II)}_{2, 0, 1, 0}}\,  {\cal H}^2  {\cal I}_2 +
{\scriptstyle C^{\rm (II)}_{2, 0, 0, 1}}\,  {\cal H}^2 {\cal I}_{12} +
{\scriptstyle C^{\rm (II)}_{1, 2, 0, 0}}\,  {\cal H} {\cal I}_1^2    \nonumber \\ & \hspace{30pt} +
{\scriptstyle C^{\rm (II)}_{1, 1, 1, 0}}\,  {\cal H} {\cal I}_1 {\cal I}_2 +
{\scriptstyle C^{\rm (II)}_{0, 3, 0, 0}}\,  {\cal I}_1^3  +
{\scriptstyle C^{\rm (II)}_{0, 2, 1, 0}}\,  {\cal I}_1^2 {\cal I}_2  \nonumber \\ & +
{\scriptstyle C^{\rm (II)}_{2, 0, 0, 0}}\,  {\cal H}^2  +
{\scriptstyle C^{\rm (II)}_{0, 2, 0, 0}}\,  {\cal I}_1^2  -
  128\,  {\cal I}_2^2 +
{\scriptstyle C^{\rm (II)}_{1, 1, 0, 0}}\,  {\cal H} {\cal I}_1 +
{\scriptstyle C^{\rm (II)}_{1, 0, 1, 0}}\,  {\cal H} {\cal I}_2  \nonumber 
\\ & \hspace{30pt} + 
{\scriptstyle C^{\rm (II)}_{1, 0, 0, 1}}\,  {\cal H} {\cal I}_{12} +
{\scriptstyle C^{\rm (II)}_{0, 1, 1, 0}}\,  {\cal I}_1 {\cal I}_2 +
{\scriptstyle C^{\rm (II)}_{0, 1, 0, 1}}\,  {\cal I}_1  {\cal I}_{12} \nonumber 
\\ &+
{\scriptstyle C^{\rm (II)}_{1, 0, 0, 0}}\,  {\cal H} +
{\scriptstyle C^{\rm (II)}_{0, 1, 0, 0}}\,  {\cal I}_1 +
{\scriptstyle C^{\rm (II)}_{0, 0, 1, 0}}\,  {\cal I}_2 +
{\scriptstyle C^{\rm (II)}_{0, 0, 0, 1}}\,  {\cal I}_{12} \ ,  \nonumber
\end{align}
in total 28 terms, denoted as $C^{\rm(II)}_{p, q, r, s} {\cal H}^p \,{\cal I}_1^q \,{\cal I}_2^r \,{\cal I}_{12}^s$, 
where the coefficients are the following:

%%%%%%%%%%%%%%%%%
\noindent
{\bf (I)}\ Coefficients in front of degrees 5 and 4:
{\footnotesize
\begin{align*}
  C^{\rm(II)}_{5, 0, 0, 0} & = 4096\,\om^3\, \Big(27 + 13 b + 3 b^2 - 8 b^3 - 928 a^5 + 4 a^3 (231 + 117 b + 35 b^2)
\\ &
   + 2 a^2 (47 - 13 b + 29 b^2 + 6 b^3) - 8 a (14 + 27 b + b^2)\Big)\ , \\
  C^{\rm(II)}_{4, 1, 0, 0} & = 8192\,\om^4\, (167 + 161 b + 31 b^2 + 104 a^2 + 8 a (11 + 4 b))\ , \\
%  C^{\rm(II)}_{4, 0, 1, 0} & = 128\ , \\
  C^{\rm(II)}_{4, 0, 0, 0} & = -32768\, \om^4\,  \Big(175 - 61 b - 315 b^2 - 66 b^3 + 32 b^4 + 3712 a^6 + 928 a^5 (1 + 4 b)  
\\ &
    - 16 a^4 (253 + 117 b + 35 b^2) -  4 a^3 (425 + 1199 b + 561 b^2 + 152 b^3) \\ &
    -  2 a^2 (-183 + 281 b + 87 b^2 + 122 b^3 + 24 b^4) + a (470 + 378 b + 214 b^2 - 60 b^3) \Big)\ , \\
  C^{\rm(II)}_{3, 1, 0, 0} & = 131072 \, \om^5\,  \Big(70 + 408 b + 336 b^2 + 66 b^3 + 464 a^5 
    - 2 a^3 (127 + 117 b + 35 b^2) 
\\ &
   +  a^2 (181 + 285 b - 29 b^2 - 6 b^3) + 2 a (217 + 311 b + 65 b^2)\Big)\ , \\
  C^{\rm(II)}_{3, 0, 1, 0} & = 1024\,\om\, (1 + 4 a + 4 b)\ ,\\
  C^{\rm(II)}_{2, 2, 0, 0} & = -98304\,\om^6\, \Big(179 + 161 b + 31 b^2 + 88 a^2 + 8 a (13 + 4 b)\Big)\ , \\
  C^{\rm(II)}_{2, 1, 1, 0} & = -2048 \,\om^2\ ,
\end{align*}
}

%%%%%%%%%%%%%%%%%

\vfil

\noindent
{\bf (II)}\ Coefficients in front of degrees 3:
{ \footnotesize
\begin{align*}
C^{\rm(II)}_{3, 0, 0, 0} & = -65536 \,\om^5\,\Big(1051 + 29696 a^7 + 3481 b - 537 b^2 - 4700 b^3 - 1008 b^4 +  128 b^5 \\ &
  - 3712 a^6 (3 - 4 b) - 32 a^5 (1361 + 352 b - 324 b^2) + 16 a^4 (17 - 1609 b - 479 b^2 - 164 b^3) \\ &
  + 4 a^3 (2905 - 5237 b - 7459 b^2 - 2208 b^3 - 608 b^4) \\ &
  + 2 a^2 (3071 - 4841 b - 13087 b^2 - 2670 b^3 - 488 b^4 - 96 b^5)  + 
  4 a (1103 + 1427 b - 2019 b^2 - 2120 b^3 - 432 b^4)\Big)\ , \\
C^{\rm(II)}_{2, 1, 0, 0} & = 131072 \, \om^6\,\Big(-1605 - 1023 b + 2927 b^2 + 2692 b^3 + 560 b^4 + 7424 a^6 \\ &
 + 1856 a^5 (1 + 4 b) - 32 a^4 (143 + 117 b + 35 b^2) -  8 a^3 (17 + 679 b + 561 b^2 + 152 b^3) \\ &
  +  a^2 (3548 + 7492 b + 3324 b^2 - 488 b^3 - 96 b^4)+ 4 a (-677 + 491 b + 1365 b^2 + 284 b^3)\Big)\ , \\
C^{\rm(II)}_{2, 0, 1, 0} & = 2048 \,\om^2\,(-9 + 4 b + 16 b^2 + 32 a^2 - 4 a (3 - 4 b))\ , \\
C^{\rm(II)}_{2, 0, 0, 1} & = 1024\, \om^2\ , \\
C^{\rm(II)}_{1, 2, 0, 0} & = -196608 \,\om^7\,\Big(331 + 1741 b + 1347 b^2 + 256 b^3  + 928 a^5 
  - 4 a^3 (55 + 117 b + 35 b^2) \\ &
    + 2a^2 (457 + 493 b - 29 b^2 - 6 b^3)+ 8 a (219 + 300 b + 64 b^2)\Big)\ , \\
C^{\rm(II)}_{1, 1, 1, 0} & = -8192 \,\om^3\,(1 + 4 a + 4 b)\ , \\
C^{\rm(II)}_{0, 3, 0, 0} & = 262144\,\om^8\, (185 + 161 b + 31 b^2 + 80 a^2 + 16 a (7 + 2 b))\ , \\
C^{\rm(II)}_{0, 2, 1, 0} & = 6144\, \om^4\ , \\
  \end{align*}
  \vfill
 }
%
%%%%%%%%%%%%%%%%%

{\bf (III)}\ Coefficients in front of degrees 2:
{\footnotesize
  \begin{align*}
  C^{\rm(II)}_{2, 0, 0, 0} & = -131072\,\om^6\, \Big(-13619 - 14278 b + 27055 b^2 + 20702 b^3 \\ &
                - 14471 b^4 - 4272 b^5 + 64 b^6 + 861184 a^{10} - 7424 a^8 (215 + 117 b + 35 b^2) \\ &
                -  3712 a^7 (79 - 13 b + 29 b^2 + 6 b^3) +  16 a^6 (49185 + 62374 b + 22243 b^2 + 8190 b^3 + 1225 b^4) \\ &
                +  16 a^5 (10165 - 9252 b + 5275 b^2 + 5156 b^3 + 1717 b^4 +  210 b^5) \\ &
                +  4 a^4 (-11991 - 108734 b - 42025 b^2 + 6018 b^3 + 4605 b^4 +  348 b^5 + 36 b^6) \\ &
                -  8 a^3 (-14829 - 10678 b + 25617 b^2 + 13450 b^3 - 229 b^4 +  112 b^5) \\ &
                -  4 a^2 (-12103 - 47682 b + 5187 b^2 + 53180 b^3 + 9227 b^4 + 118 b^5 + 48 b^6)\\ &
                -  16 a (1974 - 1223 b - 4798 b^2 + 2025 b^3 + 2847 b^4 + 552 b^5)  \Big) \ ,  \\
  C^{\rm(II)}_{0, 2, 0, 0} & = -131072 \,\om^8\, \Big(71225  + 151340 b + 106903 b^2 + 29946 b^3 +  2883 b^4
             \\ & \hspace{30pt}
             + 19008 a^4 + 384 a^3 (140 + 29 b)  +  16 a^2 (7055 + 5838 b + 1026 b^2)
              \\ & \hspace{30pt}
              + 16 a (6580 + 8093 b + 2751 b^2 + 279 b^3) \Big)\ ,  \\
  %C^{\rm(II)}_{0, 0, 2, 0} & = -128\ , \\
  C^{\rm(II)}_{1, 1, 0, 0} & = 524288 \,\om^7\,\Big(-5660  - 15395 b - 13434 b^2 - 3658 b^3 +699 b^4 + 248 b^5 \\ &
           + 74240 a^7 + 7424 a^6 (14 + 3 b) + 32 a^5 (2068 + 3499 b + 549 b^2)  \\ &
           - 32 a^4 (3405 + 2266 b + 986 b^2 + 135 b^3)-4 a^3 (34702 + 49869 b + 33584 b^2 + 9946 b^3 + 1157 b^4)  \\ &
           - 2 a^2 (10270 + 7293 b + 7366 b^2 + 5750 b^3 + 1865 b^4 +186 b^5)\\ &
           + 4 a (579 + 3791 b + 1279 b^2 - 1370 b^3 - 386 b^4)\Big)\ , \\
  C^{\rm(II)}_{1, 0, 1, 0} & = 8192 \,\om^3\,\Big(-37 - 57 b - 19 b^2 + 8 b^3 + 928 a^5 - 4 a^3 (215 + 117 b + 35 b^2)
  \\ & \hspace{30pt}
            - 2 a^2 (79 - 13 b + 29 b^2 + 6 b^3)+ 4a (13 + 26 b - 19 b^2)\Big)\ ,  
  \\
           C^{\rm(II)}_{1, 0, 0, 1} & = 4096 \,\om^3\,(1 + 4 a + 4 b)\ , 
  \\
           C^{\rm(II)}_{0, 1, 1, 0} & = -16384\, \om^4\, \Big(146 + 161 b + 31 b^2 + 80 a^2 + 8 a (14 + 3 b)\Big)\ , 
  \\
           C^{\rm(II)}_{0, 1, 0, 1} & = -6144 \,\om^4\ ,
\end{align*}
}
%
%%%%%%%%%%%%%%%%%fns
Coefficients in front of degrees 1:
{\footnotesize
  \begin{align*}
 C^{\rm(II)}_{1, 0, 0, 0} & = 4194304\,\om^7\, \Big(4932  + 16665 b + 11175 b^2 - 14314 b^3 - 17698 b^4 - 2758 b^5 + 1195 b^6 + 248 b^7 \\ &
    + 81664 a^9 + 14848 a^8 (7 + 13 b)+ 64 a^7 (-732 + 4939 b + 2461 b^2)  \\ &
   - 16 a^6 (12937 + 1873 b - 17789 b^2 - 3566 b^3) \\ &
   - 4 a^5 (34391 + 118156 b + 58145 b^2 + 886 b^3 - 163 b^4)  \\ &
   - 2 a^4 (-41701 + 71982 b + 181625 b^2 + 148168 b^3 + 59993 b^4 +  7678 b^5)  \\ &
   - 4 a^3 (-25072 - 31047 b - 7060 b^2 + 18362 b^3 + 30122 b^4 + 11927 b^5 + 1367 b^6) \\ &
   - a^2 (-13177 + 22918 b + 43651 b^2 - 34790 b^3 - 34703 b^4 +  3260 b^5 + 3730 b^6 + 372 b^7) \\ &
      + 2 a (1455 - 7845 b - 27962 b^2 - 17466 b^3 + 6066 b^4 + 4099 b^5 +  500 b^6)\Big)\ , \\
 C^{\rm(II)}_{0, 1, 0, 0} & = -4194304  \,\om^8\,  \Big(-19215 - 45724 b - 16620 b^2 + 34776 b^3 +
   33578 b^4+ 9982 b^5 + 961 b^6  \\ &  
   \hspace{30pt} + 7744 a^6 + 1408 a^5 (14 + 15 b) + 16 a^4 (1895 + 4683 b + 1593 b^2) \\ &  
   \hspace{30pt}  +
   32 a^3 (399 + 3587 b + 3374 b^2 + 643 b^3)
%   \\ & \hspace{30pt}
   - a^2 (17727 - 74802 b - 155743 b^2 - 79982 b^3 - 11409 b^4)
   \\ & \hspace{30pt}  
   - 2 a (15904 + 3733 b - 52668 b^2 - 56213 b^3 - 18606 b^4 - 1953 b^5) \Big)\ ,  \\
   C^{\rm(II)}_{0, 0, 1, 0} & = -131072\,\om^4\, \Big(-110 - 161 b + 76 b^2 + 161 b^3 + 31 b^4
   \\ & \hspace{30pt}
   + 88 a^4 + 16 a^3 (7 + 13 b)  + a^2 (25 + 385 b + 183 b^2) - 2 a (56 - 94 b - 217 b^2 - 47 b^3) \Big)\ , \\
   C^{\rm(II)}_{0, 0, 0, 1} & = 8192\,\om^4\, (146 + 80 a^2 + 161 b + 31 b^2 + 8 a (14 + 3 b))\ .
\end{align*}
}

\normalsize

%%%%%%%%%%%%%%%%%
Both double commutators (44)-(45) depend on ${\cal I}_{12}$ linearly. Syzygy has the form (\ref{syzygy}), 
being 18th order differential operator, at $\om=0$ its explicit expression reads
\begin{align}
{\cal I}_{12}^2 & =  1024 (2a+1)(2a-3) H^8  - 256 H^4\,{\cal I}_1 {\cal I}_2 
  - 4096 (2a^2 + 2 a b -a + b -9)\,H^4 {\cal I}_2  +  128  H^4\,{\cal I}_{12}   
\non \\ &
 + 256\,{\cal I}_1\,{\cal I}_2^2 - 256\,{\cal I}_2\,{\cal I}_{12}
 + 4096\,((a+b)^2 - 9)\,{\cal I}_2^2\ ,
\end{align}
it is the eighth degree polynomial in $H,{\cal I}_1,{\cal I}_2,{\cal I}_{12}$. 

In conclusion of this Section we have to note that the calculation of the double commutators 
$[{\cal I}_1,{\cal I}_{12}]$ and $[{\cal I}_2,{\cal I}_{12}]$ for the TTW at $k=4$ is 
cumbersome, it is extremely difficult technically and very lengthy. This calculation 
was carried out using specially designed code in three different symbolic programs 
MAPLE-18, MAPLE-2024 and Mathematica-13, it took several months of calculations carried 
out on the regular desktop with extended RAM. Finally, we arrived at the quintic algebra 
of integrals generated 
by the operators $H,{\cal I}_1,{\cal I}_2,{\cal I}_{12}$~\footnote{Syzygy 
has the form (\ref{syzygy}), it is not shown in the explicit form.}. This algebra is embedded into 
$g^{(4)}$ being its subalgebra. It must be emphasized that the quintic polynomial algebra 
of integrals remains for any values of the parameters $\om, a, b$ of the TTW model at $k=4$ 
even though it loses its Hermiticity, see (1). 

\section{Conclusions}
In this work we study the hidden algebra and polynomial algebra of integrals for the quantum 
TTW system for \hbox{$k=1,2,3,4$}. None of our results are reproduced in the so-called 
recurrence relation approach by Kalnins et al. \cite{Miller:2011} as particular cases, which might be 
an explicit indication that the second integral ${\cal I}_2$, constructed there, is of non-minimal 
degree~\footnote{The authors of \cite{Miller:2011} were aware that ${\cal I}_2$ 
appears in their approach in non-minimal degree and promised to fix it, to the best knowledge 
of the present authors it was never done. Their remarkable proof of superintegrability of 
the TTW system at rational index $k$ remains valid.}. 
%\newpage

\noindent
Present authors are not aware about the method which would lead to ${\cal I}_2$ as 
the differential operator of degree $2k$ for the TTW with integer index $k$ as 
it was conjectured in the original TTW paper [1].

After formidable calculations and based on the obtained results we arrived at a very simple Conjecture:

\noindent
{\it  The superintegrable and exactly solvable TTW system with integer index $k$ is characterized by 
the hidden algebra $g^{(k)}$ and 4-generated polynomial algebra of integrals of order $(k+1)$ with 
$H,{\cal I}_1,{\cal I}_2,{\cal I}_{12}$ as generating elements and with double commutators 
$[{\cal I}_1,{\cal I}_{12}]=P_{k+1}$,  $[{\cal I}_2,{\cal I}_{12}]=Q_{k+1}$, which can depend on ${\cal I}_{12}$ 
linearly; here $$Q_{k+1}= 8k^2\left((-)^k\, H^k {\cal I}_2\ -\ {\cal I}_2^2\right)+O(\om)\ .$$
The basic integrals ${\cal I}_1,{\cal I}_2$ have the form of the second order and $(2k)$-th order differential 
operators, respectively, as already conjectured in \cite{TTW:2009}.  
$H,{\cal I}_1,{\cal I}_2,{\cal I}_{12}$ are algebraically related: there exists syzygy
\[
  {\cal I}_{12}^2\ =\ {\cal R}_k(H,{\cal I}_1,{\cal I}_2,{\cal I}_{12})\ , 
\]
where ${\cal R}$ is $(4k+2)$-th order differential operator in dihedral group $I_{2k}$ invariants 
of the lowest order as variables. From another side, ${\cal R}_k$ is polynomial in $H$ of degree $2k$, where 
some terms can be predicted,
\[
   {\cal R}_k\ =\ 4\,k^4\,(2a+1)(2a-3)\,H^{2k}\ +\ (-)^{k+1} 8 k^2 H^k (2{\cal I}_1\,{\cal I}_2 -
    {\cal I}_{12}) + 
\]
\[
    + 16 k^4 (-)^{k+1} (2 a^2 + 2ab - a + b - 9) H^k {\cal I}_2 +
\]   
\[   
   + 16 k^2 \Bigg({\cal I}_1\,{\cal I}_2^2 + k^2 ((a+b)^2-9) {\cal I}_2^2 - {\cal I}_2\,{\cal I}_{12}\Bigg)\ +\ O(\om)
   \quad . 
\] 

}

\bigskip

%\noindent
It seems evident that in general the obtained results will continue to hold for 
the classical TTW system \cite{TTW:2010} for \hbox{$k=1,2,3,4$} when the Lie brackets 
are replaced by Poisson brackets and a dequantization procedure is performed: 
$\pa \rar i p$, where $p$ is classical momentum.

\section*{Author's contributions}

Both authors contributed equally to this work.

\section*{Acknowledgments}

\noindent
This work is partially supported by DGAPA grant IN113025 (Mexico).
The authors thank A.M. Escobar Ruiz for interest to the work and 
helpful discussions in the early stage of the work. We are grateful to one 
of the anonymous referees who raised interesting questions which led to 
the improvement and clarity of the presentation.
J.C.L.V. thanks the hospitality of the Mathematics Department of the
Southern Methodist University (SMU), Dallas TX, USA, where this work was completed.
He also acknowledges the support from PASPA-UNAM for his sabbatical stay at SMU.

\section*{DATA AVAILABILITY}

Data sharing is not applicable to this article as no new data were created or analyzed in this study.

%%%%%%%%%%%%%%%%%%%%%%%%%%%%%%%%%%%%

%%%%%%%%%%%%%%%%%

\newpage 

\appendix

\section{ TTW at $k=3$: the second integral  ${\cal I}_2$ coefficients, see (\ref{I2-k=3}) \label{I2k3coeffs}}

Non-vanishing coefficients for the integral ${\cal I}_2$ of TTW at $k=3$, defined in (\ref{I2-k=3}):

%%%%%%%%%%%%%%%%% 1

\begin{align*}
  C^{(2)}_{ 0, 1} &=
  - 1296 (2 a+1) (2 b+1) (a+b+2) (3 a+3 b+1) (3 a+3 b+2)
  \\ & 
+ 1296 \, \omega  \, t \, (2 a+1) (2 b+1) (3 a+3 b+2) (3 a+3 b+5)
  \\ &
  -36\, \omega ^2 \, t^2\, (2 b+1) \big(185 a^2+a (216 b+355)-27 (b-5) b+136\big)
  \\ & 
  +24\, \omega ^3 \,\big(9 b \left(u (12 a+b+11)-2 (2 b+1) t^3\right) + (a (41 a+67)+28) u\big)
\end{align*}

%%%%%%%%%%%%%%%%% 2

\begin{align*}
  C^{(2)}_{ 0, 2} & =
  648 \Big(9 (2 b+1) (2 b+3) t^3 (a+b+4) (4 a+2 b+7)
  - 2 u \big(36 a^4+18 a^3 (18 b+25)
\\ & 
  + a^2\, (18 b \, (32 b+109) + 1619)
  + a\, \big(6 b \big(54 b^2+327 b+613\big) + 2213\big)
\\ &
 + (3 b (b+9)+32) (6 b (2 b+7)+31)\big)\Big)
\\ & 
 +  324 \, \omega \, \Big(4 t u
  \big(36 a^3+6 a^2 (36 b+53)+a (72 b (3 b+11)+685)+9 b \big(4 b^2+34 b+69\big)+362\big)
  \\ & \hspace{50pt}
-9 (4 b (b+2)+3) t^4 (6 a+4 b+17) \Big)
\\ & 
+  \omega ^2 \, \Big(
2916 (4 b (b+2)+3) t^5 - 72 t^2 u \big(185 a^2+a (594 b+922)+27 b (5 b+24)+541\big)
\Big)
  \\ &
 + 648 \,\omega ^3 u \Big(2 u (a+b+1)-(2 b+1) t^3\Big)
\end{align*}

%%%%%%%%%%%%%%%%% 3

\begin{align*}
  C^{(2)}_{ 0, 3} & =
  -1296 u \Big(2 u \left(108 a^3+36 a^2 (13 b+30)+a (36 b (13 b+69)+3194)
  + 54 b (2 b (b+10)+59) \right.\\ &\left.  +2893\right)
  - 9 (2 b+3) t^3 \left(8 a^2+4 a (6 b+19)
+4 b (3 b+22)+151\right)\Big)
\\ & 
 + 3888 \,\omega\,  t u \Big(2 u \left(24 a^2+2 a (30 b+71)+6 b (4 b+23)+173\right)
  - 3 (2 b+3) t^3 (6 a+8 b+27)\Big)
\\ &
-1944 \,\omega^2 \,t^2 u \left(u (14 a+12 b+25)-6 (2 b+3) t^3\right)
%%%
+432 \,\omega^3 \,u^2 \left(u-t^3\right)
\end{align*}

%%%%%%%%%%%%%%%%% 4

\begin{align*}
  C^{(2)}_{ 0, 4} & =
  23328 u^2 \Big(t^3 \left(4 a^2+a (30 b+83)+b (26 b+179)+288\right)
\\ &
\hspace{70pt}   - 2 u \left(13 a^2+a (30 b+97)+b (13 b+97)+169\right)\Big)
\\ &
-11664 \, \omega \, t u^2 \left(t^3 (6 a+20 b+57)-4 u (5 a+5 b+16)\right)
+ 11664 \,\omega^2 \, t^2 u^2 \left(t^3-u\right)
\end{align*}

%%%%%%%%%%%%%%%%% 5

\begin{align*}
  C^{(2)}_{ 0, 5} & =
  46656 u^3 \left(t^3 (6 a+12 b+47)-2 u (6 a+6 b+25)\right) - 93312 \,\omega\, t u^3 \left(t^3-u\right)
\end{align*}

%%%%%%%%%%%%%%%%% 6

\begin{align*}
  C^{(2)}_{ 0, 6} = 186624 (t^3 - u) u^4
\end{align*}

%%%%%%%%%%%%%%%%% 7

\begin{align*}
  C^{(2)}_{ 1, 0} &= 
  8 \,\omega^2\, (3 a+3 b+1) \big(a (31 a+41)+27 (b-1) b+8\big)
  - 8 \,\omega^3 t \big(a (31 a+41)+27 (b-1) b+8\big)
\end{align*}

%%%%%%%%%%%%%%%%% 8

\begin{align*}
  C^{(2)}_{ 1, 1} &=
  - 2592 (2 a+1) (2 b+1) t (3 a+3 b+2) (3 a+3 b+5) \\&
  + 7776 \,\omega t^2 (2 a+1) (2 b+1)  (2 a+2 b+3) 
  \\ & 
  + \,\omega^2 \, \Big(48 u \left(58 a^2+a (54 b+131)
  + 27 b (2 b+5)+107\right) - 648 t^3  (2 b+1) (9 a+b+9) \Big)
  \\ &
  +216 \,\omega^3 t \Big((2 b+1) t^3-2 u (a+b+1)\Big)
\end{align*}

%%%%%%%%%%%%%%%%% 9 

\begin{align*}
  C^{(2)}_{ 1, 2} &=
  -648 \Big(4 t u \left(36 a^3+6 a^2 (36 b+53)+a (72 b (3 b+11)+685)+9 b  \left(4 b^2+34 b+69\right)+362\right)
\\ & 
  -9 (4 b (b+2)+3) t^4 (6 a+4 b+17)\Big)
\\ & 
+3888\, \omega \Big(t^2 u \left(48 a b+16 a (a+5)+12 b^2+48 b+45\right)-3 (4 b (b+2)+3) t^5\Big)
\\ & 
- 3888\, \omega^2 u \Big(t^3 (3 a+b+4)-u (a+b+3)\Big) + 432 \,\omega^3\, t u \left(t^3-u\right)
\end{align*}

%%%%%%%%%%%%%%%%% 10

\begin{align*}
  C^{(2)}_{ 1, 3} & =
  -7776 t u \Big(2 u \left(24 a^2+2 a (30 b+71)+6 b (4 b+23)+173\right)
  -3 (2 b+3) t^3 (6 a+8 b+27)\Big)
\\ & \hspace{40pt}
  +15552\, \omega \,t^2 u \left(u (8 a+6 b+13)-3 (2 b+3) t^3\right)-2592 \, \omega^2 \, u^2 \left(t^3-u\right)
\end{align*}

%%%%%%%%%%%%%%%%% 11

\begin{align*}
  C^{(2)}_{ 1, 4} & =
  23328 \,t u^2 \Big(t^3 (6 a+20 b+57)-4 u (5 a+5 b+16)\Big) - 46656 \,\omega\, t^2 u^2 \left(t^3-u\right)
\end{align*}

%%%%%%%%%%%%%%%%% 12

\begin{align*}
  C^{(2)}_{ 1, 5} = 186624\, t (t^3 - u) u^3
\end{align*}

%%%%%%%%%%%%%%%%% 13

\begin{align*}
  C^{(2)}_{ 2, 0} & =
   - 144\,\omega\, (2 a+1)  (3 a+3 b+1) (3 a+3 b+2) \\ & + 
  8 \,\omega^2 t \Big(247 a^2+a (216 b+293)+27 b (b+3)+80\Big)
     - 144\, \omega^3\, (2 a+1) t^2
\end{align*}

%%%%%%%%%%%%%%%%% 14

\begin{align*}
  C^{(2)}_{ 2, 1} &=
  -7776\, t^2 (2 a+1) (2 b+1)  (2 a+2 b+3) \\ & \hspace{30pt}
  -1296 \,\omega \left(2 u \left(4 a^2+a (6 b+9)+b (2 b+11)+7\right)-(2 b+1) t^3 (12 a+2 b+13)\right)
\\ & \hspace{50pt}
  +1944 \,\omega^2 \left(2 t u (a+b+1)-(2 b+1) t^4\right) 
\end{align*}

%%%%%%%%%%%%%%%%% 15

\begin{align*}
  C^{(2)}_{ 2, 2} & =
  -3888\, \Big(t^2 u \left(48 a b+16 a (a+5)+12 b^2+48 b+45\right)-3 (4 b (b+2)+3) t^5\Big)
\\ & \hspace{50pt}
  + 2592 \,\omega \, u \left(t^3 (12 a+6 b+19)-2 u (3 a+3 b+8)\right)-3888\, \omega^2 \, t u \left(t^3-u\right)
\end{align*}

%%%%%%%%%%%%%%%%% 16

\begin{align*} C^{(2)}_{ 2, 3} & =
   - 15552\, t^2 u \left(u (8 a+6 b+13)-3 (2 b+3) t^3\right)
   + 10368 \,\omega\, u^2 \left(t^3-u\right)
\end{align*}

%%%%%%%%%%%%%%%%% 17

\begin{align*} C^{(2)}_{ 2, 4} =
  46656 t^2 (t^3 - u) u^2
\end{align*}

%%%%%%%%%%%%%%%%% 18

\begin{align*}
  C^{(2)}_{ 3, 0} & =
   96 (2 a+1) (3 a+3 b+1) (3 a+3 b+2)
   -576 \,\omega t\,(2 a+1)  (3 a+3 b+2)
   \\ & \hspace{70pt}
  + 576\,\omega^2\, t^2 (2 a+1) 
  + 64\, \omega^3\, \left(u-t^3\right)
\end{align*}

%%%%%%%%%%%%%%%%% 19

\begin{align*}
  C^{(2)}_{ 3, 1} &=
  864 \left(2 u \left(4 a^2+a (6 b+9)+b (2 b+11)+7\right)-(2 b+1) t^3 (12 a+2 b+13)\right)
\\ & \hspace{70pt}
  - 3456 \,\omega\, \left(2 t u (a+b+1)-(2 b+1) t^4\right)
\end{align*}

%%%%%%%%%%%%%%%%% 20

\begin{align*}
  C^{(2)}_{ 3, 2} &=
- 1728 u \left(t^3 (12 a+6 b+19)-2 u (3 a+3 b+8)\right)
+  6912 \,\omega\, t u \left(t^3-u\right) 
\end{align*}

%%%%%%%%%%%%%%%%% 21

\begin{align*}
  C^{(2)}_{ 3, 3} =  -6912 (t^3 - u) u^2
\end{align*}

%%%%%%%%%%%%%%%%% 22

\begin{align*}
  C^{(2)}_{ 4, 0} & =
  288 (2 a+1) t (3 a+3 b+2) - 720 \,\omega\, t^2 (2 a+1) + 192 \,\omega^2\, \left(t^3-u\right)
\end{align*}

%%%%%%%%%%%%%%%%% 23

\begin{align*}
  C^{(2)}_{ 4, 1} & = 1728 \left(2 t u (a+b+1)-(2 b+1) t^4\right)  
\end{align*}

%%%%%%%%%%%%%%%%% 24

\begin{align*}
  C^{(2)}_{ 4, 2} = -3456 t (t^3 - u) u
\end{align*}

%%%%%%%%%%%%%%%%% 25

\begin{align*}
  C^{(2)}_{ 5, 0} = 288 (2 a+1) t^2-192 \, \omega\, \left(t^3-u\right)
\end{align*}

%%%%%%%%%%%%%%%%% 26

\begin{align*}
  C^{(2)}_{ 6, 0} =  64 (t^3 - u)
\end{align*}

%%%%%%%%%%%%%%%%%%%%%%%%%%%%%%%%%%%%%%%%%%%%%%%%%%%%%%%%%%%%%%%%%%%%%%%%%%%%%%%%%%%%%%%%%%

 %% FILE OK %%

\section{TTW at $k=3$: the commutator ${\cal I}_{12}$ coefficients, see (\ref{I12-k=3}) \label{I12k3coeffs}}

Non-vanishing coefficients for the integral (commutator) ${\cal I}_{12}$ of TTW at $k=3$, defined in (\ref{I12-k=3}):

%%%% Coefficients of I12 (TTW k=3) LaTeX form  January 10, 2025
%%%% Some terms edited manually, factoring minus signs, removing extra brackets, etc 
%%%% Corrected Typos found by Co-Scientist on Sept 03, 2026
%%%%%%%%%%%%%%% 1
\begin{align*}
  C^{(12)}_{0,1} &= 5184 (1+2 a) (1+2 b) (1+3 a+3 b) (2+3 a+3 b) (4+3
  a+3 b) (5+3 a+3 b)
\\ & 
  -15552\, {\omega} t \,(1+2 a) (1+2 b) (2+3 a+3 b) (4+3 a+3 b)
  (5+3 a+3 b) 
\\ & 
  + 15552 \, {\omega}^2 t^2 \,(1+2 a) (1+2 b) (4+3 a+3 b) (5+3 a+3 b) 
\\ & 
  +1728 \, {\omega}^3\, (1+2 b)  
  \left( -     (22+9 b+9 a (5+2 a+2 b)) t^3 +4 u\right)
\end{align*} 

%%%%%%%%%%%%%%% 2
\begin{align*}
  C^{(12)}_{0,2} &=
  -23328(2b + 3)(2b + 1)\Big(72a^3 + 162a^2b + 108ab^2 + 18b^3 + 513a^2
  + 720ab \\& + 207b^2 + 1195a + 755b + 900\Big)t^3  + \Big(1679616a^5 +
  20155392a^4b  + 52068096a^3b^2 \\ & + 52068096a^2b^3 + 20155392ab^4 +
  1679616b^5 + 27713664a^4 + 174680064a^3b \\ & + 293932800a^2b^2 +
  174680064ab^3 + 27713664b^4 + 144773568a^3 \\ & + 549094464a^2b +
  549094464ab^2 + 144773568b^3  + 338909184a^2 \\ & + 741270528ab +
  338909184b^2 + 363953088a + 363123648b + 143182080 \Big)u
\\[5pt] &
%%%%%
+ \omega\, \Big(
46656 (2 b+3) (2 b+1) (27 a^2+36 a b+9 b^2+138 a+78 b+170) t^4 \\ &
+(-1679616 a^4-15116544 a^3 b-26873856 a^2 b^2-15116544 a b^3-1679616 b^4
\\ & -22114944 a^3-98257536 a^2 b-98257536 a b^2-22114944 b^3-86173632 a^2
\\ & -198474624 a b-86173632 b^2-127666368 a-125178048 b-60030720) u t
\Big) \\ & 
+ \omega^2\Big(
-69984 (2 b+3) (2 b+1) (4 a+2 b+11) t^5+(559872 a^3+3359232 a^2b
\\ & + 3359232 a b^2+559872 b^3+5318784 a^2+13436928 a b+5318784
b^2 \\ &  +12581568 a+11337408 b+6920640) u t^2
\Big) 
+ \omega^3\,\Big(
15552 (2 b+3) (2 b+1) t^6 \\ & + (-62208 a^2-186624 a b-62208 b^2-342144 a
- 217728 b-216000) u t^3 + 13824 u^2
\Big)
\end{align*}

%%%%%%%%%%%%%%% 3
\begin{align*}
  C^{(12)}_{0,3} &=
  104976 (2 b+5) (2 b+3) (2 b+1) (6 a+4 b+25) t^6
  \\ & - 46656 (2 b+3)
  \Big(144 a^3+648 a^2 b+684 a b^2+180 b^3+1836 a^2+4500 a b
  \\ & +2052 b^2+7115
  a+7295 b+8275\Big) u t^3 \\ & +(13436928 a^4+83980800 a^3 b+141087744 a^2
  b^2+83980800 a b^3+13436928 b^4 \\ & +193155840 a^3+733992192 a^2
  b+733992192 a b^2+193155840 b^3+954488448 a^2 \\ & +2101946112 a
  b+954488448 b^2+1975601664 a+1975601664 b+1459866240) u^2
  \\ &
  -\omega\Big(
    209952 (2 b+5) (2 b+3) (2 b+1) t^7
  - 23328 (2 b+3) (216 a^2+576 a  b+252 b^2 \\ & +1824 a+1860 b+3325) u t^4
  + (10077696 a^3 + 43670016 a^2 b + 43670016 a b^2
  \\ & + 10077696 b^3 + 105255936 a^2 + 242984448 a b  + 105255936  b^2
  \\ & + 326032128 a + 326032128 b + 307463040) t u^2
  \Big)
  + \omega^2\Big(
-1119744 (2 b+3) (a+b+4) u t^5 \\& + (2239488 a^2+5598720 a b+2239488 b^2+13996800 a+13996800 b+18195840) t^2 u^2  
  \Big) \\ & 
  +\omega^3 \Big(
  (124416 b+186624) t^6 u  - 124416 (a+b+2) t^3 u^2                            
  \Big)
\end{align*}

%%%%%%%%%%%%%%% 4
\begin{align*}
  C^{(12)}_{0,4} &=
  209952 (2 b+5) (2 b+3) (18 a+20 b+103) t^6 u+  \Big(-6718464 a^3-75582720 a^2
  b \\ & -141087744 a b^2-58786560 b^3-199034496 a^2-900274176 a b-654210432
  b^2 \\& -1375418880 a -2301447168 b-2585675520 \Big) t^3 u^2 \\ &
  + \Big(41990400 a^3+159563520 a^2  b+159563520 a b^2+41990400 b^3 
  +517321728 a^2 \\& +1142138880 a b+517321728  b^2+2022630912 a+2022630912 b+2538086400\Big) u^3
\\ &
  - \omega\Big(
   1259712 (2 b+5) (2 b+3) u t^7
  -(5038848 a^2+33592320 a b+26873856 b^2\\ &  + 92938752 a +180278784 b+288567360) t^4 u^2
  +(21835008 a^2 + 50388480 a b \\ &  + 21835008 b^2  + 168521472 a + 168521472 b + 303264000) t u^3
  \Big) \\&
-279936\,\omega^2\, t^2 u^2 (4 a t^3+10 b t^3+31 t^3-10 a u-10 b u-34 u) 
  \\ & 
  + 62208\,\omega^3\, u^2 t^3 (t^3-u)
\end{align*}

%%%%%%%%%%%%%%% 5   (line 104 Typo 5 corrected: added sign + (otherwise it looks like a product)
\begin{align*}
  C^{(12)}_{0,5} &=
  1259712 (2 b+5) (12 b+6 a+53) u^2 t^6 + \Big(-30233088 a^2-131010048 a b
  \\ & -94058496b^2-482049792 a-787739904 b-1593768960 \Big) u^3 t^3
  \\ &
  + \Big(63825408 a^2+141087744 a  b+63825408 b^2+591224832 a+591224832 b+1325030400 \Big) u^4
\\&
+279936 \,\omega\, t u^2   \Big(-18 b t^6-45 t^6+48 a t^3 u+90 b t^3 u+341 t^3 u-72 a u^2-72 b
u^2-308 u^2 \Big)  \\&
  -1119744\,\omega^2\,  t^2 u^3 (t^3-u)
\end{align*}

%%%%%%%%%%%%%%% 6
\begin{align*}
  C^{(12)}_{0,6} &=
  839808 u^3 (6 a t^6+28 b t^6+109 t^6-52 a t^3 u-84 b t^3 u-392 t^3 u
 +56 a u^2+56 b u^2+288 u^2) 
\\& 
  -1679616\,\omega\, t u^3 (t^3-4 u) (t^3-u)
\end{align*}

%%%%%%%%%%%%%%% 7
\begin{align*}
  C^{(12)}_{0,7} &=6718464 \left(t^3-2 u\right)
\left(t^3-u\right) u^4\end{align*}

%%%%%%%%%%%%%%% 8
\begin{align*}
  C^{(12)}_{1,1} &=31104 (1+2 a) (1+2 b) (2+3 a+3 b) (4+3 a+3 b) (5+3 a+3 b) t
\\&
  - 77760\, {\omega}\, (1+2 a) (1+2 b) (4+3 a+3 b) (5+3 a+3 b) t^2
\\ &
+ 20736\,{\omega}^2\, (1+2 b)  \left((22+9 b+9 a (5+2 a+2 b)) t^3-4 u\right)
\\ &
+5184 \,{\omega}^3\, (1+2 b) 
  t \left(-(7+6 a) t^3 + 4 u\right)
\end{align*} 

%%%%%%%%%%%%%%% 9
\begin{align*}
  C^{(12)}_{1,2} &=
  -93312 (2 b+3) (2 b+1) (27 a^2+36 a b+9 b^2+138 a+78 b+170) t^4
  \\& +(3359232 a^4+30233088 a^3 b+53747712 a^2 b^2+30233088 a b^3+3359232 b^4
  \\& +44229888 a^3+196515072 a^2 b+196515072 a b^2+44229888 b^3+172347264 a^2
  \\& +396949248 a b+172347264 b^2+255332736 a+250356096 b+120061440) u t
  \\&
  + \omega\,\Big(
  349920 (2 b+3) (2 b+1) (4 a+2 b+11) t^5
  - ( 2799360 a^3 + 16796160 a^2 b
  \\ & +  16796160 a b^2 + 2799360 b^3 + 26593920 a^2
  +  67184640 a b + 26593920 b^2 \\& + 62907840 a + 56687040 b + 34603200) u t^2
 \Big)
\\ &
- \omega^2\Big(
  186624 (2 b+3) (2 b+1) t^6
- (746496 a^2+2239488 a b+746496 b^2 \\& + 4105728 a+2612736 b+2592000) u t^3
+ 165888 u^2
  \Big)
-10368\,\omega^3\, t u (6 a t^3+7 t^3-4 u)
\end{align*}

%%%%%%%%%%%%%%% 10
\begin{align*}
  C^{(12)}_{1,3} &=
  419904 (2 b+5) (2 b+3) (2 b+1) t^7 - 46656 (2 b+3) \Big(216 a^2+576 a b+252 b^2+1824 a
  \\& +1860 b+3325\Big) u t^4 + \Big(20155392 a^3+87340032 a^2 b+87340032 a b^2+20155392 b^3
  \\& +210511872 a^2+485968896 a b+210511872 b^2+652064256 a+652064256 b+614926080\Big) u^2 t
\\& + \omega\,\Big(
5598720 (2 b+3) (a+b+4) u t^5
- ( 11197440 a^2 + 27993600 a b \\& + 11197440 b^2 + 69984000 a
+ 69984000 b + 90979200) u^2 t^2
\Big)
\\&
+ 746496\,\omega^2\, u t^3 (-2 b t^3-3 t^3+2 a u+2 b u+4 u)
\end{align*}

%%%%%%%%%%%%%%% 11
\begin{align*}
  C^{(12)}_{1,4} &=
  2519424 (2 b+5) (2 b+3) u t^7+   \Big(-10077696 a^2-67184640 a b-53747712 b^2
  \\& -185877504 a-360557568 b-577134720 \Big) u^2 t^4+\Big(43670016 a^2+100776960 a b
  \\& +43670016 b^2+337042944 a+337042944 b+606528000 \Big) u^3 t
  \\& + 1399680\,\omega\, t^2 u^2 (4 a t^3+10 b t^3+31 t^3-10 a u-10 b u-34 u)
\\&
-746496\,\omega^2\, u^2 t^3 (t^3-u)
\end{align*} 

%%%%%%%%%%%%%%% 12
\begin{align*}
  C^{(12)}_{1,5} &=
  -559872 t u^2 \Big(-18 b t^6-45 t^6+48 a t^3 u+90 b t^3 u+341 t^3 u
  \\& -72 a u^2-72 b u^2-308 u^2\Big)
+5598720\,\omega\, t^2 u^3 (t^3-u)
\end{align*}

%%%%%%%%%%%%%%% 13
\begin{align*}
  C^{(12)}_{1,6} &=3359232\, t\, \left(t^3-4 u\right) \left(t^3-u\right) u^3
\end{align*}

%%%%%%%%%%%%%%% 14
\begin{align*}
  C^{(12)}_{2,1} &=
  77760\, (2 b+1) (2 a+1) (3 a+3 b+5) (3 a+3 b+4) t^2\\&
- 51840\,\omega\, (2 b+1) (18 a^2 t^3+18 a b t^3+45 a t^3+9 b t^3+22 t^3-4 u)\\&
+ 31104\,\omega^2\, t\, (2 b+1) (6 a t^3+7 t^3-4 u)\\&
-10368\,\omega^3\, t^2\, (t^3-u) (2 b+1)
\end{align*}

%%%%%%%%%%%%%%% 15
\begin{align*}
  C^{(12)}_{2,2} &=
  -349920 (2 b+3) (2 b+1) (4 a+2 b+11) t^5 + \Big(2799360 a^3+16796160 a^2 b+16796160
  a b^2 \\& +2799360 b^3+26593920 a^2+67184640 a b+26593920 b^2+62907840 a+56687040
  b\\ & +34603200 \Big) u t^2
  +\omega\,\Big( 466560 (2 b+3) (2 b+1) t^6
  - (1866240 a^2 +5598720 a b  \\ &  +1866240  b^2 + 10264320 a + 6531840 b + 6480000) u t^3
  +414720 u^2 \Big)
\\ &
  +62208\,\omega^2\, t u (6 a t^3+7 t^3-4 u)
-20736\,\omega^3\, t^2 u (t^3-u)
\end{align*}

%%%%%%%%%%%%%%% 16
\begin{align*}
  C^{(12)}_{2,3} &=
  - 5598720 (2 b+3) (a+b+4) u t^5 +  \Big(11197440 a^2+27993600 a b+11197440 b^2
  \\ & + 69984000 a+69984000 b+90979200 \Big) u^2 t^2
\\ & -1866240\,\omega\, u t^3 (-2 b t^3-3 t^3+2 a u+2 b u+4 u)
\end{align*}

%%%%%%%%%%%%%%% 17
\begin{align*}
  C^{(12)}_{2,4} &=
  -1399680 t^2 u^2 \Big((31+4 a+10 b) t^3-2 (17+5 a+5 b) u\Big)
+ 1866240\, {\omega} t^3 \left(t^3-u\right) u^2
\end{align*} 

%%%%%%%%%%%%%%% 18
\begin{align*}
  C^{(12)}_{2,5} &=-5598720\, t^2 \,\left(t^3-u\right) u^3\end{align*} 

%%%%%%%%%%%%%%% 19
\begin{align*}
  C^{(12)}_{3,0} &=
\omega^3\Big( -(2304 b + 1152) t^3+(2304 a+2304 b+2304) u \Big)
\end{align*}

%%%%%%%%%%%%%%% 20
\begin{align*}
  C^{(12)}_{3,1} &=
34560 (2 b+1) \Big(18 a^2 t^3+18 a b t^3+45 a t^3+9 b t^3+22 t^3-4 u \Big)
\\& -51840\,\omega\, t (2 b+1) (6 a t^3+7 t^3-4 u)
+41472\,\omega^2\, t^2 (t^3-u) (2 b+1)
 -4608\,\omega^3\, u (t^3-u)
\end{align*}

%%%%%%%%%%%%%%% 21 (Typo 6) $k=3$, $C^{(12)}_{3,2}$,  +82944\,\omega^3\, t^2 u (t^3-u) must be $\omega2$.  
\begin{align*}
  C^{(12)}_{3,2} &=
  -311040 (2 b+3) (2 b+1) t^6 + \Big(1244160 a^2+3732480 a b+1244160 b^2+6842880a
  \\ & +4354560 b+4320000 \Big) u t^3-276480 u^2 - 103680\,\omega\, t u (6 a t^3+7 t^3-4 u)
\\& +82944\,\omega^2\, t^2 u (t^3-u)
\end{align*} 

%%%%%%%%%%%%%%% 22
\begin{align*}
  C^{(12)}_{3,3} &=1244160\, t^3 u (- (3+2b) t^3 + 2 (2+a+b) u)
\end{align*} 

%%%%%%%%%%%%%%% 23
\begin{align*}
  C^{(12)}_{3,4} &=-1244160\, t^3 \left(t^3-u\right) u^2\end{align*}

%%%%%%%%%%%%%%% 24
    \begin{align*}
      C^{(12)}_{4,0} &=
      3456{\omega}^2 \left((1+2 b) t^3-2 (1+a+b) u\right)
    \end{align*} 

%%%%%%%%%%%%%%% 25    
\begin{align*}
  C^{(12)}_{4,1} &=
  25920 t (2 b+1) (6 a t^3+7 t^3-4 u) - 51840\,\omega\, t^2 (t^3-u) (2 b+1)
+13824\omega^2\, u (t^3-u)
\end{align*}

%%%%%%%%%%%%%%% 26
\begin{align*}
  C^{(12)}_{4,2} &=51840 t \Big((7+6 a) t^3-4 u\Big) u
  -103680\, {\omega}\, t^2 \left(t^3-u\right)
u\end{align*}

%%%%%%%%%%%%%%% 27
 \begin{align*}
   C^{(12)}_{5,0} &=
   3456\,{\omega}\, \left(-(1+2 b) t^3 + 2 (1+a+b) u\right)
 \end{align*}

%%%%%%%%%%%%%%% 28
\begin{align*}
  C^{(12)}_{5,1} &=20736 (1+2 b) t^2 \left(t^3-u\right)-13824 {\omega} \left(t^3-u\right)
u\end{align*} 

%%%%%%%%%%%%%%% 29
\begin{align*}
  C^{(12)}_{5,2} &=41472 t^2 \left(t^3-u\right) u\end{align*}

%%%%%%%%%%%%%%% 30
\begin{align*}
  C^{(12)}_{6,0} &=1152
\left((1+2 b) t^3-2 (1+a+b) u\right)\end{align*} 

%%%%%%%%%%%%%%% 31
\begin{align*}
  C^{(12)}_{6,1} &=4608 \left(t^3-u\right) u\end{align*}

%%%%%%%%%%%%%%&&&&&&

\section{TTW at $k=4$: the second integral ${\cal I}_{2}$ coefficients, see (\ref{I2-k=4}) \label{I2k4coeffs}}
Non-vanishing coefficients for the integral ${\cal I}_2$ of TTW at $k=4$, 
defined in (\ref{I2-k=4}):

%   TTW k=4, I2 coeffs
%   Manually reorganized   Jan 11, 2025

\begin{align*} C^{(2)}_{0,1}  &=
  -8192 (1 + 2 a) (1 + 2 b) \Big(153 + 512 a^5 + 1290 b + 
  3728 b^2 + 4672 b^3 + 2560 b^4 + 512 b^5   \\ & + 
  2560 a^4 (1 + b) +      64 a^3 (73 + 160 b + 80 b^2) + 
       16 a^2 (233 + 876 b + 960 b^2 + 320 b^3)  \\ & +  
       2 a (645 + 3728 b + 7008 b^2 + 5120 b^3 + 1280 b^4)\Big)
       \\ &
       + 262144 \,\omega t (1 + 2 a) (2 + a + b) (1 + 2 b) (1 + 2 a + 2 b) (3 + 2 a +  2 b) (3 + 4 a + 4 b)
       \\ &  
       -12288 \,\omega^2\, t^2 (1 + 2 a) (1 + 2 b) (3 + 4 a + 4 b) \big(91 + 32 a^2 + 108 b + 
       32 b^2 + 8 a (13 + 8 b)\big)
       \\ &
       +  1024\,\omega^3\,  t^3 (1 + 2 b) \Big(755 - 928 a^5 + 877 b + 259 b^2 - 8 b^3 + 
    4 a^3 (327 + 117 b + 35 b^2) \\ &  + 8 a (260 + 245 b + 63 b^2) +
    2 a^2 (879 + 435 b + 29 b^2 + 6 b^3)\Big)
    \\ &
    + 512 \,\omega^4\,  \Big(2 (1 + 2 b) (147 + 48 a + 56 a^2 + 145 b +  31 b^2) t^4  \\ & +
    \big(-815 + 928 a^5 - 1421 b - 771 b^2 - 116 b^3  
    - 4 a^3 (287 + 117 b + 35 b^2) \\ & -  4 a (307 + 275 b + 61 b^2)
    -  2 a^2 (415 + 163 b + 29 b^2 + 6 b^3)\big) u \Big)
\end{align*}  

\begin{align*} C^{(2)}_{0,2}  &=
  - 16384 \bigg(
    -\Big((3 + 8 b + 4 b^2) (36945 + 1024 a^4   + 43992 b + 19360 b^2  + 3712 b^3 + 256 b^4 \\ &
    + 128 a^3 (79 + 26 b) + 16 a^2 (2367 + 1496 b + 240 b^2)  + 
            8 a (7749 + 7154 b + 2192 b^2 + 224 b^3)) t^4\Big) \\ &  + 
       2 \Big(126153 + 512 a^6 + 366027 b + 415690 b^2 + 237376 b^3   +  71616 b^4
          + 10496 b^5  + 512 b^6 \\ & + 256 a^5 (41 + 30 b) + 
          192 a^4 (373 + 452 b + 136 b^2)   + 
          16 a^3 (14839 + 24036 b + 12992 b^2 + 2368 b^3) \\ & + 
          a^2 (416002 + 836608 b + 625920 b^2
          + 207872 b^3 +  26112 b^4)  +  \\ &
          a (366612 + 888014 b + 836560 b^2 + 384576 b^3 + 
          86784 b^4 + 7680 b^5)\Big) u
          \bigg)
    \\[3pt] &
  + 65536 \,\omega  t \Big(-2 (3 + 8 b + 4 b^2) (1827 + 96 a^3 + 1440 b + 
          376 b^2 + 32 b^3 + 4 a^2 (197 + 56 b)  \\ & + 
          a (2119 + 1164 b + 160 b^2)) t^4 + (27333 + 256 a^5 + 
          67068 b + 60140 b^2 + 24432 b^3 + 4416 b^4 \\ & + 256 b^5    + 
          192 a^4 (23 + 16 b) + 32 a^3 (765 + 886 b + 248 b^2) \\ &
          + 32 a^2 (1891 + 2945 b + 1496 b^2 + 248 b^3)   +
          4 a (16977 + 33276 b + 23548 b^2 + 7088 b^3 +   768 b^4)) u \Big)
  \\[3pt] & 
  - 8192 \,\omega^2\, t^2 \Big(-8 (3 + 8 b + 4 b^2) (442 + 56 a^2 + 206 b + 
          24 b^2 + a (309 + 76 b)) t^4  \\ & + 
       3 (10173 + 256 a^4 + 20560 b + 13792 b^2 + 3456 b^3   + 
          256 b^4 + 1152 a^3 (3 + 2 b) \\ & + 
          8 a^2 (1757 + 1948 b + 512 b^2)   + 
          2 a (10859 + 16392 b + 7824 b^2 + 1152 b^3)) u\Big)
  \\[3pt] &  
  -2048 \,\omega^3\,  t^3 \Big(32 (19 + 6 a + 4 b) (3 + 8 b + 
          4 b^2) t^4   +  (-6611 + 928 a^5 - 10157 b - 4611 b^2 - 
          504 b^3 \\ &  - 4 a^3 (327 + 117 b + 35 b^2)  - 
          8 a (1508 + 1557 b + 383 b^2) -  
          2 a^2 (2223 + 1331 b + 29 b^2 + 6 b^3)) u\Big)
  \\[3pt] &
  + 2048 \,\omega^4\,  \Big( 8 (3 + 8 b + 4 b^2) t^8 + (75 + 56 a^2 + 49 b - 
          b^2 - 16 a (3 + 4 b)) t^4 u \\ & - (191 + 56 a^2 + 161 b + 
          31 b^2 + 8 a (17 + 4 b)) u^2\Big)
\end{align*}

  \begin{align*} C^{(2)}_{0, 3}  &=   
    - 65536 \Big( 4 (15 + 46 b + 36 b^2 + 8 b^3) (501 + 32 a^2 + 176 b
      + 16 b^2 + 4 a (61 + 12 b)) t^8 \\ &   - (3 + 2 b) (295305
      + 1024 a^4 + 332096 b + 135824 b^2 + 23936 b^3 + 1536 b^4
      + 512 a^3 (36 + 13 b) \\ & + 16 a^2 (7327 + 4720 b + 736 b^2)
      + 16 a (19627 + 17498 b + 5064 b^2 + 480 b^3)) t^4 u  \\ &
      + 2 (571104 + 1280 a^5 + 911745 b + 563360 b^2 + 168160 b^3
      + 24064 b^4 + 1280 b^5 + 256 a^4 (94 + 41 b)   \\ &
      + 32 a^3 (5255 + 4040 b + 784 b^2)
      + 16 a^2 (35213 + 37202 b + 13152 b^2 + 1568 b^3) \\ &
      + a (911949 + 1209232 b + 595232 b^2 + 129280 b^3 + 10496 b^4)) u^2 \Big)
\\[3pt] &
    + 262144 \,\omega  t \Big(4 (13 + 3 a + 2 b) (15 + 46 b + 36 b^2 + 8 b^3) t^8   
      -  2 (3 + 2 b) (96 a^3 +   4 a^2 (337 + 112 b)  \\ & +  a (5849 + 3496 b + 496 b^2) + 
          4 (1938 + 1610 b + 427 b^2 + 36 b^3)) t^4 u \\ &  + (63279 + 
          512 a^4 + 83658 b + 39216 b^2 + 7648 b^3 + 512 b^4 + 
          32 a^3 (239 + 100 b) \\ &  + 32 a^2 (1227 + 913 b + 168 b^2) + 
          4 a (20973 + 21682 b + 7304 b^2 + 800 b^3)) u^2 \Big)
\\[3pt] &
    - 131072 \,\omega^2\, t^2 \Big(2 (15 + 46 b + 36 b^2 + 8 b^3) t^8 - 
      2 (3 + 2 b) (1082 + 56 a^2 + 592 b + 76 b^2  \\ &
      + a (499 + 152 b)) t^4 u +  3 (3075 + 96 a^3 + 3218 b + 1024 b^2 + 96 b^3
      +  4 a^2 (257 + 104 b)  \\ &  + a (3274 + 2384 b + 416 b^2)) u^2\Big)
\\[3pt] &
      + 131072 \,\omega^3\,  t^3 u \Big(-2 (3 + 2 b) (29 + 6 a + 8 b) t^4
        + (251 +  28 a^2 + 192 b + 32 b^2 + 4 a (51 + 20 b)) u \Big)
\\[3pt] &
    + 65536 \,\omega^4\,  t^4 u \Big((3 + 2 b) t^4 - 2 (2 + a + b) u\Big)  
  \end{align*}

  \begin{align*} C^{(2)}_{0, 4}  &=
     65536 \Big(16 (105 + 352 b + 344 b^2 + 128 b^3 + 16 b^4) t^{12}  \\ &  - 
       24 (15 + 16 b + 4 b^2) (991 + 32 a^2 + 400 b + 40 b^2 + 
          4 a (89 + 20 b)) t^8 u + (3077145 + 1024 a^4 \\ &  + 3363192 b + 
          1332928 b^2 + 227200 b^3 + 14080 b^4 + 
          128 a^3 (339 + 130 b) \\ &  + 
          16 a^2 (30887 + 20344 b + 3216 b^2) + 
          8 a (266769 + 236602 b + 67568 b^2 + 
             6240 b^3)) t^4 u^2 \\ &  - (3138489 + 10496 a^4 + 3138864 b + 
          1150112 b^2 + 182528 b^3 + 10496 b^4 + 
          256 a^3 (713 + 220 b)  \\ & + 
          32 a^2 (35941 + 20216 b + 2864 b^2) +
          16 a (196185 + 154588 b + 40432 b^2 + 3520 b^3)) u^3 \Big)
   \\[3pt] &
   + 262144 \,\omega  t u \Big(8 (53 + 9 a + 10 b) (15 + 16 b + 4 b^2) t^8 - 
       2 (47997 + 96 a^3 + 41048 b + 11152 b^2 \\ &  + 960 b^3 + 
          4 a^2 (757 + 280 b) + 
          a (22919 + 14524 b + 2176 b^2)) t^4 u \\ &  + (104727 + 
          1600 a^3 + 81684 b + 20336 b^2 + 1600 b^3 + 
          16 a^2 (1271 + 380 b) \\ &  + 
          a (81732 + 44960 b + 6080 b^2)) u^2\Big)
   \\[3pt]&
   -  131072 \,\omega^2\, t^2 u \Big(12 (15 + 16 b + 4 b^2) t^8 - 
     2 (3912 + 56 a^2 + 2374 b + 336 b^2 +  a (1069 + 380 b)) t^4 u
     \\ &  +   3 (2969 + 208 a^2 + 1632 b + 208 b^2 +  40 a (41 + 12 b)) u^2\Big)
   \\[3pt] &
    - 262144 \,\omega^3\, t^3 u^2 \Big((59 + 6 a + 20 b) t^4 - 2 (33 + 10 a + 10 b) u \Big)
    \\[3pt] &   
    +65536 \,\omega^4\,  t^4 (t^4 - u) u^2
 \end{align*}  

  \begin{align*} C^{(2)}_{0, 5}  &=
      1048576 u \Big(8 (105 + 142 b + 60 b^2 + 8 b^3) t^{12}  - 
       3 (5 + 2 b) (2349 + 32 a^2 + 1040 b + 112 b^2 \\ & + 
          4 a (145 + 36 b)) t^8 u   +  
       2 (208 a^3 + 496 a^2 (11 + 3 b) + 
          a (40965 + 20120 b + 2416 b^2)   \\ & + 
          3 (30720 + 20939 b + 4648 b^2 + 336 b^3)) t^4 u^2      - 
       2 (80847 + 704 a^3 + 50892 b \\ & + 10480 b^2 + 704 b^3   +  
       16 a^2 (655 + 156 b) + 12 a (4241 + 1880 b + 208 b^2)) u^3 \Big)
      \\[3pt] &
          + 2097152 \,\omega  t u^2 \Big(18 (5 + 2 b) (9 + a + 2 b) t^8  - (6939 + 
          112 a^2 + 3290 b + 376 b^2 \\ & + 
          2 a (961 + 252 b)) t^4 u + (6609 + 304 a^2 + 2888 b    +
          304 b^2 + 8 a (361 + 84 b)) u^2\Big)
    \\[3pt] &
    - 1048576 \,\omega^2\, t^2 u^2 (3 (5 + 2 b) t^8
    -    2 (155 + 19 a + 39 b) t^4 u  + 24 (13 + 3 a + 3 b) u^2)      
    \\[3pt]&
    - 2097152 \,\omega^3\,  t^3 (t^4 - u) u^3
  \end{align*}

  \begin{align*} C^{(2)}_{0,6}  &=
    - 2097152 u^2 \Big(-12 (35 + 24 b + 4 b^2) t^{12} + (8313 + 32 a^2 + 
          3920 b + 448 b^2 + 4 a (313 + 84 b)) t^8 u \\ & - (32223 + 
          496 a^2 + 12712 b + 1232 b^2 + 
          8 a (1051 + 224 b)) t^4 u^2 + \\ &
       4 (6321 + 208 a^2 + 2312 b + 208 b^2 + 
       8 a (289 + 56 b)) u^3\Big) \\&
        + 4194304 \,\omega  t u^3 \Big(2 (55 + 3 a + 14 b) t^8 - (681 + 84 a + 
    140 b) t^4 u + 28 (21 + 4 a + 4 b) u^2\Big)
    \\ &
    -2097152 \,\omega^2\, t^2 (t^4 - 12 u) (t^4 - u) u^3
  \end{align*}

  \begin{align*} C^{(2)}_{0, 7}  &=   
    - 33554432 u^3 \Big( - (7 + 2 b) t^{12}
   + 3 (29 + 2 a + 6 b) t^8 u - (273 + 32 a + 48 b) t^4 u^2
   + 4 (49 + 8 a + 8 b) u^3 \Big) \\[3pt] &
   + 33554432 \,\omega  t (t^4 - 4 u) (t^4 - u) u^4
  \end{align*}

  \begin{align*} C^{(2)}_{0, 8}  &=   
    16777216 (t^4 - 4 u)^2 (t^4 - u) u^4 \end{align*}

    {
      \begin{align*} C^{(2)}_{1, 0}  &=
        -128\,\omega^3\, (1 + 4 a + 4 b)
        \Big(-59 + 928 a^5 - 77 b - 3 b^2 + 8 b^3 + 
       8 a (2 + 19 b + b^2) \\ & - 4 a^3 (199 + 117 b + 35 b^2) -   
       2 a^2 (47 - 77 b + 29 b^2 + 6 b^3)\Big) 
       \\ &
       + 128\,\omega^4\,  t \Big(-59 + 928 a^5 - 77 b - 3 b^2 + 8 b^3 +    
       8 a (2 + 19 b + b^2) \\ & - 4 a^3 (199 + 117 b + 35 b^2) - 
       2 a^2 (47 - 77 b + 29 b^2 + 6 b^3)\Big)  
      \end{align*}  

      \begin{align*} C^{(2)}_{1,1}  &=
        -524288 (1 + 2 a) (1 + 2 b) \Big(18 + 16 a^4 + 81 b + 124 b^2
        + 76 b^3 + 16 b^4 + a^3 (76 + 64 b)
        \\[3pt] &
        + 4 a^2 (31 + 57 b + 24 b^2)
        + a (81 + 248 b + 228 b^2 + 64 b^3)\Big) t
        \\[3pt] & +   8192 \,\omega  t^2 (1 + 2 a) (1 + 2 b) (3 + 4 a + 4 b)
        (603 + 224 a^2 + 748 b +  224 b^2 + a (748 + 448 b))
        \\[3pt] &
        - 8192\,\omega^2\, t^3 (1 + 2 a) (1 + 2 b) (683 + 256 a^2 + 872 b + 272 b^2 + 
        24 a (35 + 22 b))
         \\[3pt] &
        -  1024 \,\omega^3\,  \Big(-8 (1 + 2 b) (79 + 56 a^2 + 36 b + 
          32 a (5 + 2 b)) t^4 + (-83 + 928 a^5 + 131 b + 61 b^2   \\ & +  
          8 b^3 + 8 a (-20 + 19 b + b^2) - 
          4 a^3 (167 + 117 b + 35 b^2) - 
          2 a^2 (-65 - 141 b + 29 b^2 + 6 b^3)) u\Big)
        \\[3pt] &
        -  8192 \,\omega^4\,  t (2 (1 + a) (1 + 2 b) t^4 + (1 + 2 a - 4 a^2 - 2 b) u)
      \end{align*}

      \begin{align*} C^{(2)}_{1,2}  &=
     -131072 \Big(-2 (3 + 8 b + 4 b^2) (1827 + 96 a^3 + 1440 b + 376 b^2 + 
          32 b^3 + 4 a^2 (197 + 56 b)  \\ &  +
          a (2119 + 1164 b + 160 b^2)) t^5 + (27333 + 256 a^5 + 
          67068 b + 60140 b^2 + 24432 b^3 \\ & + 4416 b^4 + 256 b^5 + 
          192 a^4 (23 + 16 b) + 32 a^3 (765 + 886 b + 248 b^2)  \\ &  + 
          32 a^2 (1891 + 2945 b + 1496 b^2 + 248 b^3) + 
          4 a (16977 + 33276 b + 23548 b^2 + 7088 b^3 + 768 b^4)) t u\Big)
     \\[3pt] &
         + 16384 \,\omega  t^2 \Big(-8 (3 + 8 b + 4 b^2) (1026 + 124 a^2 + 478 b + 
          56 b^2 + 5 a (145 + 36 b)) t^4 \\ & + (70173 + 1792 a^4 +   
          142320 b + 95776 b^2 + 24064 b^3 + 1792 b^4   + 
          32 a^3 (761 + 504 b) \\ & + 16 a^2 (6163 + 6836 b + 1792 b^2) +   
          2 a (75225 + 113888 b + 54544 b^2 + 8064 b^3)) u\Big)
    \\[3pt] &
        -16384 \,\omega^2\, t^3 \Big(-4 (155 + 50 a + 32 b) (3 + 8 b   + 4 b^2) t^4
          \\ &
          + (6683 + 512 a^3 + 10392 b + 4720 b^2 + 
          512 b^3 + 176 a^2 (29 + 18 b)   + 
          2 a (6335 + 6640 b + 1616 b^2)) u\Big)
        \\[3pt] & 
     + 16384 \,\omega^3\,  \Big(-12 (3 + 8 b + 4 b^2) t^8 + (223 + 56 a^2 + 
       204 b + 48 b^2 + 32 a (11 + 6 b)) t^4 u
       \\ & 
      - (19 - 4 a + 16 a^2 + 12 b) u^2 \Big)
        -16384 \,\omega^4\,  t (2 (1 + a) t^4 - u) u
      \end{align*}

      \begin{align*} C^{(2)}_{1, 3}  &=   
        -  524288 \Big(4 (13 + 3 a + 2 b) (15 + 46 b + 36 b^2 + 8 b^3) t^9 - 
       2 (3 + 2 b) (96 a^3 + 4 a^2 (337 + 112 b) \\ &   + 
          a (5849 + 3496 b + 496 b^2) + 
          4 (1938 + 1610 b + 427 b^2 + 36 b^3)) t^5 u 
          \\ &
          + (63279 + 512 a^4 + 83658 b + 39216 b^2 + 7648 b^3 + 512 b^4
          + 32 a^3 (239 + 100 b)
          \\ & + 32 a^2 (1227 + 913 b + 168 b^2) 
          + 4 a (20973 + 21682 b + 7304 b^2 + 800 b^3)) t u^2\Big)
        \\[3pt] &
        + 262144 \,\omega  t^2 \Big(6 (15 + 46 b + 36 b^2 + 8 b^3) t^8 - 
          2 (3 + 2 b) (124 a^2 + 5 a (235 + 72 b)
        \\ &
          + 4 (629 + 346 b + 45 b^2)) t^4 u
          + (21351 + 672 a^3 + 22378 b + 7136 b^2 + 672 b^3  \\ &
          + 8 a^2 (901 + 364 b) + 4 a (5701 + 4160 b + 728 b^2)) u^2 \Big)
        \\[3pt] &        
        + 262144 \,\omega^2\, t^3 u \Big((3 + 2 b) (235 + 50 a + 64 b) t^4
          - (1025 + 132 a^2 + 780 b + 128 b^2 \\ & + 8 a (107 + 42 b)) u \Big)        
        - 131072 \,\omega^3\, u \Big(
          6 (3 + 2 b) t^8 - (27 + 16 a + 12 b) t^4 u + u^2
          \Big)
      \end{align*}

      \begin{align*} C^{(2)}_{1, 4}  &=
        - 524288 t u \Big(8 (53 + 9 a + 10 b) (15 + 16 b + 4 b^2) t^8 
          - 2 (47997 + 96 a^3 + 41048 b + 11152 b^2  \\ &
          + 960 b^3  + 4 a^2 (757 + 280 b)  + 
          a (22919 + 14524 b + 2176 b^2)) t^4 u + (104727    + 
          1600 a^3 \\ & + 81684 b + 20336 b^2 + 1600 b^3 + 
          16 a^2 (1271 + 380 b) + 
          a (81732 + 44960 b + 6080 b^2)) u^2 \Big)
        \\[3pt]&
        + 262144 \,\omega  t^2 u \Big( 36 (15 + 16 b + 4 b^2) t^8
        - 2 (9156 + 124 a^2 + 5590 b + 800 b^2 \\ &
        + 25 a (101 + 36 b)) t^4 u + (20673 + 1456 a^2 + 11384 b
        + 1456 b^2 + 32 a (358 + 105 b)) u^2 \Big)
        \\ &
        +  262144 \,\omega^2\, t^3 u^2 \Big(5 (95 + 10 a + 32 b) t^4
        -  2 (267 + 84 a + 80 b) u\Big)
        \\ &
        -786432 \,\omega^3\,  t^4 (t^4 - u) u^2
      \end{align*}

      \begin{align*} C^{(2)}_{1, 5}  &=   
        - 4194304 t u^2 \Big(18 (5 + 2 b) (9 + a + 2 b) t^8
          - (6939 + 112 a^2 +   3290 b + 376 b^2
          \\ & + 2 a (961 + 252 b)) t^4 u
        + (6609 + 304 a^2 + 2888 b + 304 b^2 + 8 a (361 + 84 b)) u^2\Big)
        \\ &
                + 6291456 \,\omega  t^2 u^2 \Big(3 (5 + 2 b) t^8
          - 2 (122 + 15 a + 31 b) t^4 u 
          + 2 (121 + 28 a + 28 b) u^2 \Big)
        \\ &
        + 16777216 \,\omega^2\, t^3 (t^4 - u) u^3
      \end{align*}

      \begin{align*}
        C^{(2)}_{1, 6}  &=
        -  8388608\, t u^3 (2 (55 + 3 a + 14 b) t^8 - (681 + 84 a + 
        140 b) t^4 u + 28 (21 + 4 a + 4 b) u^2)
        \\[3pt] &
        +4194304 \,\omega  t^2 (3 t^4 - 28 u) (t^4 - u) u^3
      \end{align*}

      \begin{align*} C^{(2)}_{1,7}  &=
        -67108864 \, t u^4 (t^8 - 5 t^4 u + 4 u^2)
      \end{align*}

    \begin{align*} C^{(2)}_{2, 0}  &=   
      1024 \,\omega^2\, (1 + 2 a) (1 + 2 a + 2 b) (1 + 4 a + 4 b) (-1 + 8 a + 4 b)
      \\[3pt] &
      - 128\,\omega^3\,  t \Big( -75 + 928 a^5 - 45 b + 125 b^2 + 8 b^3
        + 8 a (10 + 75 b + 33 b^2) \\[3pt] &
        - 4 a^3 (71 + 117 b + 35 b^2)
        + a^2 (354 + 922 b - 58 b^2 - 12 b^3)\Big)
      \\[3pt]&
      +  512\,\omega^4\,  t^2 (1 + 2 a) (-1 + 8 a + 4 b)
    \end{align*}

    \begin{align*}
      C^{(2)}_{2,1}  &=
      -8192 (1 + 2 a) (1 + 2 b) \big(1809 + 896 a^3 + 4656 b
      + 3664 b^2 + 896 b^3
      \\ &
       + 16 a^2 (229 + 168 b) + 16 a (291 + 458 b + 168 b^2)\big) t^2 \\[3pt] &
      + 24576\,\omega\,  t^3 (1 + 2 a) (1 + 2 b)
      (501 + 208 a^2 + 656 b + 208 b^2 +  16 a (41 + 26 b))
           \\[3pt] &
      +  4096 \,\omega^2\, \Big(-((1 + 2 b) (767 + 560 a^2 + 392 b + 16 b^2 + 
            4 a (367 + 148 b)) t^4) \\ & + 
       2 (152 + 80 a^3 + 337 b + 144 b^2 + 16 b^3 + 
       12 a^2 (17 + 12 b) + a (213 + 304 b + 80 b^2)) u \Big)
      \\ &
            +  8192 \,\omega^3\,  t \Big(2 (1 + 2 b) (15 + 12 a + 2 b) t^4 - (27 + 
        28 a^2 + 52 b + 8 b^2 + 8 a (4 + 3 b)) u\Big)
      \\ &
      -  4096 \,\omega^4\,  t^2 \Big((1 + 2 b) t^4 - 2 (1 + a + b) u\Big)
    \end{align*}

    \begin{align*}
      C^{(2)}_{2, 2}  &=
      - 16384 \Big(-8 (3 + 8 b + 4 b^2) (1026 + 124 a^2 + 478 b + 56 b^2
      + 5 a (145 + 36 b)) t^6
        \\ &
        + (70173 + 1792 a^4 + 142320 b    + 
        95776 b^2 + 24064 b^3 + 1792 b^4 + 32 a^3 (761 + 504 b)
        \\ &
        +   16 a^2 (6163 + 6836 b + 1792 b^2) +  
        2 a (75225 + 113888 b + 54544 b^2 + 8064 b^3)) t^2 u\Big)
      \\[3pt] &
      +  49152 \,\omega  t^3 \Big(-4 (117 + 38 a + 24 b) (3 + 8 b + 
          4 b^2) t^4 + (5037 + 416 a^3 + 7856 b \\ & + 3568 b^2 + 
          384 b^3 + 16 a^2 (251 + 156 b) + 
          2 a (4813 + 5056 b + 1232 b^2)) u \Big)
      \\[3pt] &
      -  8192 \,\omega^2\, \Big(-100 (3 + 8 b + 4 b^2) t^8 + (2339 + 560 a^2 + 
          2136 b + 480 b^2 \\ & + 4 a (811 + 444 b)) t^4 u - (761 + 
          144 a^2 + 544 b + 80 b^2 + 96 a (5 + 2 b)) u^2\Big)
      \\[3pt] &
      +  98304 \,\omega^3\,  t u \Big((7 + 4 a + 2 b) t^4 - 2 (3 + a + b) u \Big)
      \\ &
            -8192 \,\omega^4\,  t^2 (t^4 - u) u
    \end{align*}

      \begin{align*} C^{(2)}_{2, 3}  &=   
      - 262144 t^2 \Big(6 (15 + 46 b + 36 b^2 + 8 b^3) t^8 -  
        2 (3 + 2 b) (124 a^2 + 5 a (235 + 72 b)  \\ &
        + 4 (629 + 346 b + 45 b^2)) t^4 u  + (21351 + 672 a^3 +    
          22378 b + 7136 b^2 \\& + 672 b^3 + 8 a^2 (901 + 364 b) + 
          4 a (5701 + 4160 b + 728 b^2)) u^2 \Big)
      \\[3pt] &
        + 786432 \,\omega  t^3 u \Big(- (3 + 2 b) (177 + 38 a + 48 b) t^4
     +  2 (387 + 52 a^2 + 294 b + 48 b^2 + 2 a (163 + 64 b)) u \Big)
      \\ &
      + 131072 \,\omega^2\, u \Big(
        25 (3 + 2 b) t^8   - 2 (69 + 37 a + 29 b) t^4 u +   
        2 (15 + 4 a + 4 b) u^2\Big)
        \\[3pt] &
        + 131072 \,\omega^3\,  t (t^4 - u) u^2
      \end{align*}

      \begin{align*}
        C^{(2)}_{2, 4}  &=   
        - 262144 t^2 u \Big( 36 (15 + 16 b + 4 b^2) t^8 - 
       2 (9156 + 124 a^2 + 5590 b + 800 b^2  \\ &  +  
          25 a (101 + 36 b)) t^4 u + (20673 + 1456 a^2 + 11384 b + 
          1456 b^2 + 32 a (358 + 105 b)) u^2 \Big)
        \\[3pt] &
        - 786432 \,\omega  t^3 u^2 \Big((357 + 38 a + 120 b) t^4
        - 2 (201 + 64 a + 60 b) u \Big)
        \\ &
       + 131072 \,\omega^2\, (25 t^4 - 4 u) (t^4 - u) u^2
      \end{align*}  

      \begin{align*}
        C^{(2)}_{2, 5}  &=
        -  6291456 t^2 u^2 \Big(3 (5 + 2 b) t^8 - 2 (122 + 15 a + 31 b) t^4 u +   
          2 (121 + 28 a + 28 b) u^2 \Big)
        \\[3pt] & 
        -37748736 \,\omega  t^3 (t^4 - u) u^3
      \end{align*}  

      \begin{align*}
        C^{(2)}_{2,6}  &=   -4194304\, t^2 u^3 (3 t^8 - 31 t^4 u + 28 u^2)
      \end{align*}

      \begin{align*}
        C^{(2)}_{3,0}  &=
        -2048\,\omega\, (1 + 2 a) (1 + 2 a + 2 b) (1 + 4 a + 4 b)(3 + 4 a +  4 b)
        \\ &
        + 8192\,\omega^2\, t (1 + 2 a) (1 + 2 a + 2 b) (2 + 5 a + 4 b) \\ &
        - 4096\,\omega^3\, t^2 (1 + 2 a) (2 + 5 a + 4 b) \\ &
        + 1024 \,\omega^4\,  t^3 (1 + 2 a)
      \end{align*}  

      \begin{align*}
        C^{(2)}_{3, 1}  &=
        -16384 (1 + 2 a)(1 + 2 b)(501 + 208 a^2
        + 656 b + 208 b^2 + 16 a(41 + 26 b)) t^3
        \\ &
        -  8192 \,\omega  \Big(- \Big((1 + 2 b) (609 + 448 a^2 + 320 b + 16 b^2 + 
          4 a (287 + 116 b)) t^4\Big) \\ & + 
          2 (159 + 48 a^3 + 319 b + 136 b^2 + 16 b^3 + 
          16 a^2 (11 + 7 b) + a (247 + 312 b + 80 b^2)) u\Big)
        \\ &
        +  16384 \,\omega^2\, t \Big(-2 (1 + 2 b) (25 + 19 a + 4 b) t^4 + (59 + 
          36 a^2 + 94 b + 16 b^2 + a (74 + 48 b)) u\Big)
        \\ &
        +  32768 \,\omega^3\,  t^2 ((1 + 2 b) t^4 - 2 (1 + a + b) u)
      \end{align*}

      \begin{align*} C^{(2)}_{3, 2}  &=  
        - 32768 \Big(-4 (117 + 38 a + 24 b) (3 + 8 b + 4 b^2) t^7 + (5037 + 
          416 a^3 + 7856 b \\ & + 3568 b^2  
          + 384 b^3  + 16 a^2 (251 + 156 b) + 
          2 a (4813 + 5056 b + 1232 b^2)) t^3 u \Big)
        \\[3pt] &
        +  16384 \,\omega  \Big(-76 (3 + 8 b + 4 b^2) t^8
        + (448 a^2 +   4 a (635 + 348 b) +  3 (631 + 576 b + 128 b^2)) t^4 u
        \\ &
        - (723 + 112 a^2 + 520 b +  80 b^2 + 8 a (61 + 24 b)) u^2 \Big)
        \\[3pt] &
        - 32768 \,\omega^2\, t u \Big((74 + 38 a + 24 b) t^4 - (67 + 24 a + 
          24 b) u \Big) \\ &
        + 65536 \,\omega^3\,  t^2 (t^4 - u) u        
      \end{align*}

      \begin{align*}
        C^{(2)}_{3,  3}  &=
          524288 t^3 u \Big(
           (3 + 2 b)(177 + 38 a + 48 b) t^4 
          - 2(387 + 52 a^2 + 294 b + 48 b^2 + 2 a (163 + 64 b)) u
          \Big)
        \\ &
        - 262144 \,\omega  u \Big(
         19 (3 + 2 b) t^8 - (111 + 58 a +  46 b) t^4 u + (29 + 8 a + 8 b) u^2
         \Big)
        \\[3pt] &
        -524288 \,\omega^2\, t (t^4 - u) u^2        
      \end{align*}

      \begin{align*}
        C^{(2)}_{3,  4}  &=
        524288 t^3 u^2 ((357 + 38 a + 120 b) t^4   -  2 (201 + 64 a + 60 b) u)
        \\[3pt] &
        -262144 \,\omega  (19 t^4 - 4 u)(t^4 - u) u^2        
      \end{align*}

      \begin{align*} C^{(2)}_{3, 5}  &=   
        25165824 t^3 (t^4 - u) u^3 \end{align*}

      \begin{align*}
        C^{(2)}_{4, 0}  &=   
        1024 (1 + 2 a) (3 + 32 a^3 + 22 b + 48 b^2 + 32 b^3
        + 48 a^2 (1 + 2 b) + a (22 + 96 b + 96 b^2))
   \\ &
   -  10240\,\omega  t \, (1 + 2 a) (1 + 2 a + 2 b) (3 + 4 a + 4 b)
   +  512 \,\omega^2\, t^2 (1 + 2 a) (53 + 80 a + 76 b)  \\ &
   -  5120 \,\omega^3\, t^3 (1 + 2 a)
   +  256 \,\omega^4\, (t^4 - u)
      \end{align*}  

      \begin{align*}
        C^{(2)}_{4,  1}  &=
          4096 \Big(- \Big((1 + 2 b) (609 + 448 a^2  + 320 b + 16 b^2
        + 4 a (287 + 116 b)) t^4 \Big)\\ &
        + 2 (159 + 48 a^3 + 319 b + 136 b^2 + 16 b^3 
        +  16 a^2 (11 + 7 b) + a (247 + 312 b + 80 b^2)) u
        \Big)
          \\[3pt] &
        + 81920 \,\omega  t \Big((1 + 2 b) (9 a + 2 (6 + b)) t^4 
          - (15 + 8 a^2 +  23 b + 4 b^2 + a (19 + 12 b)) u \Big)
        \\[3pt] &
        -77824 \,\omega^2\, t^2 ((1 + 2 b) t^4 - 2 (1 + a + b) u)
      \end{align*}  

      \begin{align*}
        C^{(2)}_{4, 2}  &=
        - 8192 \Big(-76 (3 + 8 b + 4 b^2) t^8 + (448 a^2 + 4 a (635 + 348 b)
          + 3 (631 + 576 b + 128 b^2)) t^4 u
          \\ &
          - (723 + 112 a^2 + 520 b
          + 80 b^2 + 8 a (61 + 24 b)) u^2 \Big)
        \\ &
        + 245760 \,\omega  t u (2 (6 + 3 a + 2 b) t^4 - (11 + 4 a + 4 b) u)
        -155648 \,\omega^2\, t^2 (t^4 - u) u
      \end{align*}  

      \begin{align*}
        C^{(2)}_{4, 3}  &=   
         131072 u \Big(
          19 (3 + 2 b) t^8 - (111 + 58 a + 46 b) t^4 u + (29 + 8 a + 8 b) u^2
          \Big) \\ &
        +655360 \,\omega  t (t^4 - u) u^2
      \end{align*}  

          \begin{align*} C^{(2)}_{4, 4}  &=   
            131072 u^2 (19 t^8 - 23 t^4 u + 4 u^2) \end{align*}

        \begin{align*} C^{(2)}_{5, 0}  &=   
   4096 (1 + 2 a) (3 + 8 a^2 + 10 b + 8 b^2 + 2 a (5 + 8 b)) t \\ & - 
    9216 (1 + 2 a) (3 + 4 a + 4 b) \,\omega  t^2 + 
    9216 (1 + 2 a) \,\omega^2\, t^3 - 1024 \,\omega^3\,  (t^4 - u) \end{align*}  

          \begin{align*} C^{(2)}_{5, 1}  &=   
             32768 (-((1 + 2 b) (9 a + 2 (6 + b)) t^5) + (15 + 8 a^2 + 23 b + 
            4 b^2 + a (19 + 12 b)) t u)
            \\[3pt] &
            +73728 \,\omega  t^2 ((1 + 2 b) t^4 - 2 (1 + a + b) u) 
          \end{align*}  

          \begin{align*} C^{(2)}_{5, 2}  &=   
            - 98304 t u \big(2 (6 + 3 a + 2 b) t^4 - (11 + 4 a + 4 b) u \big)
            + 147456 \,\omega  t^2 (t^4 - u) u
          \end{align*}

          \begin{align*}
            C^{(2)}_{5, 3}  &=   -262144 t (t^4 - u) u^2
          \end{align*}

        \begin{align*} C^{(2)}_{6, 0}  &=   
          3072 (1 + 2 a) (3 + 4 a + 4 b) t^2 - 7168\,\omega\,  t^3 (1 + 2 a)
          + 1536 \,\omega^2\, (t^4 - u)
        \end{align*}  

          \begin{align*} C^{(2)}_{6, 1}  &=   
            24576 (-(1 + 2 b) t^6 + 2 (1 + a + b) t^2 u)
          \end{align*}  

          \begin{align*} C^{(2)}_{6,2}  &=
            -49152 t^2 (t^4 - u) u
          \end{align*}

          \begin{align*}
            C^{(2)}_{7, 0}  &=   
            2048 (1 + 2 a) t^3 - 1024 \,\omega\,  (t^4 - u)
          \end{align*}

          \begin{align*}
            C^{(2)}_{8, 0}  &=   256 (t^4 - u)
          \end{align*}

 %% FILE OK %%

\section{TTW at $k=4$: the commutator ${\cal I}_{12}$ coefficients, see (\ref{I12-k=4}) \label{I12k4coeffs}}

Non-vanishing coefficients for the integral (commutator) ${\cal I}_{12}$ of TTW at $k=4$, 
defined in (\ref{I12-k=4}):

%   TTW k=4, I12 coeffs
%   Manually reorganized   Jan 11, 2025

\begin{align*}
  C^{(12)}_{0, 1}  &=  
   262144 (1 + 2 a) (1 + 2 b) (1 + 2 a + 2 b) (3 + 2 a + 2 b) (1 + 
   4 a + 4 b) (3 + 4 a + 4 b) (5 + 4 a + 4 b) (7 + 4 a + 4 b)
\\ &
   -  1048576\,\omega t\, (1 + 2 a) (1 + 2 b) (1 + 2 a + 2 b) (3 + 2 a + 2 b) (3 + 
   4 a + 4 b) (5 + 4 a + 4 b) (7 + 4 a + 4 b) 
\\ &
   + 786432\,\omega^2 t^2\, (1 + 2 a) (1 + 2 b) (3 + 2 a + 2 b) (3 + 4 a + 4 b) (5 + 
   4 a + 4 b) (7 + 4 a + 4 b) 
\\ &
   - 524288 \,\omega^3 t^3\, (1 + 2 a) (1 + 2 b) (3 + 2 a + 2 b) (5 + 4 a + 4 b) (7 + 4 a + 4 b) 
\\ & 
   +  65536\,\omega^4 \, (1 +  2 b) \Big(2 \Big(27 + 16 a^3 + 28 b + 8 b^2 + 32 a^2 (2 + b) +  a (79 + 72 b + 16 b^2)\Big) t^4 - 3 u\Big)
\end{align*}

  \begin{align*}
    C^{(12)}_{0, 2}  &=  
    + 524288 \Bigg(
      - (1 + 2 b) (3 + 2 b) (105 + 16 a^2 + 80 b + 16 b^2 +  16 a (5 + 2 b))\cdot \\ & 
      \cdot\big(2289 + 192 a^3 + 1762 b + 432 b^2 + 32 b^3 + 16 a^2 (85 + 26 b) + 
      4 a (781 + 448 b + 64 b^2) \big) t^4
            \\ &
     + \Big( 1519875 + 2048 a^7 + 
          4915494 b + 6477700 b^2 + 4539464 b^3 + 1833088 b^4 + 
          423680 b^5 + 50176 b^6 + 2048 b^7
          \\ &
          + 1024 a^6 (49 + 36 b) + 
          256 a^5 (1655 + 2008 b + 608 b^2)
           + 128 a^4 (14321 + 23124 b + 12536 b^2 + 2320 b^3)
          \\ &
          +  8 a^3 (567433 + 1129728 b + 845152 b^2 + 285184 b^3 + 
             37120 b^4) + 
             4 a^2 (1619425 + 3836892 b \\ & + 3602368 b^2 + 1690304 b^3
              +  401152 b^4 + 38912 b^5) \\ & + 
          4 a (1229031 + 3407690 b + 3836892 b^2 + 2259456 b^3 + 
          739968 b^4 + 128512 b^5 + 9216 b^6)\Big) u \Bigg)
      \\[4pt]  &
      + 2097152 \,\omega t \, \Bigg((3 + 8 b + 4 b^2) \Big(25515 + 640 a^4 + 27858 b + 
      11368 b^2 + 2016 b^3 + 128 b^4 + 128 a^3 (51 + 16 b)
      \\ &
      +  8 a^2 (3103 + 1884 b + 288 b^2) +  4 a (10377 + 9048 b + 2640 b^2 + 256 b^3)\Big) t^4
      \\ & 
      - \Big(168525 +  512 a^6 + 469674 b + 507700 b^2 + 274264 b^3 + 78240 b^4 + 
      10880 b^5 + 512 b^6 + 640 a^5 (17 + 12 b)
      \\ &
      + 96 a^4 (815 + 952 b + 272 b^2) + 8 a^3 (34283 + 53388 b + 27472 b^2 + 4736 b^3)
      \\ &
      + 4 a^2 (126925 + 244672 b + 174432 b^2 + 54944 b^3 +  6528 b^4)
      \\ & 
      + 16 a (29394 + 68135 b + 61168 b^2 + 26694 b^3 + 5712 b^4 +  480 b^5)\Big) u \Bigg)
      \\[4pt] &
      +  1572864 \,\omega^2 t^2 \, \Big(-((3 + 8 b + 4 b^2) (4725 + 256 a^3 +  3380 b + 816 b^2 + 64 b^3 + 48 a^2 (43 + 12 b)
      \\ &
      +  8 a (685 + 360 b + 48 b^2)) t^4) +  2 (16695 + 128 a^5 + 39699 b + 34082 b^2 + 13144 b^3
      \\ &
      + 2272 b^4 + 128 b^5 + 32 a^4 (71 + 48 b) +   8 a^3 (1643 + 1844 b + 496 b^2)
      \\ &
      +  a^2 (34082 + 51232 b + 24960 b^2 + 3968 b^3) + 2 a (20007 + 37757 b + 25616 b^2 + 7376 b^3
      +  768 b^4)) u \Big)
      \\[3pt]  &
    +1048576 \omega^3 t^3 \Big((3 + 8 b + 4 b^2) (375 + 48 a^2 + 152 b + 16 b^2 + 16 a (17 + 4 b)) t^4
      \\ &
      -  2 \Big(1470 + 32 a^4 + 2921 b + 1894 b^2 + 448 b^3 + 32 b^4 +  32 a^3 (14 + 9 b) + 2 a^2 (947 + 1016 b + 256 b^2)
      \\ &
      +  a (3026 + 4418 b + 2032 b^2 + 288 b^3) \Big) u\Big)
      \\[3pt] &
      + 131072 \,\omega^4\, \Big(-8 (6 + 2 a + b) (3 + 8 b + 4 b^2) t^8 + 
        2 (225 + 16 a^3 + 352 b + 160 b^2 + 16 b^3 
        \\ & + 32 a^2 (5 + 3 b)
        +       a (403 + 408 b + 96 b^2)) t^4 u - 3 u^2 \Big)
  \end{align*}

  \begin{align*}
    C^{(12)}_{0, 3}  &=
          + 2097152 \Big(4 (1 + 2 b) (3 + 2 b) (5 + 2 b) (11025 + 160 a^3 + 
            5852 b + 1056 b^2 + 64 b^3 \\ &
            + 48 a^2 (41 + 8 b)  +  a (8078 + 3024 b + 288 b^2)) t^8
          -  2 (3 + 2 b) (1672335 + 1536 a^5 \\ & + 2325370 b + 1253248 b^2
          + 
          327456 b^3 + 41600 b^4 + 2048 b^5 + 1280 a^4 (27 + 10 b) \\ &
          +  512 a^3 (571 + 379 b + 61 b^2)
          +  64 a^2 (18453 + 16957 b + 5094 b^2 + 504 b^3) \\ &
          +  2 a (1140845 + 1318256 b + 560176 b^2 + 104064 b^3 + 
          7168 b^4)) t^4 u
          \\ &
          + (6144 a^6 + 512 a^5 (279 + 122 b)
          +  256 a^4 (5067 + 3894 b + 760 b^2)  \\ &
          +  64 a^3 (95187 + 100130 b + 35568 b^2 + 4320 b^3)
          +  8 a^2 (1968837 + 2583720 b \\ &
          + 1277792 b^2 + 284544 b^3 +  24320 b^4)
          +   a (21265422 + 33263124 b + 20669760 b^2 \\ &
          + 6408320 b^3 +  996864 b^4 + 62464 b^5)
          \\ &
          + 3 (3895395 + 7088474 b + 5250232 b^2 + 2030656 b^3 + 
          432384 b^4 + 47616 b^5 + 2048 b^6)) u^2 \Big)
          \\[4pt] &
          + 8388608 \,\omega\, t \Big(-3 (1 + 2 b) (3 + 2 b) (5 + 2 b) (595 + 32 a^2 + 
          192 b + 16 b^2 + 12 a (23 + 4 b)) t^8
          \\ &
          + (3 + 2 b) (183015 +   640 a^4 + 202408 b + 80708 b^2 + 13664 b^3 + 832 b^4 +    128 a^3 (91 + 32 b)
          \\ &
          + 8 a^2 (9283 + 5856 b + 888 b^2) +  4 a (49447 + 43216 b + 12216 b^2 +  1120 b^3)) t^4 u
          \\ &
          - (667485 + 1280 a^5 + 1038522 b +  622424 b^2 + 179632 b^3 + 24832 b^4 + 1280 b^5 \\ & +  256 a^4 (97 + 41 b)
          + 16 a^3 (11227 + 8380 b + 1568 b^2) +  8 a^2 (77803 + 79882 b + 27312 b^2 + 3136 b^3)
          \\ &
          +  2 a (519261 + 669842 b + 319528 b^2 + 67040 b^3 +  5248 b^4)) u^2\Big)
          \\[4pt] &
          + 6291456 \,\omega^2\, t^2 \Big(
            2 (1 + 2 b) (3 + 2 b) (5 + 2 b) (55 + 12 a +  8 b) t^8
            \\ &
            - (3 + 2 b) (18425 + 256 a^3 + 15040 b + 3928 b^2 + 320 b^3 + 48 a^2 (73 + 24 b)
            \\ & + 8 a (1835 + 1080 b + 152 b^2)) t^4 u
            +  2 (35175 + 256 a^4 + 45433 b + 20696 b^2 + 3920 b^3 + 256 b^4 \\ &
            + 80 a^3 (49 + 20 b) + 8 a^2 (2587 + 1878 b + 336 b^2)
            + a (45433 + 45844 b + 15024 b^2 + 1600 b^3)) u^2 \Big)
   \\ &
   +  4194304 \,\omega^3\, t^3 \Big(-2 (1 + 2 b) (3 + 2 b) (5 + 2 b) t^8
     + (3 + 2 b) (845 + 48 a^2 + 440 b + 56 b^2
     \\ & 
     +  16 a (27 + 8 b)) t^4 u -  2 (48 a^3 + 16 a^2 (33 + 13 b) + a (1719 + 1224 b + 208 b^2)
     \\ & +   3 (560 + 573 b + 176 b^2 + 16 b^3)) u^2 \Big)
   \\ &          
   + 2097152\, \omega^4\, t^4 u\, (-(3 + 2 b) (17 + 4 a + 4 b) t^4 + (69 + 
   8 a^2 + 52 b + 8 b^2 + 4 a (13 + 5 b)) u)
  \end{align*}

  \begin{align*}
    C^{(12)}_{0, 4}  &=
   + 2097152 \Bigg(-16 (1 + 2 b) (3 + 2 b) (5 + 2 b) (7 + 2 b) (49 + 8 a +  6 b) t^{12}
     \\ &
     + 24 (3 + 2 b) (5 + 2 b) (30429 + 160 a^3 + 17696 b + 3360 b^2 +  208 b^3 + 16 a^2 (179 + 40 b)
     \\ &
     + 2 a (8253 + 3416 b + 344 b^2)) t^8 u - \Big(54993645 + 3072 a^5 + 71756970 b + 36346592 b^2 + 8936128 b^3 \\ &
     +  1067776 b^4 + 49664 b^5 + 1280 a^4 (129 + 50 b) + 256 a^3 (9909 + 6616 b + 1064 b^2)
     \\ & + 32 a^2 (522537 + 470458 b + 136992 b^2 + 12960 b^3)
     \\ &  
     + 4 a (12479505 + 13828160 b + 5591232 b^2 + 980736 b^3 + 63232 b^4) \Big) t^4 u^2
     \\ & 
     + 2 (24260880 + 15616 a^5 + 28761813 b + 13443712 b^2 + 3100448 b^3 + 352256 b^4 + 15616 b^5
     \\ &
     + 256 a^4 (1376 + 425 b)
     + 32 a^3 (96889 + 54400 b + 7760 b^2) + 32 a^2 (420116 + 329151 b + 86784 b^2 + 7760 b^3)
     \\ & 
     + a (28761813 + 28429568 b + 10532832 b^2 + 1740800 b^3 + 
     108800 b^4)) u^3\Bigg)
   \\[4pt] &
   + 8388608\, \omega\, t \Big(8 (1 + 2 b) (3 + 2 b) (5 + 2 b) (7 + 2 b) t^{12} - 
     2 (3 + 2 b) (5 + 2 b) (10087 + 288 a^2 + 3808 b + 352 b^2
     \\ &
     +   36 a (97 + 20 b)) t^8 u + (1854405 + 640 a^4 + 2016874 b + 
     790832 b^2 + 132128 b^3 + 7936 b^4 + 128 a^3 (211 + 80 b)
     \\ &
     +  8 a^2 (38323 + 24972 b + 3888 b^2) + 
     4 a (327517 + 287944 b + 81216 b^2 + 7360 b^3)) t^4 u^2
     \\ &
     -  2 (863625 + 2624 a^4 + 846129 b + 302780 b^2 + 46832 b^3 + 
     2624 b^4 + 16 a^3 (2927 + 880 b)
     \\ &
     +   4 a^2 (75695 + 41572 b + 5728 b^2) + 
     a (846129 + 652072 b + 166288 b^2 + 14080 b^3)) u^3\Big)
   \\[4pt] & 
   +     6291456\, \omega^2\, t^2 u \Big(4 (3 + 2 b) (5 + 2 b) (221 + 36 a +   40 b) t^8
     - (112305 + 256 a^3 + 96292 b + 26208 b^2 +      2240 b^3
     \\ &
     + 48 a^2 (163 + 60 b)
     +  8 a (7105 + 4488 b + 672 b^2)) t^4 u + (111195 + 1600 a^3 + 
     85108 b + 20752 b^2 + 1600 b^3 
     \\ & 
     + 16 a^2 (1297 + 380 b)     + a (85108 + 45920 b + 6080 b^2)) u^2 \Big)
   \\[4pt] & 
   +  4194304\, \omega^3\, t^3 u \Big(-12 (3 + 2 b) (5 + 2 b) t^8 + (2955 + 
          48 a^2 + 1784 b + 256 b^2 +  16 a (57 + 20 b)) t^4 u
          \\ & 
          - (3115 + 208 a^2 + 1672 b +  208 b^2 + 8 a (209 + 60 b)) u^2\Big)
   \\ &   
   +2097152 \,\omega^4\, t^4 u^2 (-2 (16 + 2 a + 5 b) t^4 +    5 (7 + 2 a + 2 b) u)
  \end{align*}

  \begin{align*}
    C^{(12)}_{0, 5}  &=
    + 8388608 u \Big(-16 (3 + 2 b) (5 + 2 b) (7 + 2 b) (125 + 16 a +  18 b) t^{12}
      \\ &
      + 12 (5 + 2 b) (102249 + 160 a^3 + 62940 b + 12496 b^2 + 800 b^3 + 48 a^2 (97 + 24 b)
      \\ &
      + 2 a (19831 + 8824 b + 944 b^2)) t^8 u
    \\ &
    - (20050065 +   6400 a^4 + 17350416 b + 5512000 b^2 + 761728 b^3 +   38656 b^4 + 768 a^3 (297 + 82 b)
    \\ &
    +  32 a^2 (82219 + 40908 b + 4992 b^2) + 16 a (768531 + 529998 b + 119680 b^2 +  8864 b^3)) t^4 u^2
    \\ & 
    + (14781585 + 21760 a^4 +  11761152 b + 3473504 b^2 + 451584 b^3 + 21760 b^4 +   21504 a^3 (21 + 5 b)
    \\ & 
    +  32 a^2 (108547 + 47968 b + 5360 b^2) +  64 a (183768 + 114853 b + 23984 b^2 + 1680 b^3)) u^3\Big)
    \\[4pt] &
    + 33554432\, \omega\, t u \Big(16 (3 + 2 b) (5 + 2 b) (7 + 2 b) t^{12} - (5 +   2 b) (22827 + 288 a^2 + 9632 b + 976 b^2
      \\ &
      + 324 a (17 + 4 b)) t^8 u  +  2 (219555 + 512 a^3 + 149017 b + 32728 b^2 + 2320 b^3 +  48 a^2 (278 + 75 b)
      \\ &
      + a (99124 + 48296 b + 5728 b^2)) t^4 u^2 - (343245 + 2816 a^3 + 212232 b + 42832 b^2 + 2816 b^3
      \\ & +   16 a^2 (2677 + 624 b)
      + 24 a (8843 + 3844 b + 416 b^2)) u^3\Big)
    \\[4pt] &
    + 50331648 \,\omega^2\, t^2 u^2 \Big(9 (5 + 2 b) (37 + 4 a + 8 b) t^8 -  2 (4080 + 72 a^2 + 1951 b + 224 b^2
      \\ &
      +     12 a (99 + 26 b)) t^4 u + (6825 + 304 a^2 + 2936 b + 304 b^2 + 8 a (367 + 84 b)) u^2\Big)
    \\[4pt] &
    + 33554432 \,\omega^3\, t^3 u^2 \Big(-3 (5 + 2 b) t^8 + (117 + 16 a + 
      30 b) t^4 u - 2 (53 + 12 a + 12 b) u^2\Big)
    \\[4pt] &
        -8388608 \omega^4 t^4 (t^4 - u) u^3
  \end{align*}

  \begin{align*}
    C^{(12)}_{0, 6}  &=
    + 67108864 u^2 \Big(-12 (85 + 8 a + 14 b) (35 + 24 b + 4 b^2) t^{12}
      + (535437 + 160 a^3 \\ &
      + 342780 b + 70752 b^2
      + 4704 b^3 + 48 a^2 (209 + 56 b) +   2 a (70315 + 33048 b + 3744 b^2)) t^8 u
      \\ &
      - 4 \Big(340074 + 656 a^3 + 191145 b + 35228 b^2 + 2128 b^3 + 12 a^2 (1429 + 308 b)
      \\ & + a  (136777 + 54504 b + 5360 b^2) \Big) t^4 u^2
      +  8 (109956 + 560 a^3 + 57199 b + 9840 b^2 + 560 b^3  \\ & +  16 a^2 (615 + 119 b) + 
          a (57199 + 20832 b + 1904 b^2)) u^3 \Big)
    \\[4pt] & 
    +  67108864 \omega t u^2 (24 (5 + 2 b) (7 + 2 b) t^{12} -  3 (8603 + 32 a^2 + 3936 b + 432 b^2
    \\ &
    + 12 a (107 + 28 b)) t^8 u + (76167 + 1200 a^2 + 29896 b + 2864 b^2 + 8 a (2517 + 532 b)) t^4 u^2
    \\ &
    -  4 (13083 + 416 a^2 + 4708 b + 416 b^2 + a (4708 + 896 b)) u^3)
    \\[4pt] &
    +  100663296 \,\omega^2\, t^2 u^3 \Big((223 + 12 a + 56 b) t^8 - (809 + 
      104 a + 168 b) t^4 u + 4 (149 + 28 a + 28 b) u^2\Big)
    \\[4pt] &
    -67108864 \,\omega^3\, t^3 u^3 (t^8 - 5 t^4 u + 4 u^2)
  \end{align*}

  \begin{align*}
    C^{(12)}_{0, 7}  &=  
     536870912 u^3 \Big(-4 (7 + 2 b) (65 + 4 a + 12 b) t^{12} 
     + 3 \Big(6111 + 32 a^2 + 2284 b + 208 b^2 + 8 a (122 + 25 b)\Big) t^8 u  \\ &
   - \Big(39123 + 528 a^2 + 13040 b +  1072 b^2 + 8 a (1173 + 208 b)\Big) t^4 u^2
   \\ &
   +  2 \Big(11417 + 272 a^2 + 3536 b + 272 b^2 + 16 a (221 + 36 b) \Big) u^3\Big)
   \\[4pt] &
    +  1073741824\, \omega\, t u^3
    \Big(2 (7 + 2 b) t^{12} -  2 (9 a + 26 (5 + b)) t^8 u \\ & 
      + (641 + 76 a + 112 b) t^4 u^2 - 2 (199 + 32 a + 32 b) u^3\Big)
    \\[4pt]  &
   +1610612736 \,\omega^2\, t^2 u^4 (t^8 - 3 t^4 u + 2 u^2)
  \end{align*}

  \begin{align*}
    C^{(12)}_{0, 8}  &=  
      - 536870912 u^4 \Big((265 + 8 a + 54 b) t^{12}
      - 6 (340 + 25 a + 57 b) t^8 u \\ & + 8 (477 + 52 a + 72 b) t^4 u^2
      -  32 (64 + 9 a + 9 b) u^3 \Big)
      \\[4pt] &
     +1073741824\, \omega\, t u^4 (t^{12} - 13 t^8 u + 28 t^4 u^2 - 16 u^3)
  \end{align*}

  \begin{align*}
    C^{(12)}_{0, 9}  &=  -2147483648 (t^4 - 4 u) (3 t^4 - 4 u) (t^4 - u) u^5
  \end{align*}

    \begin{align*}
      C^{(12)}_{1, 1}  &=  
      2097152 (1 + 2 a) (1 + 2 b) (1 + 2 a + 2 b) (3 + 2 a + 2 b) (3 + 4 a + 4 b) (5 + 4 a + 4 b) (7 + 4 a + 4 b) t
      \\ &
      - 3670016\,\omega\, t^2 (1 + 2 a) (1 + 2 b) (3 + 2 a + 2 b) (3 + 4 a + 4 b) (5 +  4 a + 4 b) (7 + 4 a + 4 b)
      \\ &
      + 4718592 \,\omega^2\, t^3(1 + 2 a) (1 + 2 b) (3 + 2 a + 2 b) (5 + 4 a + 4 b) (7 + 4 a + 4 b)
      \\ &
      + 1310720\,\omega^3\, (1 +  2 b)  \big(-2 (27 + 16 a^3 + 28 b + 8 b^2 + 32 a^2 (2 + b) +  a (79 + 72 b + 16 b^2)) t^4 + 3 u\big)
      \\ &
      + 262144\, \omega^4 t\, (1 +  2 b)  \big(2 (6 + 11 a + 4 a^2 + 2 b + 4 a b) t^4 - 3 u \big)
    \end{align*}

    \begin{align*}
      C^{(12)}_{1, 2}  &=
      4194304 t \Big(-((3 + 8 b + 4 b^2) (25515 + 640 a^4 + 27858 b +  11368 b^2 + 2016 b^3 + 128 b^4
        \\ &
        + 128 a^3 (51 + 16 b) + 8 a^2 (3103 + 1884 b + 288 b^2) + 4 a (10377 + 9048 b + 2640 b^2 +    256 b^3)) t^4)
      \\ &
      + (168525 + 512 a^6 + 469674 b +           507700 b^2 + 274264 b^3 + 78240 b^4 + 10880 b^5 + 512 b^6
      \\ &
      +     640 a^5 (17 + 12 b) + 96 a^4 (815 + 952 b + 272 b^2) +   8 a^3 (34283 + 53388 b + 27472 b^2 + 4736 b^3)
      \\ &
      +     4 a^2 (126925 + 244672 b + 174432 b^2 + 54944 b^3 + 6528 b^4) +  16 a (29394 + 68135 b + 61168 b^2
      \\ &
      + 26694 b^3 + 5712 b^4 +  480 b^5)) u \Big)
      \\[4pt] &
            +  7340032\, \omega\, t^2 \Big((3 + 8 b + 4 b^2) (4725 + 256 a^3 + 3380 b + 816 b^2 + 64 b^3 + 48 a^2 (43 + 12 b)
        \\ &
        + 8 a (685 + 360 b + 48 b^2)) t^4 -   2 (16695 + 128 a^5 + 39699 b + 34082 b^2 + 13144 b^3 +  2272 b^4
        \\ &
        + 128 b^5 + 32 a^4 (71 + 48 b) +   8 a^3 (1643 + 1844 b + 496 b^2)
        \\ & 
        +  a^2 (34082 + 51232 b + 24960 b^2 + 3968 b^3)
        \\ &
        +  2 a (20007 + 37757 b + 25616 b^2 + 7376 b^3 +  768 b^4)) u \Big)
      \\[4pt] &
      +9437184 \,\omega^2\, t^3 \Big(-\big((3 + 8 b + 4 b^2) (375 + 48 a^2 + 152 b + 16 b^2 + 16 a (17 + 4 b)) t^4\big)
        \\ &
        +  2 (1470 + 32 a^4 + 2921 b + 1894 b^2 + 448 b^3 + 32 b^4 +  32 a^3 (14 + 9 b)
        \\ &
        + 2 a^2 (947 + 1016 b + 256 b^2) +   a (3026 + 4418 b + 2032 b^2 + 288 b^3)) u \Big)
      \\ &
      +  2621440 \,\omega^3\, (8 (6 + 2 a + b) (3 + 8 b + 4 b^2) t^8 -  2 (225 + 16 a^3 + 352 b + 160 b^2 + 16 b^3
      \\ &
      + 32 a^2 (5 + 3 b) +       a (403 + 408 b + 96 b^2)) t^4 u + 3 u^2)
      \\[4pt] &      
      +   524288 \,\omega^4\, t \Big(-2 (3 + 8 b + 4 b^2) t^8 + 2 (15 + 4 a^2 + 14 b + 4 b^2 + a (23 + 12 b)) t^4 u - 3 u^2 \Big)
    \end{align*}

    \begin{align*}
      C^{(12)}_{1, 3}  &=  
         16777216 t \Big( 3 (1 + 2 b) (3 + 2 b) (5 + 2 b) (595 + 32 a^2 + 
    192 b + 16 b^2 + 12 a (23 + 4 b)) t^8
    \\ &
    - (3 + 2 b) (183015 +  640 a^4 + 202408 b + 80708 b^2 + 13664 b^3 + 832 b^4 +  128 a^3 (91 + 32 b)
    \\ &
    + 8 a^2 (9283 + 5856 b + 888 b^2) +   4 a (49447 + 43216 b + 12216 b^2 +  1120 b^3)) t^4 u
    \\ &
    + (667485 + 1280 a^5 + 1038522 b +   622424 b^2 + 179632 b^3 + 24832 b^4 + 1280 b^5 +  256 a^4 (97 + 41 b)
    \\ &
    + 16 a^3 (11227 + 8380 b + 1568 b^2) +   8 a^2 (77803 + 79882 b + 27312 b^2 + 3136 b^3)
    \\ &
    + 2 a (519261 + 669842 b + 319528 b^2 + 67040 b^3 +  5248 b^4)) u^2\Big)
    \\[4pt] & 
    +  29360128 \,\omega t^2 \Big(-2 (1 + 2 b) (3 + 2 b) (5 + 2 b) (55 + 12 a +     8 b) t^8 \\ & 
      + (3 + 2 b) (18425 + 256 a^3 + 15040 b +   3928 b^2 + 320 b^3 + 48 a^2 (73 + 24 b) + 8 a (1835 + 1080 b + 152 b^2)) t^4 u
      \\ &
      -  2 (35175 + 256 a^4 + 45433 b + 20696 b^2 + 3920 b^3 + 
          256 b^4 + 80 a^3 (49 + 20 b) +  8 a^2 (2587 + 1878 b + 336 b^2) 
          \\ &
         + a (45433 + 45844 b + 15024 b^2 + 1600 b^3)) u^2 \Big)
    \\[4pt] &
    +  37748736 \,\omega^2 t^3 \Big(2 (1 + 2 b) (3 + 2 b) (5 + 2 b) t^8
      - (3 +   2 b) (845 + 48 a^2 + 440 b + 56 b^2
      \\ &
      + 16 a (27 + 8 b)) t^4 u +  2 (48 a^3 + 16 a^2 (33 + 13 b) + a (1719 + 1224 b + 208 b^2)
      + 3 (560 + 573 b + 176 b^2 + 16 b^3)) u^2 \Big)
    \\ &
          + 41943040 \,\omega^3 t^4 u ((3 + 2 b) (17 + 4 a + 4 b) t^4 - (69 +    8 a^2 + 52 b + 8 b^2 + 4 a (13 + 5 b)) u)
    \\ &     
         +4194304 \,\omega^4 t^5 u (-((3 + 2 b) t^4) + 2 (2 + a + b) u)
    \end{align*}

    \begin{align*}
      C^{(12)}_{1, 4}  &=
            16777216\, t \Big(-8 (1 + 2 b) (3 + 2 b) (5 + 2 b) (7 + 2 b) t^{12}
            \\ & 
            +  2 (3 + 2 b) (5 + 2 b) (10087 + 288 a^2 + 3808 b + 352 b^2 + 
            36 a (97 + 20 b)) t^8 u
            \\ &
            - \Big(1854405 + 640 a^4 + 2016874 b +  790832 b^2 + 132128 b^3 + 7936 b^4 + 128 a^3 (211 + 80 b)
              \\ &
              +  8 a^2 (38323 + 24972 b + 3888 b^2) + 4 a (327517 + 287944 b + 81216 b^2 + 7360 b^3)\Big) t^4 u^2
            \\ &
            +  2 (863625 + 2624 a^4 + 846129 b + 302780 b^2 + 46832 b^3 +  2624 b^4 + 16 a^3 (2927 + 880 b)
            \\ &
            +  4 a^2 (75695 + 41572 b + 5728 b^2) + a (846129 + 652072 b + 166288 b^2 + 14080 b^3)) u^3\Big)
            \\[4pt] &
          +  29360128 \,\omega t^2 u \Big(-4 (3 + 2 b) (5 + 2 b) (221 + 36 a + 
            40 b) t^8 + \big(112305 + 256 a^3 + 96292 b \\&  + 26208 b^2 +  2240 b^3           
            + 48 a^2 (163 + 60 b) + 8 a (7105 + 4488 b + 672 b^2)\big) t^4 u \\ &
            - (111195 + 1600 a^3 +  85108 b + 20752 b^2 + 1600 b^3
            \\ &
            + 16 a^2 (1297 + 380 b) +  a (85108 + 45920 b + 6080 b^2)) u^2 \Big)
          \\[4pt] &
      +     37748736 \,\omega^2 t^3 u (12 (3 + 2 b) (5 + 2 b) t^8 - (2955 + 
          48 a^2 + 1784 b + 256 b^2 +  16 a (57 + 20 b)) t^4 u
          \\ &
          + (3115 + 208 a^2 + 1672 b +   208 b^2 + 8 a (209 + 60 b)) u^2)
          \\ &
      + 41943040 \,\omega^3 t^4 u^2 (2 (16 + 2 a + 5 b) t^4 -  5 (7 + 2 a + 2 b) u)
      -4194304 \,\omega^4\, t^5 (t^4 - u) u^2
 \end{align*}

    \begin{align*}
      C^{(12)}_{1, 5}  &=
    -  67108864\, t u \Big(16 (105 + 142 b + 60 b^2 + 8 b^3) t^{12} - (5 + 2 b) (22827 + 288 a^2 + 9632 b + 976 b^2
      \\ &
      +     324 a (17 + 4 b)) t^8 u +  2 (219555 + 512 a^3 + 149017 b + 32728 b^2 + 2320 b^3 +  48 a^2 (278 + 75 b)
      \\ &
      + a (99124 + 48296 b + 5728 b^2)) t^4 u^2 - (343245 +  2816 a^3 + 212232 b + 42832 b^2 + 2816 b^3
      \\ &
      + 16 a^2 (2677 + 624 b) +  24 a (8843 + 3844 b + 416 b^2)) u^3\Big)
    \\[4pt] &
        + 234881024 \,\omega t^2 u^2 \Big(-9 (5 + 2 b) (37 + 4 a + 8 b) t^8 + 
       2 (4080 + 72 a^2 + 1951 b + 224 b^2 \\ & + 
          12 a (99 + 26 b)) t^4 u - (6825 + 304 a^2 + 2936 b + 
          304 b^2 + 8 a (367 + 84 b)) u^2 \Big)
    \\[4pt] &
      + 301989888 \,\omega^2 t^3 u^2 \Big(3 (5 + 2 b) t^8 - (117 + 16 a + 
      30 b) t^4 u + 2 (53 + 12 a + 12 b) u^2 \Big)
    \\[4pt] &
      + 167772160 \,\omega^3 t^4 (t^4 - u) u^3
    \end{align*}

    \begin{align*}
      C^{(12)}_{1, 6}  &=  
     134217728\, t u^2 \Big(-24 (35 + 24 b + 4 b^2) t^{12} + 3 (8603 + 32 a^2 + 3936 b + 432 b^2 +  12 a (107 + 28 b)) t^8 u
     \\ &
     - (76167 + 1200 a^2 + 29896 b +      2864 b^2 + 8 a (2517 + 532 b)) t^4 u^2  \\ & +  4 (13083 + 416 a^2 + 4708 b + 416 b^2 + 
       a (4708 + 896 b)) u^3\Big)
   \\[4pt] &
         + 469762048 \,\omega t^2 u^3 \Big(
        -(223 + 12 a + 56 b) t^8 + (809 +  104 a + 168 b) t^4 u
        - 4 (149 + 28 a + 28 b) u^2\Big)
         \\[4pt] &
      + 603979776 \,\omega^2 t^3 u^3 (t^8 - 5 t^4 u + 4 u^2) 
    \end{align*}

    \begin{align*}
      C^{(12)}_{1, 7}  &=
       2147483648\, t u^3 \Big(-2 (7 + 2 b) t^{12} + 
       2 (9 a + 26 (5 + b)) t^8 u - (641 + 76 a + 112 b) t^4 u^2 + 
       2 (199 + 32 a + 32 b) u^3\Big)
      \\[4pt] &
      -7516192768 \,\omega t^2 u^4 (t^8 - 3 t^4 u + 2 u^2)
    \end{align*}

    \begin{align*}
      C^{(12)}_{1,  8}  &=  -2147483648 t (t^4 - u) u^4 (t^8 - 12 t^4 u + 16 u^2)
    \end{align*}

    \begin{align*}
      C^{(12)}_{2, 1}  &=  
      3670016\, t^2\, (1 + 2 a) (1 + 2 b) (3 + 2 a + 2 b) (3 + 4 a + 4 b) (5 + 4 a + 4 b) (7 + 4 a + 4 b)
      \\ &
      -  11010048\,\omega t^3\, (1 + 2 a) (1 + 2 b) (3 + 2 a + 2 b) (5 + 4 a + 4 b) (7 + 4 a + 4 b) 
      \\ &
      +  5898240\,\omega^2\, (1 +  2 b)  \Big(2 (27 + 16 a^3 + 28 b + 8 b^2
      + 32 a^2 (2 + b) +  a (79 + 72 b + 16 b^2)) t^4 - 3 u\Big)
      \\ &
      +  2621440\,\omega^3 t\, (1 +  2 b)  (-2 (6 + 11 a + 4 a^2 + 2 b + 4 a b) t^4 +  3 u)
            \\ & 
      +  196608\,\omega^4 t^2 \, (1 + 2 b) ((5 + 4 a) t^4 - 3 u)
    \end{align*}

    \begin{align*}
      C^{(12)}_{2, 2}  &=  
             7340032\, t^2\, \Big(-((3 + 8 b + 4 b^2) (4725 + 256 a^3 + 3380 b +  816 b^2 + 64 b^3 + 48 a^2 (43 + 12 b)
        \\ & 
        + 8 a (685 + 360 b + 48 b^2)) t^4) +  2 (16695 + 128 a^5 + 39699 b + 34082 b^2 + 13144 b^3
        \\ &
        +   2272 b^4 + 128 b^5 + 32 a^4 (71 + 48 b) 
        +  8 a^3 (1643 + 1844 b + 496 b^2) \\ &
        +  a^2 (34082 + 51232 b + 24960 b^2 + 3968 b^3)
      +  2 a (20007 + 37757 b + 25616 b^2 + 7376 b^3 +    768 b^4)) u \Big)
      \\[4pt] &
      +  22020096 \,\omega t^3 \Big((3 + 8 b + 4 b^2) (375 + 48 a^2 + 152 b +  16 b^2 + 16 a (17 + 4 b)) t^4
        \\ &
        -  2 (1470 + 32 a^4 + 2921 b + 1894 b^2 + 448 b^3 + 32 b^4 +  32 a^3 (14 + 9 b) + 2 a^2 (947 + 1016 b + 256 b^2)
        \\ &
        +   a (3026 + 4418 b + 2032 b^2 + 288 b^3)) u \Big)
      \\[4pt] &
      +     11796480 \,\omega^2 \Big(-8 (6 + 2 a + b) (3 + 8 b + 4 b^2) t^8 + 
       2 (225 + 16 a^3 + 352 b + 160 b^2 + 16 b^3 \\ & + 32 a^2 (5 + 3 b) +
       a (403 + 408 b + 96 b^2)) t^4 u - 3 u^2\Big)
    \\ & 
    + 5242880 \,\omega^3 t \Big(2 (3 + 8 b + 4 b^2) t^8 - 
       2 (15 + 4 a^2 + 14 b + 4 b^2 + a (23 + 12 b)) t^4 u + 
       3 u^2\big]
    \\[4pt] &
    + 393216 \,\omega^4 t^2 ((5 + 4 a) t^4 - 3 u) u
    \end{align*}

    \begin{align*}
      C^{(12)}_{2, 3}  &=  
     29360128 t^2 \Big(2 (1 + 2 b) (3 + 2 b) (5 + 2 b) (55 + 12 a +  8 b) t^8
      - (3 + 2 b) (18425 \\ & + 256 a^3 + 15040 b + 
      3928 b^2 + 320 b^3 + 48 a^2 (73 + 24 b) +  8 a (1835 + 1080 b + 152 b^2)) t^4 u  \\ &
      +  2 (35175 + 256 a^4 + 45433 b + 20696 b^2 + 3920 b^3 +  256 b^4 + 80 a^3 (49 + 20 b)
      \\ &
      +  8 a^2 (2587 + 1878 b + 336 b^2) +   a (45433 + 45844 b + 15024 b^2 + 1600 b^3))
      u^2\Big)
    \\[4pt] &
    + 88080384 \,\omega t^3 \Big(-2 (1 + 2 b) (3 + 2 b) (5 + 2 b) t^8 + (3 +  2 b) (845 + 48 a^2 + 440 b + 56 b^2 \\ &
      + 16 a (27 + 8 b)) t^4 u -  2 (48 a^3 + 16 a^2 (33 + 13 b) + a (1719 + 1224 b + 208 b^2)
      \\ &
      +     3 (560 + 573 b + 176 b^2 + 16 b^3)) u^2\Big)
    \\[4pt] &      
      +  188743680 \,\omega^2 t^4 u \Big(
        -(3 + 2 b) (17 + 4 a + 4 b) t^4 + (69 + 8 a^2 + 52 b + 8 b^2 + 
      4 a (13 + 5 b)) u\Big)
    \\[4pt] &
    + 41943040 \,\omega^3 t^5 u ((3 + 2 b) t^4 - 2 (2 + a + b) u) 
    \end{align*}

    \begin{align*}
      C^{(12)}_{2, 4}  &= 
        29360128 t^2 u \Big(4 (221 + 36 a + 40 b) (15 + 16 b +  4 b^2) t^8
    - (112305 + 256 a^3 + 96292 b + 26208 b^2 \\ & +  2240 b^3
    + 48 a^2 (163 + 60 b) +  8 a (7105 + 4488 b + 672 b^2)) t^4 u
          + (111195 + 1600 a^3 \\ & +  85108 b + 20752 b^2 + 1600 b^3
          + 16 a^2 (1297 + 380 b) +  a (85108 + 45920 b + 6080 b^2)) u^2\Big)
        \\[4pt] &
      + 88080384 \,\omega t^3 u \Big(-12 (3 + 2 b) (5 + 2 b) t^8
      + (2955 +   48 a^2 + 1784 b + 256 b^2 +  16 a (57 + 20 b)) t^4 u
      \\ & 
      - (3115 + 208 a^2 + 1672 b +  208 b^2 + 8 a (209 + 60 b)) u^2\Big)
      \\[4pt] &          
      + 188743680 \,\omega^2 t^4 u^2 \Big(-2 (16 + 2 a + 5 b) t^4
        +  5 (7 + 2 a + 2 b) u\Big)
      + 41943040 \,\omega^3 t^5 (t^4 - u) u^2
    \end{align*}

    \begin{align*}
      C^{(12)}_{2,  5}  &=
      234881024 t^2 u^2 \Big(9 (5 + 2 b) (37 + 4 a + 8 b) t^8 - 
       2 (4080 + 72 a^2 + 1951 b + 224 b^2  \\ & + 
          12 a (99 + 26 b)) t^4 u + (6825 + 304 a^2 + 2936 b + 
          304 b^2 + 8 a (367 + 84 b)) u^2 \Big)
      \\[4pt] &
      +  704643072 \,\omega t^3 u^2 \Big(-3 (5 + 2 b) t^8 + (117 + 16 a + 
        30 b) t^4 u - 2 (53 + 12 a + 12 b) u^2\Big)
      \\[4pt] &      
      -754974720 \,\omega^2 t^4 (t^4 - u) u^3
    \end{align*}

    \begin{align*}
      C^{(12)}_{2,  6}  &=
      469762048 t^2 u^3 ((223 + 12 a + 56 b) t^8 - (809 + 104 a + 
      168 b) t^4 u + 4 (149 + 28 a + 28 b) u^2)
      \\[4pt] &
      -1409286144 \,\omega t^3 u^3 (t^8 - 5 t^4 u + 4 u^2)
    \end{align*}

    \begin{align*}
      C^{(12)}_{2, 7}  &=  
      7516192768 t^2 (t^4 - 2 u) (t^4 - u) u^4
    \end{align*}

    \begin{align*}
      C^{(12)}_{3, 1}  &=   
       7340032 (1 + 2 a) (1 + 2 b) (3 + 2 a + 2 b) (5 + 4 a + 4 b) (7 + 4 a + 4 b) t^3
      \\[4pt] &
      +  9175040\,\omega\, (1 +  2 b)  (-2 (27 + 16 a^3 + 28 b + 8 b^2 + 32 a^2 (2 + b) + 
      a (79 + 72 b + 16 b^2)) t^4 + 3 u)
      \\[4pt] &
      + 7864320\,\omega^2 t\, (1 + 2 b)  (2 (6 + 11 a + 4 a^2 + 2 b + 4 a b) t^4 - 3 u)
      \\[4pt] &
      +  1310720\,\omega^3 t^2 (1 + 2 b)  (-((5 + 4 a) t^4) + 3 u)
      + 131072\,\omega^4 t^3\, (1 + 2 b)  (t^4 - u)  
    \end{align*}

    \begin{align*}
      C^{(12)}_{3,  2}  &=
      - 14680064 t^3 \Big((3 + 8 b + 4 b^2) (375 + 48 a^2 + 152 b + 16 b^2 +  16 a (17 + 4 b)) t^4
      \\ &
      - 2 (1470 + 32 a^4 + 2921 b + 1894 b^2 + 448 b^3 + 32 b^4 +  32 a^3 (14 + 9 b) \\ &
      + 2 a^2 (947 + 1016 b + 256 b^2) + 
          a (3026 + 4418 b + 2032 b^2 + 288 b^3)) u\Big)
      \\[4pt] &
    +     18350080 \,\omega \Big(8 (6 + 2 a + b) (3 + 8 b + 4 b^2) t^8 \\ & 
      - 2 (225 + 16 a^3 + 352 b + 160 b^2 + 16 b^3 + 32 a^2 (5 + 3 b)
      + a (403 + 408 b + 96 b^2)) t^4 u + 3 u^2\Big)
      \\[4pt] &
      + 15728640 \,\omega^2 t (-2 (3 + 8 b + 4 b^2) t^8 + 
    2 (15 + 4 a^2 + 14 b + 4 b^2 + a (23 + 12 b)) t^4 u - 3 u^2)
    \\[4pt] &
    -2621440 \,\omega^3 t^2 ((5 + 4 a) t^4 - 3 u) u
    +  262144 \,\omega^4 t^3 (t^4 - u) u
    \end{align*}

    \begin{align*}
      C^{(12)}_{3, 3}  &=  
        58720256 t^3 \Big(2 (1 + 2 b) (3 + 2 b) (5 + 2 b) t^8
      - (3 +  2 b) \big(845 + 48 a^2 + 440 b + 56 b^2 +  16 a (27 + 8 b)\big) t^4 u
   \\ & 
   +  2 (48 a^3 + 16 a^2 (33 + 13 b) + a (1719 + 1224 b + 208 b^2)
   +  3 (560 + 573 b + 176 b^2 + 16 b^3)) u^2\Big)
      \\[4pt] & 
      + 293601280 \,\omega t^4 u \Big((3 + 2 b) (17 + 4 a + 4 b) t^4
      - (69 +  8 a^2 + 52 b + 8 b^2 + 4 a (13 + 5 b)) u\Big)
      \\[4pt] &
      + 125829120 \,\omega^2 t^5 u (-(3 + 2 b) t^4 + 2 (2 + a + b) u) 
    \end{align*}

    \begin{align*}
      C^{(12)}_{3,  4}  &=
     58720256 t^3 u \Big(12 (15 + 16 b + 4 b^2) t^8 - (2955 + 48 a^2 + 
          1784 b + 256 b^2 + 16 a (57 + 20 b)) t^4 u \\ & + (3115 + 
          208 a^2 + 1672 b + 208 b^2 + 8 a (209 + 60 b)) u^2\Big)
     \\[4pt] &
      + 293601280 \,\omega t^4 u^2 (2 (16 + 2 a + 5 b) t^4  -  5 (7 + 2 a + 2 b) u)
%      \\[4pt] &
      - 125829120 \,\omega^2 t^5 (t^4 - u) u^2
    \end{align*}

    \begin{align*}
      C^{(12)}_{3, 5}  &=  
        469762048 t^3 u^2 (3 (5 + 2 b) t^8 - (117 + 16 a + 30 b) t^4 u + 
      2 (53 + 12 a + 12 b) u^2)
      \\[4pt] &
      +1174405120 \,\omega t^4 (t^4 - u) u^3
    \end{align*}

    \begin{align*}
      C^{(12)}_{3, 6}  &=  
      939524096 t^3 (t^4 - 4 u) (t^4 - u) u^3
    \end{align*}

    \begin{align*}
      C^{(12)}_{4, 0}  &=        8192 \,\omega^4 ((1 + 2 b) t^4 - 2 (1 + a + b) u)
    \end{align*}

    \begin{align*}
      C^{(12)}_{4, 1}  &=
       4587520 (1 + 2 b) (2 (27 + 16 a^3 + 28 b + 8 b^2 + 32 a^2 (2 + b)
      +    a (79 + 72 b + 16 b^2)) t^4 - 3 u)
      \\ &
      +  9175040 (1 +  2 b) \,\omega t (-2 (6 + 11 a + 4 a^2 + 2 b + 4 a b) t^4 +  3 u)
      \\ &
      +2949120 (1 + 2 b) \,\omega^2 t^2 ((5 + 4 a) t^4 - 3 u)
      \\ &
      +  655360 (1 + 2 b) \,\omega^3 t^3 (-t^4 + u) 
      +  32768 \,\omega^4\, (t^4 - u) u
    \end{align*}

    \begin{align*}
      C^{(12)}_{4, 2}  &=  
       9175040 \Big(-8 (6 + 2 a + b) (3 + 8 b + 4 b^2) t^8 \\ & + 
       2 (225 + 16 a^3 + 352 b + 160 b^2 + 16 b^3 + 32 a^2 (5 + 3 b) +
       a (403 + 408 b + 96 b^2)) t^4 u - 3 u^2\Big)
       \\[4pt] &
       + 18350080 \,\omega t (2 (3 + 8 b + 4 b^2) t^8 - 
       2 (15 + 4 a^2 + 14 b + 4 b^2 + a (23 + 12 b)) t^4 u +  3 u^2)
       \\[4pt] &
       + 5898240 \,\omega^2 t^2 ((5 + 4 a) t^4 - 3 u) u
      - 1310720 \,\omega^3 t^3 (t^4 - u) u
    \end{align*}

    \begin{align*}
      C^{(12)}_{4, 3}  &=
      - 146800640 t^4 u ((3 + 2 b) (17 + 4 a + 4 b) t^4 - (69 + 8 a^2 + 
      52 b + 8 b^2 + 4 a (13 + 5 b)) u) \\[4pt] &
      + 146800640 \,\omega t^5 u ((3 + 2 b) t^4 - 2 (2 + a + b) u) 
    \end{align*}

    \begin{align*}
      C^{(12)}_{4, 4}  &=  
      -146800640 t^4 u^2 (2 (16 + 2 a + 5 b) t^4 -     5 (7 + 2 a + 2 b) u)
      +146800640 \,\omega t^5 (t^4 - u) u^2
    \end{align*}

    \begin{align*}
      C^{(12)}_{4,  5}  &=  -587202560 t^4 (t^4 - u) u^3
    \end{align*}

    \begin{align*}
      C^{(12)}_{5, 0}  &=  
      32768 \,\omega^3 (-(1 + 2 b) t^4 + 2 (1 + a + b) u)
    \end{align*}

    \begin{align*}
      C^{(12)}_{5, 1}  &=  
      3670016 (1 + 2 b) t (2 (6 + 11 a + 4 a^2 + 2 b + 4 a b) t^4 - 3 u)
      \\[4pt] &
      +  2752512 (1 + 2 b) \,\omega t^2 (-((5 + 4 a) t^4) + 3 u)
      + 1179648 (1 + 2 b) \,\omega^2 t^3 (t^4 - u) \\ & - 131072 \,\omega^3 (t^4 - u) u
    \end{align*}

    \begin{align*}
      C^{(12)}_{5,  2}  &=
       7340032 (-2 (3 + 8 b + 4 b^2) t^9 + 2 (15 + 4 a^2 + 14 b + 4 b^2
      + a (23 + 12 b)) t^5 u - 3 t u^2)\\[4pt] &
      - 5505024 \,\omega t^2 ((5 + 4 a) t^4 - 3 u) u
      + 2359296 \,\omega^2 t^3 (t^4 - u) u
      \\ & 
    \end{align*}

    \begin{align*}
      C^{(12)}_{5, 3}  &=  
      58720256 t^5 u (-(3 + 2 b) t^4 + 2 (2 + a + b) u)
    \end{align*}

    \begin{align*}
      C^{(12)}_{5, 4}  &=  -58720256 t^5 (t^4 - u) u^2
    \end{align*}

    \begin{align*}
      C^{(12)}_{6, 0}  &=  
      49152 \,\omega^2 ((1 + 2 b) t^4 - 2 (1 + a + b) u)
    \end{align*}

    \begin{align*}
      C^{(12)}_{6, 1}  &=  
      917504 (1 + 2 b) t^2 ((5 + 4 a) t^4 - 3 u)
      - 917504 \,\omega (1 + 2 b)  t^3 (t^4 - u)
      + 196608 \,\omega^2 (t^4 - u) u
    \end{align*}

    \begin{align*}
      C^{(12)}_{6, 2}  &=  
   1835008 t^2 ((5 + 4 a) t^4 - 3 u) u - 
   1835008 \,\omega t^3 (t^4 - u) u
    \end{align*}

    \begin{align*}
      C^{(12)}_{7, 0}  &=    32768 \,\omega (-(1 + 2 b) t^4 + 2 (1 + a + b) u)
      \end{align*}

    \begin{align*}
      C^{(12)}_{7, 1}  &=  
      262144 (1 + 2 b) t^3 (t^4 - u) - 131072 \,\omega (t^4 - u) u
    \end{align*}

    \begin{align*}
      C^{(12)}_{7,2}  &=  524288 t^3 (t^4 - u) u
    \end{align*}

    \begin{align*}
      C^{(12)}_{8, 0}  &=  
      8192 ((1 + 2 b) t^4 - 2 (1 + a + b) u)
    \end{align*}

    \begin{align*}
      C^{(12)}_{8, 1}  &=        32768 (t^4 - u) u
    \end{align*}

\end{document}